\def\br{{\bf r}}

\def\kb{{\bf k}}
\def\qb{{\bf q}}

\def\rb{{\bf r}}

\def\Gb{{\bf G}}

\def\Z3{Z_1^3}

\documentstyle[aps,prb]{revtex}
\include{epsf}
\begin{document}
\draft
\twocolumn[\hsize\textwidth\columnwidth\hsize\csname
@twocolumnfalse\endcsname

\newcommand{\y}{\'{\i}}

\title{Theory of inelastic lifetimes of low-energy electrons in metals}
\author{P. M. Echenique$^{1,4}$, J. M. Pitarke$^{2,4}$, E. V.
Chulkov$^{1,4}$,
and A.
Rubio$^3$}
\address{$^1$Materialen Fisika Saila, Kimika Fakultatea, Euskal Herriko
Unibertsitatea,\\  1072 Posta kutxatila, 20080 Donostia, Basque Country,
Spain\\
$^2$Materia Kondentsatuaren Fisika Saila, Zientzi Fakultatea, Euskal
Herriko
Unibertsitatea,
\\ 644 Posta kutxatila, 48080 Bilbo, Basque Country, Spain\\
$^3$Departamento de F\'\i sica Te\'{o}rica, Universidad de
Valladolid, Valladolid 47011, Spain\\
$^4$Donostia International Physics Center (DIPC) and Centro Mixto
CSIC-UPV/EHU}

\date\today

\maketitle

\begin{abstract}
Electron dynamics in the bulk and at the surface of solid materials are
well known to play a key role in a variety of physical and chemical phenomena. In
this article we describe the main aspects of the interaction of low-energy
electrons with solids, and report extensive calculations of inelastic lifetimes
of both low-energy electrons in bulk materials and image-potential states at
metal surfaces. New calculations of inelastic lifetimes in a homogeneous
electron gas are presented, by using various well-known representations of the
electronic response of the medium. Band-structure calculations, which have been
recently carried out by the authors and collaborators, are reviewed, and future work
is addressed. 
\end{abstract}
]

\tableofcontents
 
\section{Introduction}

Over the years, electron scattering processes in the bulk and at the
surface of solid materials have been the subject of a great variety 
of experimental and theoretical investigations.\cite{Ritchie1,Ritchie2,Petek1} 
Electron inelastic mean free paths (IMFP) and attenuation lengths 
have been shown to play a key role in photoelectron spectroscopy and 
quantitative surface analysis.\cite{Kanter,Powell1,Sernelius1} Linewidths
of bulk excited electron states in metals have also been measured, with
the use of inverse photoelectron
spectroscopy.\cite{Pendry,Plummer0,Levinson,Goldmann,Himpsel0,Ortega}
More recently, with the advent of time-resolved two-photon photoemission
(TR-2PPE)\cite{Bokor,Haight} and  ultrafast laser technology, time domain 
measurements of the lifetimes of photoexcited electrons with energies below 
the vacuum level have been performed.
In these experiments, the lifetimes of both hot electrons in bulk
materials\cite{exp1,Knoesel,Aes96,Petek0,exp4,exp6,Goldm,Cao,Reuter98,Ludeke93,Aes98,exp5,Aes98.2,Petek2,Hertel}
and image-potential states at metal 
surfaces\cite{Hertel,Wolf,Knoesel0,Lingle,Hofer,Harris,Shumay} have been probed.

These new and powerful experimental techniques, based on high resolution
direct and inverse photoemission as well as time-resolved 
measurements, have addressed aspects related to the lifetime of excited electrons and
have raised many fundamental questions. The ultrafast laser technology has
allowed to probe fast events at surfaces in real time and, therefore, extract
information about elementary electronic processes (with time scales
from pico to femtoseconds) that are relevant for potential technological
applications. In general, the two-photon photoemission spectroscopy is 
sensitive to changes of geometries, local work functions, and surface 
potentials during layer formation. The interaction of excited electrons 
and the underlying substrate governs the cross-section and branching ratios 
of all electronically induced adsorbate reactions at surfaces, such as 
dissociation or desorption, and influences the reactivity of the surfaces 
as well as the kinetics of growth.\cite{Plummer} Hot-electron lifetimes have
long been invoked to give valuable information about these processes.

Inelastic lifetimes of excited electrons with energies larger than $\sim
1\,{\rm eV}$ above the Fermi level can be attributed to
electron-electron (e-e) inelastic scattering, other processes such as
electron-phonon and electron-imperfection interactions being, in general, of
minor importance.\cite{note1} A self-consistent calculation of the interaction 
of low-energy electrons with an electron gas was first carried out by Quinn and
Ferrell.\cite{QF} They performed a self-energy calculation of e-e scattering
rates near the Fermi surface, and derived a formula for the inelastic lifetime
of hot electrons that is  exact in the high-density limit. These
free-electron-gas (FEG) calculations  were extended by Ritchie\cite{Ritchie59}
and Quinn\cite{Quinn62} to include,  within the first-Born and random-phase
approximations, energies away from  the Fermi surface, and by Adler\cite{Adler}
and Quinn\cite{Quinn63} to take account of the effects of the presence of a
periodic lattice and, in particular, the effect of virtual interband
transitions.\cite{Quinn63} Since  then, several FEG calculations of e-e
scattering rates have been  performed, with inclusion of exchange and
correlation (XC) effects,\cite{Ashley,Kleinman,Penn0,kk,Penn1} chemical
potential renormalization,\cite{Lundqvist,Shelton} plasmon
damping,\cite{Ashley0} and  core polarizability.\cite{Tung2} In the case of
free-electron materials,  such as aluminum, valence electrons were described
within the FEG model and atomic generalized oscillator strengths were used for
inner-shell ionization.\cite{Tung2,Tung1} For the description of the IMFP in
non-free-electron metals, Krolikowski and Spicer\cite{Spicer} employed a 
semiempirical approach to calculate the energy dependence of the IMFP 
from the knowledge of density-of-state distributions, which had been
deduced from photoelectron energy-distribution measurements. Tung {\it
et al}\,\cite{Tung3} used a statistical approximation, assuming that the
inelastic scattering of an electron in a given volume element of the solid 
can be represented by the scattering appropriate to a FEG with the electron 
density in that volume element. This approximation was found to predict 
IMFPs for electrons in Al that are in good agreement with predictions from 
an electron gas model plus atomic inner-shell contributions, and these
authors\cite{Tung3} went further to evaluate IMFPs and energy losses in
various noble and transition metals. Later on, new methods were
proposed\cite{Ashley1,Leckey,Liljequist,Salvat,Penn2} for
calculating the IMFP, which were based on a model dielectric function
whose form was motivated by the use of optical data. Though high-energy 
electron mean free paths now seem to be well understood,\cite{Tanuma,Ding,Powell2} 
in the low-energy domain electrons are  more sensitive to the details 
of the band structure of the solid, and a treatment of the electron dynamics
that fully includes band structure effects is necessary for quantitative
comparisons with experimentally determined attenuation lengths and relaxation 
times. {\it Ab initio} calculations of these quantities in which both the 
electronic Bloch states of the probe electron and the dielectric response 
function of the medium are described from first
principles have been performed only very recently.\cite{Igor0,Igor}

The self-energy formalism first introduced by Quinn and Ferrell for the
description of the lifetime of hot electrons in a homogeneous electron 
gas was extended by Echenique {\it et al}\,\cite{Echenique2,Echenique3,Uranga} 
to quantitatively evaluate the lifetime of image-potential
states\cite{Rundgren,Echenique0,Smith,Himpsel,Borstel,Echenique90,Fauster,Osgood}
at metal surfaces. Echenique {\it et al}\,\cite{Echenique2,Echenique3,Uranga}
used hydrogenic-like states to describe the  image-state
wave functions, they introduced a step model potential to calculate the bulk 
final-state wave functions, and used simplified free-electron-gas models to
approximate the screened Coulomb interaction. A three band model was used by Gao
and Lundqvist\cite{Gao} to describe the band structure of the (111) surfaces of
copper and nickel.  They calculated, in terms of Auger transitions, the decay of
the first  image state on these surfaces to the $n=0$ crystal-induced surface
state,  neglecting screening effects. Self-consistent calculations
of the linewidths of image states on copper surfaces have been reported
recently,\cite{Chulkov1,Osma,Silkin,Sarria} and good agreement with
experimentally determined decay times has been found. These calculations 
were performed by going beyond a free-electron description of the metal 
surface. Single-particle wave functions were obtained by solving the
Schr\"odinger equation with a realistic one-dimensional model
potential,\cite{Chulkov2} and the screened interaction was evaluated 
in the random-phase approximation (RPA).

This review includes an overview of inelastic lifetimes of low-energy
electrons in the bulk and at the surface of solid materials, as derived within
the first-Born approximation or, equivalently, linear response theory. In the
framework of linear response theory, the inelastic energy broadening or
lifetime-width of probe particles interacting with matter is found to be
proportional to the square of the probe charge. Extensions that include the
quadratic response to external perturbations have been discussed by various 
authors,\cite{Sung,Zaremba,Esbensen,Pitarke1,Pitarke2,Bergara,Wang} in order
to give account of the existing dependence of the energy loss and 
the IMFP on the sign of the projectile charge.\cite{Barkas,Andersen}

Section II is devoted to the study of electron scattering processes in a 
homogeneous electron gas, employing various representations of the electronic
response of the medium. In section III, a general self-energy formulation 
appropriate for the  description of inhomogeneous many-body systems is
introduced. This formulation is applied in sections IV and V to review
theoretical investigations  of lifetimes of both hot electrons in bulk
materials and image-potential states at metal surfaces. Future work is
addressed in Section VI.

Unless otherwise is stated, atomic units are used throughout, i.e.,
$e^2=\hbar=m_e=1$. The atomic unit of length is the Bohr radius,
$a_0=\hbar^2/m_e^2=0.529{\rm\AA}$, the atomic unit of
energy is the Hartree, $1\,{\rm Hartree}=e^2/a_0=27.2\,{\rm eV}$, and
the atomic unit of velocity is the Bohr velocity, $v_0=\alpha\,
c=2.19\times 10^8{\rm cm\,s^{-1}}$,
$\alpha$ and $c$ being the fine structure constant and the velocity of 
light, respectively.

\section{Scattering theory approach}

We take a homogeneous system of interacting electrons, and consider an excited
electron interacting through individual collisions with electrons in the Fermi
sea. Hence, we calculate  the probability $P_{i,i'}^{f,f'}$ per unit time
corresponding  to the process by which the probe particle is scattered from a 
state $\phi_i(\rb)$ of energy $E_i$ to some other state 
$\phi_f(\rb)$ of energy $E_f$, by carrying one electron of the
Fermi sea from an initial state $\phi_{i'}(\br)$ of energy 
$E_{i'}$ to a final state $\phi_{f'}(\rb)$ of energy $E_{f'}$, 
according to a dynamic screened interaction $W(\rb-\rb';E_{i}-E_{f})$ 
(see Fig. 1). By using the 'golden rule' of time-dependent perturbation 
theory and keeping only terms of first order in the screened interaction, 
one writes:\cite{Shiff}
\begin{eqnarray}\label{eqone}
P_{i,i'}^{f,f'}=&&2\pi\left|\left[W(\rb-\rb';E_{i}-E_{f})\right]_{i,i'}^{f,f'}
\right|^2\cr\cr
&&\times\delta(E_i-E_f+E_i'-E_f'),
\end{eqnarray}
where
\begin{eqnarray}
&&\left[W(\rb-\rb';\omega)\right]_{i,i'}^{f,f'}=\cr\cr
&&\int{\rm d}\rb\int{\rm
d}\rb'\phi_i^*(\rb)\phi_{i'}^*(\rb')
W(\rb-\rb';\omega)\phi_f(\rb)\phi_{f'}(\rb ').
\end{eqnarray}

Using plane waves for all initial and final states,
\begin{equation}\label{eq36}
\phi_\kb(\rb)={1\over\sqrt\Omega}{\rm e}^{{\rm i}\kb\cdot\rb},
\end{equation}
with energy $\omega_\kb=k^2/2$ and $\Omega$ being the normalization 
volume, one finds
\begin{eqnarray}\label{eq37}
P_{i,i'}^{f,f'}=&&{2\pi\over\Omega^2}\,|W_{{\bf k}_i-{\bf
k}_f,\omega_{\kb_i}-\omega_{\kb_f}}|^2\,\delta_{\kb_i-\kb_f-\kb_f'+\kb_i'}\cr\cr
&&\times\delta(\omega_{\kb_i}-\omega_{\kb_f}-\omega_{\kb_f'}+\omega_{\kb_i'}),
\end{eqnarray}
where $W_{\qb,\omega}$ represents the Fourier transform of the screened interaction
$W(\rb-\rb';\omega)$. The Kroenecker delta and the  Dirac delta function on the
right-hand side of Eq. (\ref{eq37}) allow for wave-vector and energy conservation,
respectively.

By summing the probabilities $P_{i,i'}^{f,f'}$ of Eq. (\ref{eq37}) over
all possible states $\kb_i'$ ($k_i'<q_F$, $q_F$ being the Fermi momentum), 
$\kb_f'$ ($k_f'>q_F$) and $\kb_f$, and noting that each allowed
$\kb_i'$ leads to two one-electron states (one for each spin), the 
total scattering rate of the probe electron in the state
$\kb_i$ is found to be given by the following expression:
\begin{eqnarray}\label{eq38}
\tau^{-1}=&&{4\pi\over\Omega^2}\,\sum_\qb{'}\sum_{\kb_i'}|
W_{\qb,\omega}|^2\,n_{ {\bf
k}_i'}(1-n_{\kb_i'+\qb})\cr\cr
&&\times\delta(\omega-\omega_{\kb_i'+\qb}+\omega_{\kb_i'}),
\end{eqnarray}
where
\begin{equation}\label{oc}
n_{\bf k}=\theta(q_F-|{\bf k}|)
\end{equation}
represents the occupation number. We have set the energy transfer
$\omega=\omega_{\kb_i}-\omega_{\kb_i-\qb}$, and the
prime in the summation indicates that the momentum transfer
is subject to the condition $0<\omega<\omega_{\kb_i}-E_F$ ($E_F$ is 
the Fermi energy), accounting for the fact that the probe electron
cannot make transitions to occupied states in the Fermi sea.

With the interaction $W_{\qb,\omega}$ described by the bare Coulomb
interaction, that is, $W_{\qb,\omega}=v_\qb$, the summation over $\qb$ in Eq.
(\ref{eq38}) would be severely divergent, thereby resulting in an infinite 
damping rate. Instead, we assume that the Coulomb interaction is 
dynamically screened,
\begin{equation}\label{eq39}
W_{\qb,\omega}=\epsilon^{-1}_{\qb,\omega}\,v_\qb,
\end{equation}
where $\epsilon_{\qb,\omega}$ is taken to be the dielectric function of
the medium.\cite{Echenique1,Pines}

For $\omega>0$, the
imaginary part of the RPA dielectric function\cite{Lindhard,Pines2} is simply a measure
of the number of states available  for real transitions involving a given momentum
transfer $\qb$ and energy transfer $\omega$:
\begin{eqnarray}\label{eq42}
{\rm
Im}\,\epsilon_{\qb,\omega}^{RPA}=2\pi\,\Omega^{-1}\,v_\qb\,\sum_\kb&&\,n_\kb(1-n_{
\kb+\qb})\cr\cr
&&\times\delta(\omega-\omega_{\kb+\qb}+\omega_\kb).
\end{eqnarray}
In the limit that the volume of the system $\Omega$ becomes infinite,
one can replace sums over states by integrals with the following relation
\begin{equation}
\sum_{\bf k}f({\bf k})\to{\Omega\over (2\pi)^3}\int{\rm d}{\bf
k}\,f({\bf k}),
\end{equation}
and after introduction of Eq. (\ref{eq42}) into Eq. (\ref{eq38}), one
finds
\begin{equation}
\tau^{-1}=2\,\int'{{\rm d}\qb\over (2\pi)^3}\,v_\qb\,{{\rm
Im}\,\epsilon_{\qb,\omega}^{RPA}\over\left|\epsilon_{\qb,\omega}\right|^2},
\end{equation}
where the prime in the integration indicates that the momentum transfer
$\qb$ is subject to the same condition as in Eq. (\ref{eq38}). With the screened
interaction $W_{\qb,\omega}$ of Eq. (\ref{eq39}) described within RPA, one
writes
\begin{equation}\label{eq46}
\tau^{-1}=2\,\int'{{\rm d}\qb\over (2\pi)^3}v_\qb\,{\rm
Im}\left[-\epsilon_{\qb,\omega}^{-1}\right],
\end{equation}
with $\epsilon_{\qb, \omega}$ being the RPA dielectric
function,\cite{Lindhard,Pines2} i.e.,
$\epsilon_{\qb,\omega}=\epsilon_{\qb,\omega}^{RPA}$.

In the more general scenario of many-body theory and within the first Born
approximation,\cite{Fetter} one finds the damping rate of an excited electron
in the state $\kb_i$ to also be given by Eq. (\ref{eq46}), but with the exact
inverse dielectric function $\epsilon_{\qb,\omega}^{-1}$, as defined in
Appendix A. This is the result obtained independently by Quinn and
Ferrell\cite{QF} and by Ritchie\cite{Ritchie59}. Quinn and
Ferrell\cite{QF} demonstrated, within a self-energy formalism, that the damping rate of
holes below the Fermi level is also given by Eq. (\ref{eq46}), with the energy
transfer
$\omega=\omega_{\kb_i-\qb}-\omega_{\kb_i}$ and with the prime in the
integration indicating that the momentum transfer $\qb$ is subject to the
condition $0<\omega<E_F-\omega_{\kb_i}$. For small values of
$|\omega_{\kb_i}-E_F|$, holes inside the Fermi sea ($\omega_{\kb_i}<E_F$) are
found to damp out in the same way as electrons outside ($\omega_{\kb_i}>E_F$), as
shown in Fig. 2.

If one is to go beyond RPA and introduce, through the factor 
$(1-G_{\qb,\omega})$ (see Appendix A), the reduction
in the e-e interaction due to the existence of a local XC hole around
electrons in
the Fermi sea, the dielectric function entering Eq. (\ref{eq46}) is
\begin{equation}\label{di1}
\epsilon_{\qb,\omega}=1+{\epsilon_{\qb,\omega}^{RPA}-1\over
1-G_{\qb,\omega}(\epsilon_{\qb,\omega}^{RPA}-1)},
\end{equation}
where $G_{\qb,\omega}$ is the so-called local-field
factor, first
introduced by Hubbard.\cite{Hubbard}

If one accounts, through the factor $(1-G_{\qb,\omega})$, for the existence of
a local XC hole around electrons in the Fermi sea and also around the probe
electron, the dielectric function entering Eq. (\ref{eq46}) is the so-called
test$\_$charge-electron dielectric function\cite{Kleinman68,HL} (see Appendix
A):
\begin{equation}\label{di2}
\epsilon_{\qb,\omega}=
\epsilon_{\qb,\omega}^{RPA}-G_{\qb,\omega}(\epsilon_{\qb,\omega}^{RPA}-1).
\end{equation}

Finally, we note that the inelastic mean free path (IMFP) is directly connected
to the lifetime
$\tau$ through the relation
\begin{equation}
\lambda = v\,\tau.
\end{equation}

\section{Self-energy formalism}

In the framework of many-body theory,\cite{Fetter} the damping rate of an 
electron with energy $\varepsilon_{i}>E_F$ is obtained from the
imaginary part of the electron self-energy:
\begin{equation}\label{eq92}
\tau^{-1}=-2\int{\rm d}{{\bf r}}\int{\rm d}{{\bf
r}'}\,\phi_{i}^*({\bf r}){\rm Im}\,\Sigma({\bf r},{\bf
r}';\varepsilon_i)
\phi_{i}({\bf r}'),
\end{equation}
where $\phi_{i}({\bf r})$ represents a suitably chosen one-electron orbital of energy
$\varepsilon_i$ (see Appendix B).

In the GW approximation,\cite{Hedin,GW} one considers only the first-order term 
in a series expansion of the self-energy in terms of the screened
interaction $W({\bf r},{\bf r'},\omega)$. This is
related to the density-response function $\chi({\bf r},{\bf
r}',\omega)$ of Eq. (\ref{eqa2}), as follows
\begin{eqnarray}\label{eq94}
W({\bf r},{\bf r}';\omega)=&&v({\bf r}-{\bf r}')+\int{\rm d}{{\bf
r}_1}\int{\rm d}{{\bf r}_2}\,v({\bf r}-{\bf r}_1)\cr\cr
&&\times\chi({\bf r}_1,{\bf r}_2,\omega)v({\bf r}_2-{\bf r}'),
\end{eqnarray}
where $v({\bf r}-{\bf r}')$ represents the bare Coulomb potential.

Within RPA, the
density-response function satisfies and integral equation (see Eq. (\ref{eq98})),
and is obtained from the knowledge of the density-response function of
noninteracting electrons. If, to the same order of approximation, one replaces the
exact one-particle Green function by its noninteracting counterpart, the imaginary
part of the self-energy can be evaluated explicitly:
\begin{equation}\label{eq97}
{\rm Im}\,\Sigma({\bf r},{\bf
r}';\varepsilon_i>E_F)=\sum_f{'}\phi_f^*({\bf r}')
{\rm Im}\,W({\bf r},{\bf r}';\omega)\phi_f({\bf r}),
\end{equation}
where $\omega=\varepsilon_i-\varepsilon_f$, and the prime in the summation
indicates that states $\phi_f(\rb)$ available for real transitions are subject
to the condition that $0<\omega<\varepsilon_i-E_F$. Introduction of
Eq. (\ref{eq97}) into Eq. (\ref{eq92}) yields
\begin{eqnarray}\label{eq97p}
\tau^{-1}=-2\sum_f{'}&&\int{\rm d}{{\bf r}}\int{\rm d}{{\bf
r}'}\,\phi_{i}^*({\bf r})\phi_{f}^*({\bf r}')\cr\cr
&&\times{\rm Im}\,W({\bf r},{\bf r}';\omega)
\phi_{i}({\bf r}')\phi_{f}^*({\bf r}).
\end{eqnarray}

In the so-called ${\rm GW\Gamma}$
approximation,\cite{Rice,Sernelius2,Mahan1,Mahan2} which includes XC effects not
present in the GW-RPA, the self-energy and damping rate of the excited electron
are of the GW form, i.e., they are given by Eqs. (\ref{eq97}) and (\ref{eq97p}),
respectively, but with an effective screened interaction
\begin{eqnarray}\label{eq101}
W({\bf r},{\bf r}';\omega)&=&v({\bf r}-{\bf r}')+\int{\rm d}\rb_1\int{\rm
d}\rb_2\left[v({\bf r}-{\bf r}_1)\right.\cr\cr &+&\left.K^{xc}({\bf r},{\bf r}_1)\right]
\chi({\bf r}_1,{\bf r}_2,\omega)v({\bf r}_2-{\bf r}'),
\end{eqnarray}
the density-response function now being given by Eq. (\ref{eq98p}). The kernel
$K^{xc}({\bf r},{\bf r}')$ entering Eqs. (\ref{eq101}) and (\ref{eq98p}) 
accounts for the reduction in the e-e interaction due to 
the existence of short-range XC effects associated to the probe
electron and to screening electrons, respectively.

\subsection{Homogeneous electron gas}

In the case of a homogeneous electron gas, single-particle wave
functions are simply plane waves, as defined in Eq. (\ref{eq36}). 
By introducing these orbitals into Eq. (\ref{eq97p}), the damping 
rate of an electron in the state $\kb_i$ is found 
to be given by Eq. (\ref{eq46}) with the dielectric function of either Eq.
(\ref{di1}) or Eq. (\ref{di2}), depending on weather the screened interaction of
Eq. (\ref{eq94}) or Eq. (\ref{eq101}) is taken in combination with the
density-response function of Eq. (\ref{eq98p}).\cite{note2} This is an expected
result, since these calculations have all  been performed to lowest order in the
screened interaction.

\subsection{Bounded electron gas}

In the case of a bounded electron gas that is translationally invariant
in the
plane of the surface, single-particle wave functions are of the form
\begin{equation}\label{n1}
\phi_{\kb_\parallel,i}(\rb)={1\over\sqrt A}\,\phi_i(z)\,{\rm e}^{{\rm
i}{\kb_\parallel}\cdot{\rb_\parallel}},
\end{equation}
with energies
\begin{equation}\label{n2}
\varepsilon_{\kb_\parallel,i}=\varepsilon_i+{\kb^2_\parallel\over 2},
\end{equation}
where the $z$-axis has been taken to be perpendicular to the surface.
Hence, the wave
functions $\phi_i(z)$ and energies $\varepsilon_i$ describe motion
normal to the
surface, $\kb_\parallel$ is a wave vector parallel to the surface, and
$A$ is the
normalization area.

Introduction of Eq. (\ref{n1}) into Eqs. (\ref{eq92}) and (\ref{eq97p}) yields
the following expressions for the damping rate of an electron in the
state $\phi_{\kb_\parallel,i}(\rb)$ with energy $\varepsilon_{\kb_\parallel,i}$:
\begin{eqnarray}\label{b0}
\tau^{-1}=-2\,\int&&{\rm d}{z}\int{\rm d}{z'}
\int{{\rm d}{\bf q}_\parallel\over(2\pi)^2}\phi_{i}^*(z)\cr\cr
&&\times{\rm Im}\Sigma(z,z';{\bf
q}_\parallel,\varepsilon_{\kb_\parallel,i})\phi_{i}(z')
\end{eqnarray}
and
\begin{eqnarray}\label{b1}
\tau^{-1}=-2\,\sum_f{'}\int&&{\rm d}{z}\int{\rm d}{z'}
\int{{\rm d}{\bf q}_\parallel\over(2\pi)^2}\phi_{i}^*(z)\phi_f^*(z')\cr\cr
&&\times{\rm Im}W(z,z';{\bf q}_\parallel,\omega)\phi_f(z)\phi_{i}(z'),
\end{eqnarray}
respectively, where
$\omega=\varepsilon_{\kb_\parallel,i}-\varepsilon_{\kb_\parallel-\qb_\parallel,f}$.
Here, $\Sigma(z,z';{\bf q}_\parallel,\omega)$ and $W(z,z';{\bf
q}_\parallel,\omega)$ represent the two-dimensional Fourier transforms of the
electron self-energy $\Sigma(\rb,\rb';\omega)$ and the screened interaction
$W(\rb,\rb';\omega)$.
 
\subsection{Periodic crystals}

For periodic crystals, single-particle wave functions are Bloch states
\begin{equation}
\phi_{\kb,i}(\rb)={1\over\sqrt\Omega}{\rm e}^{{\rm
i}\kb\cdot\rb}u_{\kb,i}(\rb),
\end{equation}
and one may introduce the following Fourier expansion of the screened
interaction:
\begin{eqnarray}\label{eq102}
W(\rb,\rb';\omega)=\Omega^{-1}&&\int_{BZ}{\rm
d}\qb\,\sum_{\Gb}\sum_{\Gb'}{\rm
e}^{{\rm i}(\qb+\Gb)\cdot\rb}\cr\cr
&&\times{\rm e}^{-{\rm i}(\qb+\Gb')\cdot\rb'}W_{\Gb,\Gb'}(\qb,\omega),
\end{eqnarray}
where the integration over $\qb$ is extended over the first Brillouin
zone (BZ), and the vectors $\Gb$ and $\Gb'$ are reciprocal lattice 
vectors. Introducing this Fourier representation into 
Eq. (\ref{eq97p}), one finds the following 
expression for the damping rate of an electron in the
state $\phi_{\kb,i}(\rb)$ with energy $\varepsilon_{\kb,i}$:
\begin{eqnarray}\label{eq103}
\tau^{-1}=-2\,\sum_f{'}
&&\int_{\rm BZ}{{\rm d}{\bf q}\over(2\pi)^3}\sum_{\bf G}\sum_{{\bf G}'}
B_{if}^*({\bf q}+{\bf G})
\cr\cr
&&\times B_{if}({\bf q}+{\bf G}'){\rm Im}\,W_{{\bf G},{\bf G}'}({\bf q},\omega),
\end{eqnarray}
or, equivalently,
\begin{eqnarray}\label{eq105}
\tau^{-1}={1\over \pi^2}\sum_f{'}
\int_{\rm BZ}{{\rm d}{\bf q}}\sum_{\bf G}\sum_{{\bf G}'}&&
{B_{if}^*({\bf q}+{\bf G})B_{if}({\bf q}+{\bf G}')\over
\left|{\bf q}+{\bf G}\right|^2}\cr\cr
&&\times{\rm Im}\left[-\epsilon_{{\bf G},{\bf G}'}^{-1}({\bf
q},\omega)\right],
\end{eqnarray}
where $\omega=\varepsilon_{\kb,i}-\varepsilon_{\kb-\qb,f}$,
and
\begin{equation}\label{eq104}
B_{if}({\bf q})=\int{\rm d}{\bf r}\,\phi_{\kb,i}^{\ast}({\bf r})\,{\rm
e}^{{\rm i}{\bf q}\cdot{\bf r}}\,\phi_{\kb-\qb,f}({\bf r}).
\end{equation}
$W_{{\bf G},{\bf G}'}({\bf q},\omega)$ are the Fourier coefficients 
of the screened interaction, and $\epsilon_{{\bf G},{\bf G}'}^{-1}({\bf
q},\omega)$ are the Fourier coefficients of the inverse dielectric function.

Within RPA, one writes
\begin{equation}\label{eq102pp}
\epsilon_{{\bf G},{\bf G}'}({\bf q},\omega)=\delta_{{\bf G},{\bf
G}'}-v_{{\bf G}}({\bf q})\,\chi^0_{{\bf G},{\bf G}'}({\bf q},\omega),
\end{equation}
where $v_{{\bf G}}({\bf q})$ represent the Fourier coefficients
of the bare Coulomb potential,
\begin{equation}
v_{{\bf G}}({\bf q})={4\pi\over|\qb+\Gb|^2},
\end{equation}
and $\chi_{{\bf G},{\bf G}'}^0({\bf q},\omega)$ are
the Fourier coefficients of the density-response function of noninteracting
electrons,
\begin{eqnarray}\label{eq103pp}
\chi_{{\bf G},{\bf G}'}^0({\bf q},\omega)=\Omega^{-1}&&\int_{BZ}{\rm d}\kb
\sum_{n}\sum_{n'}\cr\cr
&&\times{f_{{\bf k},n}-f_{{\bf k}+{\bf q},n'}\over\varepsilon_{{\bf
k},n}-\varepsilon_{{\bf k}+{\bf q},n'} +(\omega+{\rm i}\eta)}\cr\cr
&&\times\langle\phi_{{\bf k},n}|e^{-{\rm i}({\bf q}+{\bf G})\cdot{\bf
r}}|\phi_{{\bf k}+{\bf q},n'}\rangle\cr\cr
&&\times\langle\phi_{{\bf k}+{\bf q},n'}|e^{{\rm i}({\bf q}+{\bf
G}')\cdot{\bf
r}}|\phi_{{\bf k},n}\rangle,
\end{eqnarray}
where $\eta$ is a positive infinitesimal. The sums run over the band structure for each wave
wave vector
$\kb$ in the first BZ, and $f_{\kb,n}$ are Fermi factors
\begin{equation}
f_{\kb,n}=\theta(E_F-\varepsilon_{{\bf k},n}).
\end{equation}

Couplings of the wave vector $\qb+\Gb$ to wave vectors $\qb+\Gb'$ with
$\Gb\neq\Gb'$ appear as a consequence of the existence of electron-density
variations in real solids. If these terms, representing the  so-called
crystalline local-field effects, are neglected, one can write
\begin{eqnarray}\label{eq7p}
\tau^{-1}=\frac{1}{\pi^2}\sum_f{'}
&&\int_{\rm BZ}{{\rm d}{\bf q}}\sum_{\bf G}
{\left|B_{if}({\bf q}+{\bf G})\right|^2\over\left|{\bf q}+{\bf
G}\right|^2}\cr\cr
&&\times{{\rm Im}\left[\epsilon_{{\bf G},{\bf G}}({\bf q},\omega)\right]\over
|\epsilon_{{\bf G},{\bf G}}({\bf q},\omega)|^2}.
\end{eqnarray}
The imaginary part of $\epsilon_{{\bf G},{\bf G}}({\bf q},\omega)$
represents a measure of the number of states available for real
transitions involving a given momentum and energy transfer $\qb+\Gb$ 
and $\omega$, respectively, and the factor
$\left|\epsilon_{{\bf G},{\bf G}}({\bf q},\omega)\right|^{-2}$ accounts
for the screening in the interaction with the probe electron. Initial 
and final states of the probe electron enter through the coefficients 
$B_{if}({\bf q}+{\bf G})$.

If one further replaces in Eq. (\ref{eq7p}) the probe electron initial
and final states by plane waves, and the matrix coefficients 
$\epsilon_{{\bf G},{\bf G}}({\bf q},\omega)$ by
the dielectric function of a homogeneous electron gas,
\begin{equation}
\epsilon_{{\bf G},{\bf G}}({\bf q},\omega)\to\epsilon(|\qb+\Gb|,\omega),
\end{equation}
then Eq. (\ref{eq7p}) yields the damping rate of excited electrons in a
FEG, as given by Eq. (\ref{eq46}).

We note that the hot-electron decay in real solids depends on both the
wave vector ${\bf k}$ and the band index $i$ of the initial Bloch state. 
As a result of the symmetry of these states, one finds that 
$\tau^{-1}(S{\bf k},i)=\tau^{-1}({\bf k},i)$, with
$S$ representing a point group symmetry operation in the periodic
crystal. Hence, for each value of the hot-electron energy the scattering
rate $\tau^{-1}(E)$ is defined by averaging $\tau^{-1}({\bf k},i)$ over
all wave vectors ${\bf k}$ lying in the irreducible element of the 
Brillouin zone (IBZ), with the same energy, and also over the band 
structure for each wave vector.

\section{Lifetimes of hot electrons in metals}

\subsection{Jellium model}

Early calculations of inelastic lifetimes and mean free paths of excited
electrons in metals were based on the 'jellium' model of the solid. Within 
this model, valence electrons are described by a homogeneous assembly of 
electrons immersed in a uniform background of positive charge and volume 
$\Omega$. The only parameter in this model is the valence-electron density $n_0$,
which we represent in terms of the so-called electron-density parameter
$r_s$ defined by the relation $1/n_0=(4/3)\,\pi\,(r_s\,a_0)^3$, $a_0$ being the
Bohr radius. Hence, the damping rate of a hot electron of energy
$E=\omega_{\kb_i}$ is obtained, within this model, from Eq.
(\ref{eq46}) with $\omega=\omega_{\kb_i}-\omega_{\kb_i-\qb}$.

In the high-density limit ($r_s\to 0$), XC effects as well as high-order terms
in the expansion of the scattering probability in terms of the screened interaction
are negligible. Thus, in this limit the damping rate of hot electrons is obtained
from Eq. (\ref{eq46}) with use of the RPA dielectric function.

Now we focus on the scattering of hot electrons just above the Fermi level, i.
e., $E-E_F<<E_F$. As the energy transfer $\omega$ cannot exceed the value
$E-E_F$, the frequency entering ${\rm Im}\left[\epsilon_{\qb,\omega}^{-1}\right]$
is always small, one can take
\begin{equation}\label{eqomega}
{\rm Im}\left[-\epsilon^{-1}_{\qb,\omega}\right]={{\rm
Im}\left[\epsilon_{\qb,\omega}\right]\over|\epsilon_{\qb,0}|^2}\to{2\over
q^3}\,\epsilon_{\qb,0}^{-2}\,\omega,
\end{equation}
and the $(E-E_F)^2$ quadratic scaling of the hot-electron damping
rate is predicted. If one further replaces, within the high-density limit
($q_F\to\infty$), the static dielectric function $\epsilon_{\qb,0}$ by the
Thomas-Fermi approximation, and extends, at the same time, the maximum momentum
transfer ($q\sim 2\,q_F$) to infinity, then one finds
\begin{equation}\label{qf1}
\tau^{-1}= \frac{(\pi/q_F)^{3/2}}{16}\frac{(E - E_F)^2}{k_i}.
\end{equation}
If we replace $k_i\to q_F$ in Eq. (\ref{qf1}), then 
the damping rate of Quinn and Ferrell\cite{QF} is obtained, $\tau_{QF}^{-1}$, as
given by Eq. (\ref{eq112}). For the lifetime, one writes\cite{note3}
\begin{equation}\label{eq114}
\tau_{QF}=263\,r_s^{-5/2}\,(E-E_F)^{-2} {\rm fs}\,{\rm eV}^{2}.
\end{equation}

In Eq. (\ref{eqomega}), ${\rm Im}\,\epsilon_{\qb,\omega}$ represents a measure of
the number of states available for real transitions, whereas the denominator
$|\epsilon_{\qb,0}|^2$ accounts for the screening in the interaction
between the hot electron and the Fermi sea. Hence, the hot-electron lifetime is
determined by the competition between transitions and screening. Though
increasing the electron density makes the density of states (DOS) larger,
momentum and energy conservation prevents, in the case of a FEG, the sum over
available states from any dependence on $r_s$, as shown by Eq. (\ref{eqomega}). As
a result, the scattering rate of hot electrons in a FEG only depends on the
electron-density parameter through the screening and the
initial momentum $\kb_i$. High densities make the interaction weaker
[the integration of $|\epsilon_{\qb,\omega}|^{-2}$ scales, in the high-density
limit, as $q_F^{-3/2}$] and
momenta of excited electrons larger [$1/k_i\to q_F^{-1}$], which results in the 
$r_s^{-5/2}$ scaling described by Eq. (\ref{eq114}).

In Fig. 3 we represent the ratio $\tau/\tau_{QF}$,
versus $E-E_F$, for an electron density equal to that of valence
electrons in copper
($r_s=2.67$), as obtained from Eq. (\ref{eq46}) with the full RPA
dielectric function (solid line) and from Eq. (\ref{qf1}) (dashed line). Though
in the limit $E\to E_F$ the available phase space for real transitions is simply
$E-E_F$, which yields the $(E-E_F)^2$ quadratic scaling of Eqs. (\ref{qf1}) and
(\ref{eq114}), as the energy increases momentum and energy conservation prevents
the available phase space from being as large as $E-E_F$. Hence, the actual
lifetime departures from the $k_i/(E-E_F)^2$ scaling predicted
for electrons in the vicinity of the Fermi surface, differences between full RPA
calculations (solid line) and the results predicted by Eq. (\ref{qf1}) (dashed line)
ranging from $\sim 2\%$ at $E\approx E_F$ to $\sim 35\%$ at $E-E_F=5\,{\rm
eV}$. For comparison, also represented in this figure is the ratio
$\tau/\tau_{QF}$ obtained from the approximations of Eqs. (\ref{eq106}) (dotted line)
and (\ref{Quinn}) (dashed-dotted line).

The result of going beyond the RPA has been discussed  by various
authors.\cite{Ashley,Kleinman,Penn0,kk,Penn1,Lundqvist,Shelton,Ashley0}  In an
early paper, Ritchie and Ashley\cite{Ashley} investigated the  simplest exchange
process in the scattering between the probe electron  and the electron gas. Though
this exchange contribution to the e-e  scattering rate is of a higher order in
the electron-density parameter $r_s$ than the direct term, it
was found to  yield, for $r_s=2.07$ and $E\sim E_F$, a $\sim 70\%$
increase with respect to the RPA lifetime, and an even larger
increase in the case of metals with $r_s>2$. This reduction  of 
the e-e scattering rate appears as a consequence of the exclusion principle 
keeping two electrons of parallel spin away from the
same point, thereby reducing their  effective interaction.

Neither the effect of Coulomb correlations between the probe  electron
and the electron
gas, which also influence the e-e mutual interaction, nor XC effects
between pairs of
electrons within the Fermi sea were included by Ritchie and
Ashley.\cite{Ashley}
Kleinman\cite{Kleinman} included not only XC between the incoming
electron and an
electron from the Fermi sea but also XC between pairs  of electrons
within the Fermi
sea, and found a result which reduced the $\sim 70\%$  increase obtained
by Ritchie
and Ashley for Al to a $\sim 1\%$ increase. Alternative  approximations
for the XC
corrected e-e interaction were derived by Penn\cite{Penn0} and by
Kukkonen and
Overhauser.\cite{kk} From an evaluation of the test$\_$charge-electron
dielectric function of Eq. (\ref{di2}) and with use of a static local-field
factor, Penn\cite{Penn1} concluded that the introduction of exchange and
correlation
has little effect on the lifetime of hot electrons, in agreement with
early
calculations by Kleinman.\cite{Kleinman}

As we are interested in the low-frequency ($\omega\to 0$) behaviour of
the
electron gas, we can safely approximate the local-field factor by the
static limit,
$G_{\qb,0}$,  which we choose to be given by Eq. (\ref{eq32}). Our
results, as obtained
from Eq. (\ref{eq46}) with the dielectric function of either Eq. (\ref{di1}) or Eq.
(\ref{di2}) are presented in Figs. 4 and 5 by dashed and dotted lines,
respectively, as a function of $r_s$ for hot electrons with
$E-E_F=1\,{\rm eV}$ (Fig. 4), and as a function of $E-E_F$ with $r_s=2.67$ (Fig.
5). Solid lines represent RPA calculations, as obtained with the local-field
factor $G_{\qb,\omega}$ set equal to zero. We note from these figures that
local-field corrections in the screening reduce the lifetime of hot electrons in a
FEG with an electron density equal to that of valence ($4s^1$) electrons
in Cu
($r_s=2.67$) by $\sim20\%$. However, this reduction is slightly more than
compensated by the
large enhancement of the lifetime
produced by the existence of local-field corrections in the interaction
between
the probe electron and the electron gas. As a consequence, RPA
calculations (solid line) produce lifetimes that are shorter than more realistic
results obtained with full inclusion of XC effects (dotted line) by $\sim 5\%$.

Instead of calculating the damping rate $\tau^{-1}$ {\it
on-the-energy-shell} ($E=\omega_{\kb_i}$), Lundqvist\cite{Lundqvist} expanded the
electron self-energy in the deviation of the actual excitation energy $E$ from
the independent-particle result, showing that near the {\it energy-shell}
($E\sim\omega_{\kb_i}$) interactions renormalize the damping rate by the
so-called renormalization constant $Z_{\kb_i}$. Based on Lundqvist's
calculations, Shelton\cite{Shelton} derived IMFPs for various values of $r_s$ and
for electrons with energies between $E_F$ and
$\sim25\,E_F$. The
resulting IMFPs were larger than those obtained by Quinn\cite{Quinn62}
 by roughly $5-20\%$, depending  on $r_s$
and the electron energy.

In the case of excited electrons near the Fermi level the
renormalization
constant, as obtained within the GW-RPA, is nearly real and
$k$-independent. In the
metallic density range ($r_s\sim 2-6$) one finds $Z\sim 0.8-0.7$, and
the resulting lifetimes are, therefore, larger than those obtained
from Eq. (\ref{eq46}) by $\sim 20\%$.

\subsection{Statistical approximations}

In order to account for the inelastic scattering rates of non-free
electron materials, Tung {\it et al\,}\cite{Tung3} applied a statistical
approximation first developed by Lindhard {\it et al\,}.\cite{Lindhard2}
This approximation is based on the assumption that the inelastic electron
scattering of electrons in a small volume element ${\rm d}\rb$ at
$\rb$ is the same as that of electrons in a FEG with
density equal to the local density.

Within the statistical approximation of Ref.\onlinecite{Tung3}, for a
given density
distribution $n(\rb)$ one finds the total scattering rate $\tau^{-1}$
by averaging the corresponding local quantity
$\tau^{-1}\left[n(\rb)\right]$ over the volume $\Omega$ of the solid:
\begin{equation}\label{eq115}
\langle\tau^{-1}\rangle=\Omega^{-1}\int{\rm
d}\rb\,\tau^{-1}\left[n(\rb)\right].
\end{equation}
By calculating spherically symmetric electron density distributions
$n(r)$ in a
Wigner-Seitz cell,\cite{Ashcroft0} the total scattering rate is 
obtained from
\begin{equation}\label{eq117}
\langle\tau^{-1}\rangle=4\,\pi\,\Omega_{WS}^{-1}\int_0^{r_{WS}}{\rm
d}r\,r^2\,\tau^{-1}\left[n(r)\right],
\end{equation}
where $\Omega_{WS}$ and $r_{WS}$ represent the 
volume and the radius of the Wigner-Seitz sphere of the solid.

Alternatively, following the idea of using optical data in IMFP
calculations,\cite{Powell1} a number of approaches were
developed\cite{Ashley1,Leckey,Liljequist,Salvat} to compute a model
energy-loss
function ${\rm Im}\left[-\epsilon^{-1}_{\qb,\omega}\right]$ for real
solids and then
obtain inelastic scattering rates from Eq. (\ref{eq46}). In these
approaches the
model energy-loss function is set in the limit of zero wave vector equal
to the
imaginary part of the measured optical inverse dielectric
function,\cite{Palik} ${\rm
Im}\left[-1/\epsilon^{opt}_{\omega}\right]$, and it is then extended
into the non-zero
wave vector region by a physically motivated recipe.

Combining the statistical method of Ref.\onlinecite{Tung3} with 
the use of optical data, Penn\cite{Penn2} developed an improved 
algorithm to evaluate the dielectric function of the material. 
The Penn algorithm is based on a model dielectric function
in which the momentum dependence is determined by averaging the
energy-loss function of a FEG, ${\rm
Im}\left[-1/\epsilon^{FEG}_{\qb,\omega}\right]$, as follows
\begin{equation}\label{s6}
{\rm Im}\left[-\epsilon^{-1}_{\qb,\omega}\right]=\int_0^\infty{\rm
d}\omega_p\,G(\omega_p){\rm
Im}\left[1/\epsilon^{FEG}_{\qb,\omega}(\omega_p)\right],
\end{equation}
where
\begin{equation}
G(\omega)={2\over\pi\omega}{\rm Im}\left[-1/\epsilon^{op}_{\omega}\right].
\end{equation}

The Penn algorithm has been employed by Tanuma {\it et
al\,}\cite{Tanuma} to
calculate IMFPs for $50$ to $2000\,{\rm eV}$ in a variety of materials
comprising
elements, inorganic compounds, and organic compounds. Recently, several
other
groups\cite{Ding} have calculated IMFPs from optical data in a manner
similar to
that proposed by Penn,\cite{Penn2} with some differences in approach,
and
high-energy IMFPs now seem to be well understood\cite{Powell2}.

The effect of $d$-electrons in noble metals has been recently
investigated by Zarate {\it et al\,},\cite{Zarate} by including the $d$-band
contribution to the measured optical inverse dielectric function into a
FEG description of the $s-p$ part of the response.

In order to account for the actual DOS in real materials, early IMFP calculations
were carried out by Krolikowski and Spicer,\cite{Spicer} with the explicit
assumption that the matrix elements of the screened e-e interaction entering Eq.
(\ref{eq38}) are momentum independent. This so-called "random" $k$
approximation\cite{Berglund64,Kane67} has proved to be useful in cases
where the DOS plays a key role in the determination of scattering rates, as in
the case of ferromagnetic materials,\cite{Penn85,Drouhin97} thereby
allowing to explain the existence of spin-dependent hot-electron
lifetimes.\cite{exp5,Aes98.2} Although this method, due to its simplicity, cannot
provide full quantitative agreement with the experiment, it provides a useful
tool for the analysis of experimental data, thus allowing to isolate the effects
that are directly related to the DOS.
 
Figs. 6 and 7 show the lifetime versus energy, for representative free-electron-like
and
non-free-electron-like materials, Al (Fig. 6) and Cu (Fig. 7),
respectively.
First of all, we consider a relatively free-electron-like solid such as
Al (see Fig. 6). The contribution to the inelastic scattering of low-energy
electrons in Al coming
from the excitation of core electrons is negligible. Hence, statistical
approximations yield  results that nearly coincide with the FEG
calculation with $r_s=2.07$. However, the effective number of valence
electrons in Al is $3.1$ rather than $3$ (the actual number of valence
electrons), and
lifetimes calculated from the statistical model of
Ref.\onlinecite{Tung3} are,
therefore, slightly larger than those obtained within a FEG description. At higher
energies, new contributions to the inelastic scattering come from
the excitation of core electrons, and FEG lifetimes would, therefore, be much
longer than those obtained from the more realistic statistical approximations.

For non-free-electron-like materials such as Cu, the role of $d$ states
in the
electron relaxation process is of crucial importance, even in the case
of very-low-energy
electrons. The effective number of valence electrons in Cu that
contribute through the average of Eq. (\ref{eq117}), at
low electron energies, to the inelastic scattering ranges from $\sim
2.5$ far from
atomic positions to $\sim 7.5$ in a region where the binding energy is
already too
large. Since an enhanced electron density results in a stronger
screening and,
therefore, a longer lifetime (see, e.g., Eq. (\ref{eq114})),
the statistical approximation yields lifetimes that are longer than those
obtained within
a FEG model with the electron density equal to that of valence
($4s^1$) electrons
in Cu ($r_s=2.67$), but shorter than those obtained within a FEG
model with the
electron density equal to that of all $4s^1$ and $3d^{10}$ electrons in
Cu. We note
that the theory of Penn\cite{Penn2} gives shorter lifetimes than the
theory of Tung
{\it et al\,},\cite{Tung3} which is the result of spurious contributions
to the average energy-loss function of Eq. (\ref{s6}) from ${\rm
Im}\left[1/\epsilon_\omega^{opt}\right]$ at very low-frequencies.

For comparison with the 'universal' relationship
$\tau^{-1}=0.13\,(E-E_F)$ proposed by Goldmann {\it et al\,}\cite{Goldmann} for Cu, on
the basis of experimental angle-resolved inverse
photoemission spectra, lifetime-widths $\tau^{-1}$ of high-energy electrons in Cu
are represented in Fig. 8. Solid and dashed-dotted lines represent results obtained
from Eqs. (\ref{eq46}) and (\ref{eq117}). Dashed and dashed-dotted-dotted-dotted
lines represent the result of introducing into Eq. (\ref{eq46}) the model
energy-loss function of Eq. (\ref{s6}) with either the recipe described by Salvat
{\it et al\,} (dashed line) or with the measured optical response function taken
from Ref.\onlinecite{Palik} (dashed-dotted-dotted-dotted line), and the empirical
formula of Goldmann {\it et al\,}\cite{Goldmann} is represented by a dotted line.
We note from this figure that while at low electron energies
$\tau^{-1}$
increases quadratically with $E-E_F$, a combination of inner-shell and
plasmon
contributions results in lifetime-widths that approximately reproduce,
for electron
energies in the range $\sim 10-50\,{\rm eV}$ above the Fermi level, the empirical
prediction\cite{Goldmann} that the lifetime-width increases linearly
with increasing distance from $E_F$.

High-energy lifetime-widths and IMFPs seem to be well described by model
dielectric
functions, by assuming that the probe wave functions
are simply
plane waves. Nevertheless, in the case of low-energy electrons band
structure
effects are found to be important, even in the case of
free-electron-like metals
such as Al, and a a treatment of the electron dynamics that fully
includes band
structure effects is necessary for quantitative comparisons with the
experiment.

\subsection{First-principles calculations}

{\it Ab initio} calculations of the inelastic lifetime of hot electrons
in metals
have been carried out only very recently.\cite{Igor0,Igor} In this
work,\cite{Igor0,Igor} Bloch states were first expanded in a plane-wave basis,
and the Kohn-Sham equation of density-functional theory (DFT)\cite{Kohn1,Kohn2}
was then solved by invoking the local-density approximation (LDA). The
electron-ion interaction was described by means of a non-local, norm-conserving
ionic pseudopotential,\cite{Troullier} and the one-electron Bloch states were
then used to evaluate both the
$B_{if}$ coefficients and the dielectric matrix
$\epsilon_{\Gb,\Gb'}$ entering Eq. (\ref{eq105}).

First-principles calculations of the average lifetime $\tau(E)$ of hot electrons
in real Al,\cite{Igor0} as obtained from Eq. (\ref{eq105}) with full inclusion of
crystalline local field effects, are presented in Fig. 9 by solid circles. As an Al
crystal does not present strong electron-density gradients nor special
electron-density directions (bondings), contributions
from the so-called crystalline local-field effects are found to be
negligible. On the other hand, band-structure effects on the imaginary part of the
inverse dielectric matrix are approximately well described with the use of a
statistical approximation, as obtained from Eq. (\ref{eq117}) (dotted line),
thereby resulting in lifetimes that are just slightly larger than those of hot
electrons in a FEG with $r_s=2.07$ (solid line). Therefore, differences between
full {\it ab initio} calculations (solid circles) and FEG calculations (solid
line) are mainly due to the sensitivity of hot-electron initial and final wave
functions on the band structure of the crystal. When the hot-electron energy
is well above the Fermi level, these orbitals are very nearly plane-wave states
and the lifetime is well described by FEG calculations.
However, in the case of hot-electron energies near the Fermi level,
initial and
final states strongly depend on the actual band structure of the
crystal. Due to
the opening, at these energies, of interband transitions, band
structure effects
tend to decrease the inelastic lifetime by a factor that varies from
$\sim 0.65$
near the Fermi level ($E-E_F=1\,{\rm eV}$) to a factor of $\sim 0.75$
for
$E-E_F=3\,{\rm eV}$.

{\it Ab initio} calculations of the average lifetime $\tau(E)$ of hot electrons
in real Cu,\cite{Igor} the most widely studied
metal by TR-2PPE, are exhibited in Fig. 10 by solid circles, as obtained from Eq.
(\ref{eq105}) with full inclusion of crystalline local field effects and by
keeping all $4s^1$ and $3d^{10}$ Bloch states as valence
electrons in the pseudopotential generation. The lifetime of hot electrons in a
FEG with the electron density equal to that of valence ($4s^1$) electrons in Cu
($r_s=2.67$) is represented by a solid line, and the statistically averaged
lifetime, as obtained from Eq. (\ref{eq117}), is represented by a dotted line.
These calculations indicate that the lifetime of hot electrons in real Cu is,
within RPA, larger than that of electrons in a FEG with $r_s=2.67$, this
enhancement varying from a factor of
$\sim 2.5$ near the Fermi level ($E-E_F=1.0\,{\rm eV}$) to a factor of
$\sim 1.5$
for $E-E_F=3.5\,{\rm eV}$. {\it Ab initio} calculations of the lifetime
of hot electrons in Cu, obtained by just keeping the $4s^1$ Bloch states as
valence electrons in the pseudopotential generation, were also performed, and they
were found to nearly coincide with the FEG calculations. Hence, $d$-band states
play a key role in the hot-electron decay mechanism.

In order to address the various aspects of the role that localized
$d$-bands play on
the lifetime of hot electrons in Cu, now we neglect crystalline local-field
effects and present the result of evaluating hot-electron lifetimes from Eq.
(\ref{eq7p}). First, we replace hot-electron initial and final states
in $|B_{if}(\qb,\Gb)|^2$ by plane waves, and the dielectric function in
$\left|\epsilon_{\Gb,\Gb}(\qb,\omega)\right|^{-2}$ by that of a FEG with
$r_s=2.67$. If we further replaced ${\rm
Im}\left[\epsilon_{\Gb,\Gb}(\qb,\omega)\right]$ by that of a FEG, i.e., Eq.
(\ref{eq42}), then we would obtain the FEG calculation represented by a solid line.
Instead, Ogawa {\it et al\,}\cite{Petek0} included the effect of $d$-bands on
the lifetime by computing the actual number of states available for real
transitions, within Eq. (\ref{eq42}), from the band structure of Cu, and they
obtained a result that is for $E-E_F>2\,{\rm eV}$ well below the FEG
calculation.\cite{note4} However, if one takes into account, within a full
description of the band structure of the crystal in the evaluation of ${\rm
Im}\left[\epsilon_{\Gb,\Gb}(\qb,\omega)\right]$ (see Eqs. (\ref{eq102pp}) and
(\ref{eq103pp})), couplings between the states participating in real
transitions, then one obtains the result represented in Fig. 10 by open circles.
Since the states just below the Fermi level, which are available for real
transitions, are not those of free-electron states, localization results in
lifetimes of hot electrons with
$E-E_F<2\,{\rm eV}$ (open circles) that are slightly larger than
predicted within
the FEG model of the metal. At larger energies this band-structure
calculation (open
circles) predicts a lower lifetime than within the FEG model, due to
opening of the
$d$-band scattering channel dominating the DOS with energies from $\sim
2\,{\rm eV}$
below the Fermi level. Thus, this calculation shows at $E-E_F\sim
2\,{\rm eV}$ a
slight deviation from the quadratic scaling predicted within the FEG
model, in qualitative
agreement with experimentally determined decay times in Cu.

While the excitation of $d$ electrons diminishes the lifetime of hot
electrons with
energies $E-E_F>2\,{\rm eV}$, $d$ electrons also give rise to additional
screening,
thus increasing the lifetime of all hot electrons above the Fermi level. That this
is the case is obvious from the band-structure calculation exhibited by full
triangles in Fig. 10. This calculation is the result obtained from  Eq.
(\ref{eq7p}) by still replacing  hot-electron initial and final states in
$\left|B_{0f}({\bf q}+{\bf G})\right|^2$ by plane waves (plane-wave calculation) but
including the full band structure of the crystal in the evaluation of both
${\rm Im}\left[\epsilon_{{\bf G},{\bf G}}({\bf q},\omega)\right]$ and $\left|\epsilon_{{\bf
G},{\bf G}}({\bf q},\omega)\right|^{-2}$. The effect of virtual interband
transitions giving rise to additional screening is to increase, for hot-electron
energies under study, the lifetime by a factor of $\approx 3$, in qualitative
agreement with the approximate prediction of Quinn\cite{Quinn63} and
with the use of
the statistical average of Ref.\onlinecite{Tung3}.

Finally, band-structure effects on hot-electron energies
and wave functions are investigated. Full band-structure
calculations of Eq. (\ref{eq105})
with and without (see also Eq. (\ref{eq7p})) the inclusion of crystalline local field
corrections were carried out,\cite{Igor} and these corrections were found to be
negligible for
$E-E_F>1.5\,{\rm eV}$, while for energies very near the Fermi level neglection of
these corrections resulted in an overestimation of the lifetime of less than $5\%$.
Therefore, differences between the full (solid circles) and plane-wave (solid
triangles) band-structure calculations come from the sensitivity of hot-electron
initial and final states on the band structure of the crystal. When the
hot-electron energy is well above the Fermi level, these states are very nearly
plane-wave states for most of the orientations of the wave vector, and the
lifetime is well described by plane-wave calculations (solid circles and triangles
nearly coincide for $E-E_F>2.5\,{\rm eV}$). However, in the case of hot-electron
energies near the Fermi level initial and final states strongly depend on the
orientation of the wave vector and on the shape of the Fermi surface. For most
orientations, flattening of the Fermi surface tends to increase the hot-electron
decay rate,\cite{Adler} while the existence of the so-called necks on the Fermi
surface of noble metals results in very small scattering rates for a few
orientations of the wave vector. After averaging $\tau^{-1}({\bf
k},n)$ over all orientations, Fermi surface shape effects tend to decrease the
inelastic lifetime.

Scaled lifetimes, $\tau\times(E-E_F)^2$, of hot electrons in Cu are represented in
Fig. 11, as a function of $E-E_F$. Results obtained, within RPA, from Eqs. (\ref{eq46})
and (\ref{eq117}) are represented by solid and dashed-dotted lines, respectively, the
{\it ab initio} calculations of Ref.\onlinecite{Igor} are represented by solid circles,
and the dashed line represents the calculations described in Ref.\onlinecite{Zarate}.
These model calculations\cite{Zarate} show that above the $d$-band threshold, at
$\sim-2\,{\rm eV}$ relative to the Fermi level, $d$-band electrons can only
participate in the screening, thereby producing longer lifetimes, while at larger
energies lower lifetimes are expected, due to opening of the $d$-band scattering
channel that dominates the DOS with energies $\sim 2{\rm eV}$ below the Fermi
level. For comparison, the empirical formula proposed by Goldmann {\it et al\,}
is represented by a dotted line.
 
Scaled lifetimes of hot electrons in Cu, determined from a variety of
experiments,\cite{Aes96,Petek0,exp4,exp6} are represented in Fig. 12, as a
function of
$E-E_F$. The energy dependence of both {\it ab initio} calculations and experimentally
determined relaxation times shows deviations with respect to
the $(E-E_F)^{-2}$ scaling. On the other hand, though there are large
discrepancies among results obtained in different laboratories, most
experiments
give lifetimes that are considerably longer than predicted within a
free-electron
description of the metal, in agreement with first-principles
calculations. Measurements of hot-electron lifetimes have also been performed for
other noble
and transition metals,\cite{Knoesel,Aes96,Cao,Reuter98,Ludeke93}
simple metals,\cite{Aes98} ferromagnetic solids,\cite{exp5,Aes98.2} and high-$T_c$
superconductors.\cite{Petek2} (see Table \ref{tabla0})

\section{Lifetimes of image-potential states at metal surfaces}

\subsection{Concept and development of image states}

A metal surface generates electron states that do not exist in a bulk
metal.
These states can be classified into two groups, according to their charge
density localization relative to the surface atomic layer: intrinsic surface states and
image-potential states. The so-called intrinsic surface
states, predicted by Tamm\cite{tazp32} and Shockley,\cite{shpr39} are
localized mainly at the surface atomic layer. Image-potential states
\cite{Rundgren,Echenique0,Smith,Himpsel,Borstel,Echenique90,Fauster,Osgood}
appear in the vacuum region of metal surfaces with a band gap near the vacuum level,
as a result of the self-interaction of the electron with the polarization charge it
induces at the surface. Far from the surface, into the vacuum, this potential
well approaches the long-range classical image potential, $-1/4z$, $z$ being the
distance from the surface, and it gives rise
to a series of image-potential states localized outside the metal.

In a hydrogenic
model, with an infinitely high  repulsive surface barrier, these states form a 
Rydberg-like series with energies\cite{Echenique0}
\begin{equation}
E_n = \frac{-0.85 eV}{n^2},
\label{en}
\end{equation}
converging towards the vacuum level $E_v = 0$. The
corresponding eigenfunctions are given by the radial solutions of an
$s$-like state
of the hydrogen atom
\begin{equation}
\phi_n(z)\propto z\,R_n^{l=0}(z/4).
\label{varphi}
\end{equation}
The lifetime of these states scales asymptotically with the quantum
number $n$, as follows \cite{Echenique0}
\begin{equation}
\tau_n \propto n^3.
\label{tau}
\end{equation}

For a finite repulsive surface barrier, as is the case for real metal
surfaces, Eq. (\ref{en}) may be transformed into
\begin{equation}
E_n = \frac{-0.85 eV}{(n+a)^2},
\label{ena}
\end{equation}
where $a$ is a quantum defect depending on both the energy-gap
position and
width and also on the position of the image state relative to the
gap.\cite{Rundgren,Echenique0}

After demonstration of the resolubility of the image-state series on
metal surfaces,\cite{Echenique0} these states were found
experimentally.\cite{josmprb83,doalprl84,sthiprl84}
Binding energies of these states have been measured by inverse
photoemission (IPE),\cite{josmprb83,doalprl84,sthiprl84,hiorprb92}
two-photon photoemission (2PPE),\cite{gihaprl85,Fujimoto,scfiprb92,wafass97} and
time-resolved two-photon photoemission
(TR-2PPE).\cite{Hertel,Wolf,Knoesel0,Lingle,Hofer,Harris,Shumay}  These
measurements have provided  highly accurate data of image-state binding energies at
the surfaces of many  noble  and transition metals, as shown, e.g., in
Ref.\onlinecite{Fauster}. Along
with the measurements of  image-state energies, the dispersion of these
states has also been  measured, and it has been found that only on a few surfaces
such as Ag(100),  Ag(111), and Ni(111) the first image state is characterized by
an effective mass that exceeds the free-electron mass.\cite{Fauster} At the same
time, theoretical efforts have  been directed to create relatively simple models that
reproduce the experimentally observed binding energies and effective masses of image
states, and also to evaluate the image-plane position.
\cite{Smith,Himpsel,Borstel,Echenique90,wehuprl85,orecprb86,pelass86,lesuprb87,liwaprb89,smchprb89,faap94,justss97,chsitbp1}
First-principles calculations of image states have also been carried
out,\cite{Chulkov2,hujoprb86,neinel92,egheprl92,necrprl93,sichpss94,ligaprb94,heflprb98}
with various degrees of sophistication.

This intensive work on image states has resulted in an understanding of 
some of the key points of the physics of these states and of the relatively 
extensive data-base of their energies on noble and transition metals.
 
\subsection{Lifetimes of image states}

\subsubsection{Introduction}

In contrast to the relatively simple spectroscopic problem of
determining
the position of spectral features that directly reflect the density of
states
and which may be, in principle, calculated within a one-electron theory,
the study of spectral widths or lineshapes is essentially a many-body
problem.
\cite{Kevan} These spectral
widths appear as a result of electron-electron,
electron-defect, electron-phonon, and electron-photon
interactions,\cite{Kevan,wereap,mcbaprb95,thmaprb97,reshprl99} and they are also
influenced by phonon-phonon interactions.\cite{Knoesel0,chsitbp2} Accurate
and systematic measurements of the linewidth of image states on metal
surfaces were carried out with the use of 2PPE spectroscopy (for a review see,
e.g., Ref.\onlinecite{Fauster}). These experiments gave smaller values for the
image-state lifetime than the ones obtained in recent very-high resolution TR-2PPE
measurements.\cite{Hertel,Wolf,Knoesel0,Lingle,Hofer,Harris,Shumay}
The reason for this discrepancy is that the 2PPE linewidth contains not
only an energy
relaxation contribution (intrinsic lifetime), but also contributions tha arise
from phase-relaxation processes.\cite{reshprl99}

The first estimation\cite{Echenique0} of the lifetime of image states
used simple wave-function arguments, to show that the lifetime of image
states asymptotically increases with the quantum number $n$, as in Eq. (\ref{tau}).
Nearly twenty years later, this prediction was confirmed experimentally for the
(100) surface of Cu, for which lifetimes of the first six image states were measured
with the use of quantum-beat spectroscopy.\cite{Hofer}  The first quantitative evaluation
of the lifetime of image states, as obtained within  the self-energy formalism, was
reported in Ref.\onlinecite{Echenique2}. In this calculation hydrogenic-like states
were used, with no penetration into the solid, to describe the image-state wave
functions, a step model potential was introduced to calculate the  bulk final-state
wave functions, and a simplified  free-electron-gas (FEG) model was utilized to 
approximate the screened
Coulomb interaction.  More realistic wave functions, allowing
for penetration of the
electron into the crystal, were introduced in subsequent
calculations.\cite{Echenique3,Uranga} In these evaluations the linewidth of the
first image state at the
$\overline{\Gamma}$ point was shown to be directly proportional to the 
penetration, and the prediction of Eq. (\ref{tau}) was confirmed.

The penetration of an image state into the bulk is defined as
\begin{equation}
p_n=\int_{bulk}{\rm d}z\,\phi_n^*(z)\phi_n(z),
\label{pnint}
\end{equation}
thereby giving a measure of the coupling of this state to bulk electronic
states.
This coupling, weighted by the screened interaction, is
responsible for the
decay of image states through electron-hole pair creation.
Intuitively, it seems clear that the larger the penetration the stronger
the coupling and,
therefore, the smaller the lifetime. This idea was exploited to qualitatively
explain the linewidth of image
states on various surfaces,\cite{Fauster} and also the temperature-dependence of
the linewidth of the n=1 state on  Cu(111).\cite{Knoesel0}
In this heuristic approximation, the
linewidth of an image state is determined by
\begin{equation}
\Gamma(E_n) = p_n\Gamma_b(E_n),
\label{gam1}
\end{equation}
where $\Gamma_b(E_n)$ is the linewidth of a bulk state
corresponding to the energy $E_n$. The $\Gamma_b(E_n)$ value can be 
obtained either from first-principles
calculations or from the experiment.
In many angle-resolved photoemission experiments a linear dependence of
the linewidth of bulk $s-p$ and $d$ states is observed for energies in the range 5-50
eV above the Fermi level,
\cite{Pendry,Plummer0,Levinson,Goldmann,Himpsel0,Ortega}
\begin{equation}
\Gamma_b(E_n) = b\,(E_n-E_F),
\label{gam2}
\end{equation}
while the linewidth of bulk states in a FEG shows a $(E_n-E_F)^2$ quadratic scaling for
energies near the Fermi level, as discussed in section IV.A and in Appendix C. Image
states on noble and transition metal surfaces have energies in the range 4-5 eV above
the Fermi level, so that Eqs. (\ref{gam1}) and (\ref{gam2}) have been applied in
Ref.\onlinecite{Fauster}, for an estimate of 
the lifetime broadening, with use of the experimentally determined coefficient
$b=0.13$ for Cu and Ag.\cite{Goldmann} For Au one also uses $b=0.13$, and for
Ni and Fe $b$ is taken to be $0.18$\cite{Plummer0} and $0.6$,\cite{Himpsel0}
respectively.

Recent TR-2PPE measurements
have shown that the intrinsic linewidths of the n=1 image state on
Cu(111)\cite{Knoesel0} and Cu(100)\cite{Hofer,Shumay} are 30 meV and 16.5 meV,
respectively, while accurate model potential calculations\cite{Chulkov2} yield
penetrations $p_1=0.22$ and $p_1=0.05$, respectively. Accordingly, image-state
linewidths cannot be explained by simply applying Eq. (\ref{gam1}); instead,
contributions to the image-state decaying mechanism coming from either the evanescent
tails of bulk states outside the crystal or the existence of intrinsic surface
states must also be taken into account, together with an accurate description of
surface screening effects. Here we give the results obtained within a
theory that incorporates these effects and that has been used recently
to evaluate intrinsic linewidths or, equivalently, lifetimes of image states on metal
surfaces.\cite{Chulkov1,Osma,Silkin,Sarria}

\subsubsection{Model potential}

It is  well known\cite{Echenique0,Echenique90,hujoprb86} that
image-state wave functions lie mainly in the vacuum side of the
metal surface, the electron moving, therefore, in a region with
little potential
variation parallel to the surface. Hence, these wave functions can be described, with a
reasonable accuracy, by using a one-dimensional potential that reproduces the key
properties of image states, namely, the position and width of the energy gap
and, also,
the binding energies of both intrinsic and image-potential states at
the $\overline{\Gamma}$ point. Such a one-dimensional potential has recently been
proposed for a periodic-film model with large vacuum intervals between the solid
films:\cite{Chulkov2}
\begin{equation}
V(z)=\cases{
A_{10}+A_1\,\cos(2\pi z/a_s), & $z < D$\cr\cr
A_{20}+A_2\,\cos\left[\beta (z-D)\right], & $D<z<z_1$\cr\cr
A_3\,\exp\left[-\alpha(z-z_1)\right], & $z_1<z<z_{im}$\cr\cr
\displaystyle{\exp\left[-\lambda(z-z_{im})\right]-1\over 4(z-z_{im})}, &
$z_{im}<z$,\cr\cr} 
\label{Vz}
\end{equation}
where the $z$-axis is taken to be perpendicular to the surface.
$D$ is the halfwidth of the film, $a_s$ is the interlayer
spacing, $z_{im}$ represents the image-plane position, and the origin is chosen
in the middle of the film. This one-dimensional potential has
ten
parameters, $A_{10}$, $A_1$, $A_{20}$, $A_2$, $A_3$, $\alpha$, $\beta$, $z_1$,
$\lambda$, and $z_{im}$, but only four of them are independent. $A_{10}, A_1, A_2,$
and
$\beta$
are chosen as adjustable parameters, the other six parameters being
determined
from the requirement of continuity of the potential and its first
derivative
everywhere in space. The parameters $A_1$ and $A_{10}$ reproduce the
width and position
of the energy gap, while $A_2$ and $\beta$ reproduce experimental or
first-principles energies $E_0$ and $E_1$ of the $n=0$ $s$-$p$ like surface state
at the $\overline{\Gamma}$ point and
the $n=1$ image
state, respectively. This potential is shown schematically in
Fig. \ref{figpot}. To illustrate the
good quality of the image-state wave functions obtained with this model
potential,
we compare such wave functions with those obtained with the use of first-principles
calculations. Probability amplitudes
of the $n=1$ image state
on Li(110), as obtained from either a self-consistent pseudopotential
calculation or the one-dimensional model potential of Eq. (\ref{Vz}) are
represented in Fig. \ref{figpsi}, showing that the agreement between
the two curves is excellent. The probability amplitude of the $n=1$ image
state on Cu(100), as obtained from the one-dimensional model potential of Eq.
(\ref{Vz}), also shows very good agreement with the result obtained with the use of a
FLAPW calculation\cite{hujoprb86} (see Fig. \ref{figpsi}b).

Assuming that corrugation effects, i.e., effects associated with spatial
variations of the potential in the plane parallel to the surface, are not
important and that the three-dimensional potential
can be described by the $(x,y)$-plane average, one-electron
wave functions
and energies are taken to be given by Eqs. (\ref{n1}) and (\ref{n2}),
respectively. Within a many-body self-energy formalism, the linewidth of the $n$ image
state with energy
$\varepsilon_{\kb,n}$ is then obtained from Eq. (\ref{b0}) or, within either the
GW or the ${\rm GW\Gamma}$ approximation, from Eq. (\ref{b1}).

\subsubsection{Results and discussion}

First of all, we present results obtained with use of the one-dimensional model potential
(MP) of Eq. (\ref{Vz}), and compare with experimental and other theoretical results. A
summary of experimental results for image-state lifetimes in noble and transition metal
surfaces, as obtained from either 2PPE or TR-2PPE measurements, is presented in
Table \ref{tabla1}, together with the result of theoretical calculations at the
$\overline\Gamma$ point ($\kb_{\parallel}=0$). We note that there are large differences
between 2PPE and TR-2PPE experimental results for copper and silver surfaces, the
lifetime broadening derived from recent very-high resolution TR-2PPE measurements being
smaller than that obtained from 2PPE experiments by nearly a factor of $2$.  
 
Theoretical calculations presented in Table \ref{tabla1} can be
classified into two groups. First, there is the heuristic approximation of
Eqs. (\ref{gam1}) and (\ref{gam2}), which was carried out by
Fauster and Steinmann\cite{Fauster} for a variety of metal surfaces. This approach
results in a semiquantitative  agreement with 2PPE measurements, except for the (111)
surfaces of noble metals and the (100) surface of Ni. Similar computations were
performed in Ref.\onlinecite{Chulkov2} for the $n=1$ and $n=2$ image states on Cu
and Ag surfaces, with use of the penetration of the image-state wave function that
results from the one-dimensional model potential of Eq. (\ref{Vz}). Though an
accurate description of the penetration of the $n=1$ image-state wave function
yields better agreement, in the case of Cu(111), with the experiment, this
heuristic approach is still in semiquantitative agreement with TR-2PPE
measurements. A quantitative agreement was found for the $n=2$ image-state
linewidth in Cu(111).

In the other group of calculations the many-body self-energy
formalism described in Section III was used for the evaluation of the lifetime of image
states,\cite{Chulkov1,Osma,Silkin,Sarria} resulting in a quantitative agreement with
TR-2PPE measurements of the lifetime of image states on Cu surfaces and showing,
therefore, that the present state of the theory enables to accurately predict the
broadening of image states on metal surfaces.

To illustrate the importance that an accurate
description of the self-energy might have on the evaluation of the linewidth, we show in
Fig. \ref{figself} ${\rm Im}\left[-\Sigma(z,z';\kb_\parallel=0,E_1)\right]$ of the
$n=1$ image-state electron at the 
$\overline{\Gamma}$ point ($\kb_\parallel=0$) on the (111) and (100) surfaces of Cu.
${\rm Im}\left[-\Sigma(z,z';\kb_\parallel=0,E_1)\right]$ is represented in this
figure as a function of $z$ for a fixed value of $z'$ in the vacuum side of the
surface (upper panel), within the bulk (middle panel), and at the surface (lower
panel). It is obvious from this figure that the imaginary part of the self-energy
is highly nonlocal,\cite{degprb97} and strongly depends on the
$z$ and $z'$ coordinates. We note that ${\rm
Im}\left[-\Sigma(z,z';k_\parallel=0,E_1)\right]$ presents a maximum at $z=z'$ when
$z$ is located at the surface, and surface states can, therefore, play an important
role in the decay mechanism of image states. The magnitude of this maximum is
plotted, as a function of $z'$, in Fig. \ref{figmax}, showing that it is an
oscillating function of $z$ within the bulk,\cite{note5} and reaches its highest
value at the surface.

It is interesting to note from Fig. \ref{figmax}a that the magnitude of ${\rm
Im}\left[-\Sigma(z,z';\kb_\parallel=0,E_1)\right]$ is larger at the surface for
Cu(111) than for Cu(100). Though the (100) surface of Cu only presents an intrinsic
surface resonance, in the case of the (111) surface of Cu there is an intrinsic
surface state just below the Fermi level. This intrinsic surface state provides a
new channel for the decay of image states, thereby enhancing the imaginary part of
the self-energy and the linewidth. The role that the intrinsic surface state on
Cu(111) plays in the decaying mechanism of image states is obvious from Fig.
\ref{figmax}b, where contributions to the maximum self-energy coming from
transitions to the intrinsic surface state and from transitions to bulk states have
been plotted separately by dashed and dotted lines, respectively. The intrinsic
surface state provides a $\sim 75\%$ of the decay mechanism at the surface. The
intrinsic-surface-state contribution to the total linewidth of the $n=1$ image
state on Cu(111) was found to be of about
$40\%$.\cite{Chulkov1,Osma} Similarly, lower lying image states can give noticeable
contributions to the linewidth of excited, i.e., $n=2,3,...$ image states. 
For example, the  decay from the $n=2$ to the $n=1$ image state on Cu(100) yields a
linewidth of  0.5 meV, i.e. a
$10\%$ of the total $n=2$ image-state linewidth. The decay from the $n=3$ to the $n=1$
image state on Cu(100) yields a linewidth of 0.17 meV, and the decay from the $n=3$ to
the $n=2$ image state on Cu(100) yields a linewidth of 0.05 meV, i.e. 
$\sim 10-15\%$ of the total linewidth.

Coupling of image states with the crystal occurs through the penetration
of the image state wave function and, also, through the evanescent
tails of bulk and surface states outside the crystal. To illustrate this 
point, the linewidth $\Gamma=\tau^{-1}$ can be split as follows
\begin{equation}
\Gamma = \Gamma_{bulk} + \Gamma_{vac} + \Gamma_{inter},
\label{tauu}
\end{equation}
where $\Gamma_{bulk}$, $\Gamma_{vac}$, and $\Gamma_{inter}$ represent bulk, 
vacuum and interface contributions, respectively. They are obtained by 
confining the integrals in Eq. (\ref{b1}) to either
bulk ($z<0$, $z'<0$), vacuum ($z>0$, $z'>0$), or vacuum-bulk ($z<0,z'>0$;
$z>0, z'<0$) coordinates. These separate contributions 
to the linewidth of image states on Cu surfaces are shown in Table \ref{tabla2}. We
note that the contribution to 
$\Gamma$ coming from the interference term $\Gamma_{inter}$ is 
comparable in magnitude and opposite in sign to both
bulk and vacuum contributions. This is a consequence of the behaviour of
the imaginary part of the two-dimensional Fourier transform of the
self-energy, as discussed in Ref.\onlinecite{Osma}. The contributions $\Gamma_{vac}$ and
$\Gamma_{inter}$ almost completely compensate each other, and, in a
first approximation, the total linewidth $\Gamma$ can be represented by the bulk
contribution $\Gamma_{bulk}$ within an accuracy of $\sim 30\%$.

The linewidth of image states can vary as a function of the
two-dimensional momentum ${\bf k}_{\parallel}$. In
Fig. \ref{figstr} we show schematically the projection of the bulk band structure
onto the (111) and (100) surfaces of Cu. In the case of the (111) surface of Cu, the
$n=1$ image state becomes a resonance at $|{\bf k}_{\parallel}|\sim
0.114\,{\rm a_0^{-1}}$, thereby the image-state wave function presenting a larger
penetration into the bulk and an enhanced linewidth. Table \ref{tabla3} shows the
linewidth of the $n=1$ image state on Cu(111), as obtained for three different values of
${\bf k}_{\parallel}$ in the range $0-0.114\,{\rm a_0^{-1}}$.\cite{note6} In these
calculations two approaches for the $n=1$ image-state wave function  have been used.
First, it has been obtained as an eigenfunction of the model potential
of Eq. (\ref{Vz}), with the parameters chosen so as to reproduce the width and position
of the energy gap and the binding energies at the $\bar\Gamma$ point (${\bf
k}_\parallel=0$). Secondly, the $n=1$ image-state wave function has been obtained with
use of the model potential of Eq. (\ref{Vz}), but with the parameters chosen so as to
reproduce the width and position of the energy gap and the binding energies at the
corresponding values of ${\bf k}_{\parallel}$, thus allowing for the penetration of
the image-state wave function into the bulk to increase with ${\bf k}_{\parallel}$.
Though both approximations yield an image-state linewidth that increases with ${\bf
k}_{\parallel}$, it increases very slowly within the first approach and more rapidly
within the second approach, showing the key role that the penetration of the
image-state wave function plays in the decay mechanism, i.e., as the coupling of
the image-state wave function with bulk and intrinsic-surface states increases,
the image-state linewidth is enhanced.

As all theoretical calculations presented in Tables II, III and IV have been obtained
with use of the wave functions of Eq. (\ref{n1}), surface corrugation has not been
taken into account. Estimating the influence of surface corrugation on the
image-state broadening requires the use of three-dimensional wave functions for the
evaluation of initial and final states and, also, for the evaluation of the
screened interaction. Work along these lines is now in progress.

An approximate way of including surface-corrugation effects
on the final-state wave functions is to account for the actual effective mass of the
intrinsic-surface and all bulk states, as obtained from the theoretically or
experimentally determined dispersion of these states. In Table \ref{tabla4} we
compare the results of this calculation\cite{Osma,Sarria} with recent accurate
TR-2PPE measurements of the lifetime of image  states on Cu surfaces, showing very good
agreement between theory and experiment.

The present state of a theory that uses the self-energy formalism in combination with an
accurate description, within a quasi one-dimensional model, of the key aspects of image
states, namely, the position and width of the energy gap and the binding energies of the
intrinsic and image-potential surface states has been shown to give quantitative account of
the lifetime of image states on metal surfaces.  Moreover, this theory also gives a
linewidth of the Shockley surface state on Cu(111)\cite{chsitbp2} at the $\overline\Gamma$
point that is in excellent agreement with  recent very-high-resolution angle-resolved
photoemission measurements.\cite{mcbaprb95,thmaprb97}
The calculated inelastic linewidth has been found to be of $26$ meV, while measurements
give $30$ meV\cite{mcbaprb95} and $21\pm5$ meV.\cite{thmaprb97} This
calculation\cite{chsitbp2} emphasizes the  extremely important role that the intrinsic
surface state plays in the decaying mechanism of this state at the $\overline{\Gamma}$
point, resulting in a contribution of $\sim 70\%$ of the total linewidth. This
surface-state contribution explains the difference between the experimental
data\cite{mcbaprb95,thmaprb97} and theoretical results obtained within a bulk
description of the broadening mechanism.

If dielectric layers are grown on the metal 
substrate, one can analyze the layer growth by simply looking at the energetics and
lifetimes of image-state electrons.\cite{Padowitz} Deposition of an overlayer on a metal
substrate can change drastically the properties of image states, such as
binding energies, wave functions, and lifetimes. This change depends on weather the
adsorbate is a transition metal, an alkali metal, or a noble
gas atom. In particular, deposition of alkali metal adlayer
on Cu(111) decreases the work function by nearly a factor of 2.\cite{Fischer94} The
linewidths of image states on a single layer of  Na and K on Cu(111), Fe(110) and
Co(0001) were measured with 2PPE by  Fisher {\it et al\,}.\cite{Fischer93} A large
value of 150 meV was obtained for the first image state, which is quite close to the
linewidth of the $n=1$ image-state on Fe(110) and Co(0001) but
much larger than the linewidth of the $n=1$ image state on
Cu(111).\cite{Fauster} All these values were measured with 2PPE, and they include
both energy and phase relaxation contributions.  Additionally, these experiments
showed the presence of the $n=0$ intrinsic surface state generated by the Na/K
layer, which replaces the intrinsic surface state on Cu(111). More accurate
measurements of the intrinsic linewidth can be obtained with use of TR-2PPE
spectroscopy. Nevertheless, the influence of impurities and imperfections on the
linewidth remains to be evaluated. For more realistic estimates  of the intrinsic
linewidth, accurate  models and/or first-principles calculations are necessary. The
same  applies to other metal overlayers.\cite{Fischer96}.
        
Overlayers of Xe and Kr on Ag(111), Cu(111), and Ru(0001)
have been studied recently with use of 2PPE \cite{mejoss93,Neill96} and TR2PPE
\cite{Wolf,Knoesel0,Wolf97,Harris97,Berthold}
spectroscopies. All these measurements have shown that the lifetime of the
$n=1$ state increases significantly upon deposition of the noble atom
adlayer on all metal substrates of interest. Qualitatively, this increase
can be explained by the fact that the interaction of the image-state
electron with the closed Xe or Kr valence-shell is repulsive and,
therefore, the probability amplitude of this state moves away from
the crystal, as compared to the simple case of clean metal surfaces. Therefore, the
coupling  to the substrate decreases and the lifetime increases. The same qualitative
argument can also explain the decrease of the binding energy of image states
upon deposition of Xe or Kr on metal substrates. Moreover, Harris {\it et
al\,}\cite{Harris,Neill96} have studied the evolution of image states  as  a
function  of the number of deposited atomic layers of Xe on Ag(111). They have
found that with increase of a number of Xe layers the $n=2$ and $n=3$ image states
evolve  into quantum-well states of the overlayer. A qualitative interpretation of
this behaviour of image states has been given, within a macroscopic dielectric 
continuum model.\cite{Neill96} Unfortunately, no microscopic investigation of the
image-state evolution of adlayers on metal surfaces that takes into account the band
structure of both the substrate and the overlayer has been carried out.

Defects on the surface or adsorbed particles cause electron scattering processes that
lead to phase relaxation of the wave function. This can be monitored in real
time, in order to extract relevant information. In fact, measurements of the
$n=1$ and $n=2$ image-potential states of CO adsorbed in Cu(100)
indicate a decreasing dephasing time when the CO molecules form an
ordered c(2x2) structure on the surface.\cite{wereap,reshprl99}
Furthermore, measurements on Cu(100)\cite{wereap} have shown correlation between
decay and dephasing, on the one hand, and the existence of surface defects, on the other
hand. A first-principles description of this problem is still lacking, due to intrinsic
difficulties in dealing with the loss of two-dimensional translational symmetry.

Another important field of research is the understanding of the processes
leading to the electronic relaxation in magnetic materials.
The spin-split image states on magnetic 
surfaces\cite{Donath94,Passek95,Nekovee93,Rossi96} offer the possibility of extracting
information about the underlying surface magnetism.
These spin-split states can decay in different ways
and, therefore, their linewidths can be different.
In particular, spin-resolved inverse
photoemission experiments on Fe(110)\cite{Passek95} give
an intrinsic linewidth of 140 (70) meV
for the first minority (majority) image state.
The difference in the lifetime is of the order
of the total linewidth of the $n=1$ image state on other metal
surfaces. At the same time, as the spin-splitting is only of $\sim 8\%$ of the total
binding energy ($E_1=-0.73{\rm eV}$), it is unlikely that this splitting is
responsible for the large difference between linewidths. Hence, one has to resort to
details of the phase space of final states and to the screened Coulomb
interaction as responsible for this effect. Work along these lines is now in 
progress.

All these problems are of technological relevance and pose technical and
theoretical questions that need to be answered in order to make a correct
interpretation of what is really being measured.
One technique is based on the {\it ab initio} description of the
fast-dynamics of a wave-packet of excited electrons in front of the
surface.\cite{Jacinto} The time evolution will pick
up all the relevant information concerning scattering processes and 
electronic excitations that can be mapped directly
with experiments.\cite{Hofer}
On a more complex and fundamental level, there is the theoretical
description of coupled electron-ion dynamics, which is relevant in many experiments.

\section{Future}

We present here a brief summary of on-going and future work in the field of
inelastic electron scattering in solids and, in particular, in the investigation
of electron and hole inelastic lifetimes in bulk materials and low-dimensional
structures. The advance in our knowledge is closely linked to the experimental
developments that combine state-of-the-art angle-resolved 2PPE with ultrafast
laser technology. These investigations might be relevant for potential
technological applications, such as the control of chemical reactions in surfaces
and the developing of new materials for opto-electronic devices.

A theoretical and experimental challenge is the description 
of the reactivity at surfaces. Experiments are being performed nowadays directed to
get a deeper understanding of the electronic processes involved. We note that
electronic excitation is the initial step  in a chemical reaction, and the
energetics and lifetimes of these processes directly govern the reaction
probability. For example, we can achieve  chemical selectivity through a femtosecond
activation of the chemical  reaction.\cite{Diau} This shows clearly that nonrandom
dissociation  exists in polyatomic molecules on the femtosecond time-scale, by
exciting  the reactant to high energies (well above the threshold for
dissociation)  and sampling the products on time scales that are shorter than the
rate for  intramolecular vibrational energy-distribution (this concept is relevant 
in chemical reactivity and assumes ergodicity or, equivalently, that the  internal
energy is statistically redistributed). The idea of ergodicity  has to be revisited
in this short time-scale.

Very recently, it has been shown that selective adsorption of
low-energy electrons into an image-potential state, followed by
inelastic scattering and desorption, can provide information on the
interaction between these states and the substrate.\cite{Petaccia} A deep
theoretical analysis of this interaction, as well as the role of the
substrate/adsorbate band structure, is still lacking and is needed in order to
interpret the experimental data.

So far, we have concentrated our attention to the investigation of bulk and
surfaces, i.e., extended systems. From a technological point of view, and due to
the fast  miniaturization of the magneto- and opto-electronic components in current
devices, the study of the electron dynamics in nanostructures is of
relevance. For example, alkali metals that have image states as resonances
would have,  in a finite piece of material,
a well-defined image state with a long lifetime. These states are
spatially located outside the nanostructure and, at least in principle, could be used in
a possible self-assembling mechanism to build controlled structures made of
clusters, and also as an efficient external probe for chemical
characterization.  Measurements on negatively charged clusters would be able 
to assess this effect, as well as its size dependence. Experiments
performed on a Na$_{91}^-$ cluster have looked at two decay mechanisms for 
the collective excitation, namely, electron and atom emission. The estimated
electronic escape time is of the order of $1\,{\rm fs}$.\cite{Reiners}
The relaxation time for two-electron collisions in small sodium clusters
has been estimated theoretically at the level of a time-dependent
local-density-functional approach (TDLDA).\cite{Domps} The computed values are in the range
of
$3-50\,{\rm fs}$, which are between the direct electron emission and the ionic motion
($>100\,{\rm fs}$).  These values compete with the scale for
Landau damping (coupling of the collective excitation to
neighbouring particle-hole states). 
A first non-perturbative approach to the
quasiparticle lifetime in a quantum dot has been presented in Ref.\onlinecite{BLA},
where localized (quasiparticle states are single-particle-like states)
and delocalized (superposition of states) regimes are identified.  
Furthermore, if we wish
to use these nanostructures in devices, we need to understand the 
scattering mechanism that controls
the electronic transport at the nanoscale level. We expect new physical
phenomena to appear in detailed time-resolved experiments in these systems,
related to quantum confinement. In summary, the investigation of electron-electron
interactions in nanostructures is still in its infancy, and much work is
expected to be done in the near future. In particular, we are planning to
investigate electron lifetimes in fullerene-based materials, such as C$_{60}$
and carbon nanotubes.

Asymmetries in electron lifetimes arise from the 
different nature and localization of electrons; in this sense, 
noble and transition metals offer a valuable framework to deal 
with different type of electrons that present various degrees of localization.  
New theoretical techniques should be able to address the calculation 
of excitations and inelastic electron lifetimes, including to some extent
electron-phonon couplings [which might be important and even dominant for high
enough temperatures and very-low-energy electrons] and also both
impurity and grain-boundary scattering.  Final-state effects have been neglected in
most practical implementations, and they might be important when there is strong
localization, as in the case of transition and rare earth materials. In the case of
semiconductors, electron-hole interactions (excitonic renormalization) strongly
modify the single-particle optical absorption profile, and they need
to be included in the electronic response.\cite{excitons} Although similar 
interactions are expected to be present in metals, the large screening in 
these systems makes their 
contribution less striking as compare to the case of semiconductors. However,
in the case of low-dimensional structures they might play an important role in the
broadening mechanism of excited electrons and holes.

All calculations presented in sections IV and V stop at the first iteration of the 
GW approximation. Although going beyond this approximation is possible,
this has to be done with great care, since higher-order corrections tend to
cancel out the effects of selfconsistency\cite{shirley,holm,schone} 
(see Appendix B). As we start from an RPA-like screening, the net effect of including the
so-called vertex corrections for screening electrons would be a reduction of the
screening. Furthermore, a simpler and important effect to be included in the
present  calculations is related to the renormalization of the excitation
spectral  weight due to changes in the self-energy close to the Fermi surface. We
know that this renormalization could be as large as 0.5  for
Ni\cite{Aryasetiawan,Gerhardt} and of the order of 0.8 for Si.\cite{Louie} This
modifies the energy of the excitation and, therefore, the lifetime.  We aim to
include such effects in the calculation of the inelastic electronic  scattering
process in noble and transition metals\cite{Jorge}, along the lines  described in
the Appendix B. The main idea is to work directly with the Green function in an 
imaginary-time/energy representation. The choice of representing the 
time/energy dependence on the imaginary rather than the real axis allows us to deal
with smooth, decaying quantities, which give faster convergence. Only after
the full imaginary-energy dependence of the expectation values of the
self-energy operator has been established do we use a fitted model function
(whose sophistication may be increased as necessary with negligible expense), 
which we then analytically continue to the real energy axis in order to compute
excitation spectra and lifetimes.\cite{Rojas,Blase,imtime,Lutz} Furthermore, this
technique is directly connected with finite-temperature many-body Green
functions, and can be used to directly address temperature effects on the lifetime
that can be measured experimentally.

An interesting aspect in the theory of inelastic electron
scattering appears when one looks at the energy dependence of the electron 
lifetime in layered materials. In a semimetal as graphite,
the lifetime has been found to be inversely proportional to the energy above
the Fermi level\cite{Xu}, in contrast to the quadratic behaviour predicted for
metals with the use of Fermi-liquid theory (see Section IV). This behaviour has been
interpreted  in terms of electron-plasmon interaction in a layered electron gas;\cite{Xu}
however, this is not consistent with the fact that a layered Fermi-liquid shows
conventional electron lifetimes.\cite{Zheng} A different interpretation based on the
particular band structure of graphite (with a nearly point-like Fermi surface) yielding a
reduction of the screening can explain the linear dependence of the lifetime.\cite{Guinea} A
similar linear dependence of the inelastic lifetime has been found for other
semiconducting-layered compounds  as SnS$_2$.\cite{Xu1} We are presently working on the
evaluation, within the GW approximation, of electron lifetimes in these layered
compounds.\cite{MiguelAngel} The special band structure of graphite has 
also been invoked as responsible for the peculiar plasmon dispersion and
damping of the surface plasmon.\cite{Palmer}
Therefore, a careful analysis of 
the layer-layer interaction and broadening of the Fermi 
surface needs to be included, in order to understand
this behaviour. We note that in a metal like Ni the imaginary part of the
self-energy shows a quadratic Fermi-liquid behaviour, which  becomes linear very 
quickly.\cite{Aryasetiawan,Gerhardt}

Together with the self-energy approach discussed in this review,
an alternative way of computing the excitation spectra of a many-body system,
which is based on information gleaned from an ordinary ground-state calculation,
is the time-dependent density-functional theory
(TDDFT).\cite{Runge,Gross2,Gross3,tddft} In this approach, one
studies how the system behaves under an external perturbation. The response of the system is
directly related to the $N$-particle excited states of an $N$-particle system, in a similar
manner as the one-particle Green function is related to the $(N+1)$- and $(N-1)$-particle
excited states of the same system. TDDFT is an ideal tool for studying the dynamics of
many-particle systems, which is based on a complete representation of the
XC kernel,
$K^{xc}$, in time and space. One computes the time-evolution of the 
system\cite{Flocard,review,Yabana,rby} without resorting 
to perturbation theory and dealing, therefore, with an 
external field of arbitrary strength. The fact that the evolution of the wave function is
mapped for a given time-interval helps one to extract useful information 
on the dynamics and electron relaxation of many-electron 
systems. The method does not stop on the linear response and includes, in
principle, higher-order nonlinear response as well as multiple
absorption and emission processes.

On a more pure theoretical framework, the connection between TDDFT
and many-body perturbation theory is needed, in order to get further insight into
the form of the frequency-dependent and non-local XC 
kernel $K^{xc}$. If one were able to design an XC kernel that
works for excitations as the LDA does for
ground-state properties, then one could handle many interesting
problems that are related to electron dynamics of many-electron systems.

In summary, many theoretical and experimental challenges related to the
investigation of lifetimes of low-energy electrons in metals and
semiconductors are open, and even more striking theoretical and
experimental advances are ready to come in the near future. Lifetime
measurements can be complementary to current spectroscopies for the attainment of
information about general properties (structural, electronic, dynamical, ....) of a given
system.

\acknowledgments
The authors would like to thank I. Campillo, M. A. Cazalilla, J. Osma,
I. Sarria, V. M. Silkin, and E. Zarate, for their contributions to 
some of the results that are reported here, and 
M. Aeschlimann, Th. Fauster, U. H\"ofer, and M. Wolf, for enjoyable discussions.
Partial support by the  University of the Basque Country, the Basque Unibertsitate
eta Ikerketa Saila, the Spanish Ministerio de Educaci\'on y Cultura, and Iberdrola S. A.
is gratefully acknowledged.

\appendix

\section{Linear response}

Take a system of interacting electrons exposed to an external potential
$V^{ext}(\rb,\omega)$. According to time-dependent perturbation theory and
keeping only terms of first order in the external perturbation
$V^{ext}(\rb,\omega)$, the charge density induced in the electronic system is
found to be
\begin{eqnarray}\label{eqa1}
\rho^{ind}({\bf r},\omega)=\int{\rm d}{\bf r}'\,
\chi({\bf r},{\bf  r'};\omega)\,V^{ext}({\bf r'},\omega),
\end{eqnarray}
where $\chi({\bf r},{\bf  r'};\omega)$ represents the so-called linear density
response function
\begin{eqnarray}\label{eqa2}
\chi({\bf r}, {\bf r'},\omega)=\sum_n \, && 
\rho^{*}_{n0}({\bf r}) \rho_{n0}({\bf r}') 
\left[ {1\over\omega-\omega_{n0}+{\rm i}\eta} \nonumber \right. \\
&& \left. -{1\over\omega+\omega_{n0}+{\rm i}\eta}\right].
\end{eqnarray}
Here, $\omega_{n0}=E_n-E_{0}$ and
$\rho_{n0}({\bf r})$ represent matrix elements taken between the unperturbed
many-particle ground state $|\Psi_{0}\rangle$ of energy $E_0$ and the
unperturbed many-particle excited state $|\Psi_{n}\rangle$ of energy $E_n$:
\begin{equation}\label{eqa3}
\rho_{n0}({\bf r})=\langle \Psi_n|\rho({\bf r})|\Psi_{0} \rangle, 
\end{equation}
$\rho({\bf r})$ being the charge-density operator,
\begin{equation}\label{eqa4}
\rho(\rb)= - \sum_{i = 1}^N\delta(\rb-\rb_i),
\end{equation}
and $\rb_i$ describing electron coordinates.

In a time-dependent Hartree or random-phase approximation, the electron
density induced by the external potential, $V^{ext}(\rb,\omega)$, is
approximated by  the electron density induced in a noninteracting electron gas
by the total field $V^{ext}(\rb,\omega)+ V^{ind}(\rb,\omega)$:
\begin{eqnarray}\label{eqa1p}
\rho^{ind}({\bf r},\omega)=\int{\rm d}{\bf r}'\,
&&\chi({\bf r},{\bf  r'};\omega)\,\left[V^{ext}({\bf r'},\omega)+\right.\cr\cr
&&\left.V^{ind}({\bf r'},\omega)\right].
\end{eqnarray}
This approximation for the induced electron
density can be written in the form of Eq. (\ref{eqa1}), with
\begin{eqnarray}\label{eq98}
\chi^{RPA}({\bf r},&&{\bf r}';\omega)=\chi^0({\bf r},{\bf r}';\omega)
+\int{\rm d}{\bf r}_1\int{\rm d}{\bf r}_2\cr\cr
&&\times \chi^0({\bf r},{\bf r}_1;\omega)\,v({\bf r}_1-{\bf r}_2)\,\chi^{RPA}({\bf
r}_2,{\bf r}';\omega),
\end{eqnarray}
where $\chi^0({\bf r},{\bf r}';\omega)$ is the density-response function of
noninteracting electrons,
\begin{eqnarray}\label{eq99}
\chi^0(\rb,\rb';\omega)=&&2\sum_{i,j}{\theta(E_F-\omega_i)-\theta(E_F-\omega_j)\over
\varepsilon_i-\varepsilon_j+(\omega+{\rm i}\eta)}\cr\cr
&&\times\phi_i(\rb)\phi_j^*(\rb)\phi_j(\rb')\phi_i^*(\rb'),
\end{eqnarray}
$\phi_i(\rb)$ representing a set of single-particle states of energy
$\varepsilon_i$.

In the framework of time-dependent density-functional
theory,\cite{Runge,Gross2,Gross3,tddft} the theorems of which generalize
those of the usual density-functional theory,\cite{Kohn1,Kohn2} the density
response function satisfies the integral equation
\begin{eqnarray}\label{eq98p}
\chi({\bf r},&&{\bf r}';\omega)=\chi^0({\bf r},{\bf r}';\omega)
+\int{\rm d}{\bf r}_1\int{\rm d}{\bf r}_2\,\chi^0({\bf r},{\bf r}_1;\omega)\cr\cr
&&\times\left[v({\bf r}_1-{\bf r}_2)
+K^{xc}(\rb_1,\rb_2;\omega)\right]
\chi({\bf r}_2,{\bf r}';\omega),
\end{eqnarray}
the kernel $K^{xc}(\rb,\rb';\omega)$ representing the reduction in the e-e
interaction due to the existence of short-range XC effects. In the static limit
($\omega\to 0$), DFT shows that\cite{tddft}
\begin{equation}\label{eq31}
K^{xc}(\rb,\rb';\omega\to 0)=\left[{\delta^2E_{xc}[n]\over\delta n(\rb)\delta
n(\rb')}\right]_{n_0(\rb)},
\end{equation}
where $E_{xc}[n]$ represents the XC energy functional and $n_0(\rb)$ is
the actual density of the electron system.

In the case of a homogeneous electron gas, one introduces Fourier transforms
and writes
\begin{equation}\label{eqa6}
\rho^{ind}_{\qb,\omega}=\chi_{\qb,\omega}V^{ext}_{\qb,\omega}.
\end{equation}

Within RPA,
\begin{equation}\label{eq22}
\chi_{{\bf q},\omega}^{RPA}=\chi_{{\bf q},\omega}^0+\chi_{{\bf 
q},\omega}^0v_{\bf q}\chi_{{\bf
q},\omega}^{RPA},
\end{equation}
where
\begin{eqnarray}\label{eqa9}
\chi^0_{{\bf q},\omega} = \frac{2}{\Omega} 
\sum_{\bf k}\,
&& n_{\bf k}(1-n_{{\bf k}+{\bf q}}) 
\left[ 
\frac{1}{ \omega + \varepsilon_{\bf k} - \varepsilon_{{\bf k}+{\bf q}} + 
{\rm i}\eta } \right. \cr\cr
&&
-\left.
\frac{1}{\omega+\varepsilon_{{\bf k}+{\bf q}} - 
\varepsilon_{\bf k} + {\rm i}\eta} \right],
\end{eqnarray}
$v_\qb=4\pi/q^2$ is the Fourier transform of the Coulomb
potential, $\varepsilon_{\bf k} = {\bf k}^2/2$, and $n_{\bf k}$ are occupation
numbers, as given by Eq. (\ref{oc}).

In the more general scenario of TDDFT,
\begin{eqnarray}\label{eq28}
\chi_{{\bf q},\omega}&=&\chi_{{\bf q},\omega}^0+\chi_{{\bf 
q},\omega}^0\left(v_{\bf
q}+K^{xc}_{\qb,\omega}\right)\chi_{{\bf q},\omega}\cr\cr
&=&\chi_{{\bf q},\omega}^0+\chi_{{\bf q},\omega}^0v_{\bf
q}\left(1-G_{\qb,\omega}\right)\chi_{{\bf q},\omega},
\end{eqnarray}
where $K^{xc}_{\qb,\omega}$ is the Fourier transform of the XC kernel
$K^{xc}(\rb,\rb',\omega)$. In Eq. (\ref{eq28}), we have set
\begin{equation}\label{eq29}
K^{xc}_{\qb,\omega}=-v_\qb\,G_{\qb,\omega},
\end{equation}
$G_{\qb,\omega}$ being the so-called
local-field factor. In the local-density approximation, which is rigorous in
the long-wavelength limit
($q\to 0$), it follows from Eqs. (\ref{eq31}) and (\ref{eq29}) that
\begin{equation}\label{eq32}
G_{\qb,0}^{LDA}=A\left({q\over q_F}\right)^2,
\end{equation}
where
\begin{equation}\label{eq33}
A={1\over 4}-{4\pi\over q_F^2}\,{d^2E_{c}\over d\,n_0^2},
\end{equation}
$E_c(n_0)$ being the correlation contribution to the ground-state energy of the 
uniform
electron gas. This quantity has been extensively studied, ranging from the 
simple Wigner
interpolation formula\cite{Pines2,Wigner} to accurate
parametrizations\cite{Vosko,Perdew} based on Monte Carlo calculations by Ceperley and
Alder.\cite{CA}

Diffusion Monte Carlo calculations of the static density-response function of
the uniform electron gas\cite{Alder,Ceperley} have shown that the LDA static
local-field factor of Eq. (\ref{eq32}) correctly reproduces the static response 
for all $q\le 2\,q_F$, as long as the exact correlation energy is used to
calculate $A$. For  larger values
of $q$ both exact and RPA density-response functions decay as $1/q^2$, their difference
being of order
$1/q^4$, and fine details of $G_{\qb,0}$ are expected to be of little 
importance.\cite{Singwi,Ichimaru,Gold}

Calculations of the frequency dependence of the local-field factor were carried out
by Brosens and coworkers\cite{Brosens} and, more recently, by Richardson and
Ashcroft.\cite{Ashcroft} Local-field factors are in general expected to be complex
for nonzero frequencies, and the importance of their frequency dependence is
reflected, e.g., in the finite lifetime of the volume plasmon. For aluminum,
measurements of $G_{\qb,\omega}$ have shown\cite{Larson} a weak $\omega$ dependence of
the local-field factor for energies below $\sim 35\,{\rm eV}$.

If the homogeneous electron gas is exposed to an external test charge of density
$\rho^{ext}(\rb,t)$, one writes
\begin{equation}\label{eq14}
V^{ext}_{\qb,\omega}=v_\qb\,\rho^{ext}_{\qb,\omega},
\end{equation}
and with the aid of Eq. (\ref{eqa6}) one finds the Fourier transform of the total
change of the charge density
$\rho^{tot}(\rb,t)=\rho^{ext}(\rb,t)+\rho^{ind}(\rb,t)$ to be given by the
following expression:
\begin{equation}\label{eq1pp}
\rho^{tot}_{\qb,\omega}=\epsilon^{-1}_{\qb,\omega}\,\rho^{ext}_{\qb,\omega},
\end{equation}
where $\epsilon_{\qb,\omega}$ is the so-called test$\_$charge-test$\_$charge
dielectric function:
\begin{equation}\label{eps1}
\epsilon^{-1}_{\qb,\omega}=1+v_\qb\,\chi_{\qb,\omega}.
\end{equation}
Hence, this dielectric function screens the potential both generated and 'felt' by a
test charge.

If the external potential is that generated by an electron, then one
writes
\begin{equation}\label{eq14p}
V^{ext}_{\qb,\omega}=v_\qb\,(1-G_{\qb,\omega})\,\rho^{ext}_{\qb,\omega},
\end{equation}
and with the aid of Eq. (\ref{eqa6}) one finds Eq. (\ref{eq1pp}), but now
$\epsilon_{\qb,\omega}$ being the test$\_$charge-electron dielectric function:
\cite{HL}
\begin{equation}\label{eq40}
\epsilon^{-1}_{\qb,\omega}=1+v_\qb(1-G_{\qb,\omega})\chi_{\qb,\omega}.
\end{equation}

By combining Eq. (\ref{eq28}) with either Eq. (\ref{eps1}) or Eq. (\ref{eq40}), one
can easily obtain the following expressions for the test$\_$charge-test$\_$charge
and the test$\_$charge-electron dielectric functions,
\begin{equation}\label{eq30}
\epsilon_{\qb,\omega}=1-{v_\qb\chi^0_{\qb,\omega}\over 1+v_\qb
\,G_{\qb,\omega}\,\chi^0_{\qb,\omega}}
\end{equation}
and
\begin{equation}\label{eqa17}
\epsilon_{\qb,\omega}^{-1}=1+v_\qb (1-G_{{\bf q},\omega})\chi_{\qb,\omega},
\end{equation}
respectively. If the local-field factor $G_{\qb,\omega}$ is set equal to zero,
both Eqs. (\ref{eqa17}) and (\ref{eqa13}) yield the RPA dielectric
function\cite{Lindhard,Pines2}
\begin{equation}\label{eqa13}
\epsilon^{RPA}_{\qb,\omega}=1-v_\qb\,\chi^0_{\qb,\omega}.
\end{equation}
In terms of $\epsilon^{RPA}_{\qb,\omega}$,
Eqs. (\ref{eqa17}) and (\ref{eqa13}) yield Eqs. (\ref{di1}) and (\ref{di2}),
respectively.

\section{GW approximation and beyond}

Let us introduce the many-body Green function:\cite{Nozieres} 
\begin{eqnarray}\label{eqb2}
{\cal G}({\bf r}\; t, {\bf r'}\; t') = &&  
-{\rm i} \theta(t-t')\;  \langle \Psi^N_{0} | \psi({\bf r}, t)
\psi^{\dagger}({\bf r'},t') | \Psi^N_{0} \rangle  \nonumber \\
&& + {\rm i}  \theta(t'-t) \; \langle \Psi^N_{0} | \psi^{\dagger}({\bf 
r'}, t') \psi({\bf r},t) | \Psi^N_{0} \rangle.
\end{eqnarray}
In this equation, $\psi^{\dagger}({\bf r},t)| \Psi^N_{0} \rangle$ 
stands for a $(N+1)$-electron state in which an electron
has been added to  the system at point ${\bf r}$ and time $t$.
When $t'<t$, the many-body Green function
gives the probability amplitude to detect
an electron at point ${\bf r}$ and time $t$ when a
(possibly different) electron has been added to
the system at point ${\bf r'}$ and time $t'$.
When $t'>t$, the Green function 
describes the propagation of a many-body state in which one electron has 
been removed at point ${\bf r}$ and time $t$, that is, the propagation of a hole. 

For a system of interacting 
electrons, there is little hope of calculating
${\cal G}({\bf r}, {\bf r'}, \omega)$ exactly.   
One usually has to resort to perturbation theory, starting
from a suitably chosen one-electron problem with a Hamiltonian 
$H_0({\bf r})$, eigenfunctions $\phi_{i}({\bf r})$, and
eigenenergies $\varepsilon_i$. Hence, the noninteracting
Green function ${\cal G}^0({\bf r}, {\bf r'}, \omega)$ is
given by the following expression:\cite{Hedin}
\begin{equation}\label{eqb3}
{\cal G}^0({\bf r}, {\bf r'}, \omega) =
  \sum_{i} \frac{\phi^{*}_i({\bf r}) \phi_i({\bf r'})}
{\omega - \varepsilon_i + {\rm i} \eta\; {\rm sgn}
(\varepsilon_i - E_F)}. 
\end{equation}
In usual practice the LDA\cite{Kohn2} is considered, which provides a local
one-electron potential, $u_{LDA}({\bf r})$.

The exact Green function obeys the following Dyson's equation,\cite{Hedin} 
\begin{eqnarray}\label{eqb4}
&&  \left( \omega  + {1 \over 2} \nabla^2_{\bf r} - u({\bf r}) \right) 
  {\cal G}({\bf r}, {\bf r'} , \omega)  +  \nonumber \\
&& \int d{\bf r''}  \Sigma({\bf r}, {\bf r''}, \omega) {\cal G}({\bf r''}, 
{\bf r}, \omega) =    {\bf \delta}({\bf r} - {\bf r'}), 
\end{eqnarray}
where the integral kernel $\Sigma({\bf r}, {\bf r'}, \omega)$ is known
as the self-energy. It can be understood as the complex non-local
energy-dependent potential felt by the electron added to the 
system at point ${\bf r'}$ and time $t'$. This potential arises from the 
response of the rest of electrons to the presence of the additional
electron. However, one must be careful with this 
interpretation, since the  many-body Green  function for $t' < t$ 
not only describes the propagation of the additional electron, 
but also that of the whole $(N+1)$-electron system. 
This means, for instance, that
the self-energy also accounts for the exchange processes that
occur in a system of indistinguishable particles.

To obtain the inelastic lifetime of one-electron-like 
excitations, called quasiparticles, one seeks for the poles of the many body Green
function. A good estimate of the position of these poles can be obtained by 
projecting Eq. (\ref{eqb4}) onto the chosen basis of
one-electron orbitals, and neglecting the off-diagonal terms 
in the self-energy, i.e.,
\begin{equation}\label{seq11}
\omega - \varepsilon_i - \Delta\Sigma_{ii}(\omega)
 \approx 0,
\end{equation}
where 
\begin{eqnarray}\label{seq12}
\Delta \Sigma_{ii}(\omega)=\int{\rm d}{\bf r}\int{\rm d}{\bf r'} \,
&&\phi^{*}_{i}({\bf r})\left[\Sigma({\bf r},{\bf r'};\omega)\right.\cr\cr
&&\left.-u_{LDA}({\bf r}){\bf \delta}({\bf r} - {\bf r'})\right] 
\phi_{i}({\bf r'}). 
\end{eqnarray}
In general, Eq.(\ref{seq11})  has complex solutions at 
$\omega = E-{\rm i}\,\Gamma/2$, where
$E$ is the excitation energy and $\Gamma$ the 
linewidth of the quasiparticle. Near the {\it energy-shell}
($\omega\sim\varepsilon_i$, $\varepsilon_i$ being the free-particle energy), one
finds for the linewidth $\tau^{-1}=\Gamma$:
\begin{equation}\label{seq13}
\tau^{-1}=-2\,Z_i\, 
{\rm Im} \, \Delta \Sigma_{ii} (\varepsilon_i),
\end{equation}
where
\begin{equation}\label{eq74}
Z_i=\left[1-\left.{\partial{\rm Re}\Delta\Sigma_{ii}(\omega)\over\partial 
\omega}\right|_{\omega=\varepsilon_i}\right]^{-1}
\end{equation}
is a {\it renormalization} constant. On the {\it energy-shell}
($Z_i\sim 1$), one writes
\begin{equation}\label{seq16}
\tau^{-1}=-2\,{\rm Im}\,\Sigma_{ii}(\varepsilon_i), 
\end{equation}
and after noting the reality of 
the matrix elements of the LDA potential one finds Eq. (\ref{eq92}).

Within many-body perturbation theory,\cite{Fetter} it is possible to
obtain $\Sigma({\bf r},{\bf r'}, \omega)$  as a series in the
Coulomb interaction $v({\bf r}-{\bf r'})$. Due to the long range of this interaction, such a
perturbation series is expected to contain  divergent terms. However, it has been known for
a long time that when the polarization induced in the system by the added electron 
is taken into account the series is free of 
divergences. Thus, the perturbation series for the self-energy 
can be rewritten in terms of the so-called screened interaction $W({\bf r},
{\bf r'};\omega)$.

To lowest order in the screened interaction, the self-energy reads:
\begin{equation}\label{eqb5}
\Sigma({\bf r}, {\bf r'}, \omega) = 
{\rm i} \int \frac{d\omega'}{2 \pi} \, e^{{\rm i}\eta \omega'} 
{\cal G}({\bf r}, {\bf r'},  \omega + \omega') 
 {W}({\bf r}, {\bf r'}, \omega ),
\end{equation}
which is the so-called GW approximation. The  screened interaction is given by Eq.
(\ref{eq94}), in terms of the exact time-ordered density-response function of interacting
electrons, or, equivalently, by the following integral equation
\begin{eqnarray}\label{eqb6}
&& {W}({\bf r}, {\bf r'}, \omega) =  v({\bf r}-{\bf r'}) + \nonumber \\
 && \mbox{    } \int d{\bf r}_1\int d{\bf r}_2 \; 
    v({\bf r}-{\bf r}_1) \Pi({\bf r}_1, {\bf r}_2, \omega)
               {W}({\bf r}_2, {\bf r'}, \omega),
\end{eqnarray} 
where $\Pi({\bf r}, {\bf r'}, \omega)$ represents the polarization propagator.
In the GW approximation,
\begin{equation}\label{eqb7}
\Pi^{GW}({\bf r}, {\bf r'}, \omega) 
 = -{\rm i} \int \frac{d\omega'}{2\pi} \; 
   {\cal G}({\bf r}, {\bf r'}, \omega') 
    {\cal G}({\bf r'}, {\bf r}, \omega' -\omega ).
\end{equation}
If  $\cal G$ is replaced in this expression by ${\cal G}^0$ ($\Pi^{GW})\to\Pi^0$), one
easily finds
\begin{equation}
{\rm Re }\; \Pi^0({\bf r},{\bf r'};\omega) = {\rm Re} \; \chi^0({\bf r},{\bf 
r'};\omega) 
\end{equation}
and
\begin{equation}
{\rm Im} \; \Pi^0({\bf r},{\bf r'};\omega) = {\rm sgn} \omega \, {\rm Im} 
\, \chi^0({\bf r},{\bf r'};\omega),
\end{equation}
where $\chi^0({\bf r},{\bf r'};\omega)$ represents the retarded density-response function
of noninteracting electrons, as defined by Eq. (\ref{eq99}). For positive frequencies
($\omega>0$), both $\Pi^0({\bf r},{\bf r'};\omega)$ and $\chi^0({\bf r},{\bf r'};\omega)$
coincide.

The GW approximation gives a comparatively simple expression for the
self-energy operator, which allows the  Green function of an
interacting many-electron system to be computed by simply starting from the Green
function ${\cal G}^0({\bf r},{\bf r'};\omega)$ of a fictitious system with an
effective  one-electron potential.
The GW approximation has been shown to be physically well motivated, specially
for metals where the Hartree-Fock approximation leads to unphysical results.

Eqs. (\ref{eqb4})-(\ref{eqb7}) form a set of equations which 
must be solved self-consistently for ${\cal G}({\bf r}, {\bf r'}, \omega)$. 
This means that the Green function used to calculate the self-energy must be
found to coincide with the Green function obtained from the Dyson
equation with  the very same self-energy. However, there is some 
evidence\cite{GW} supporting the idea that introducing the noninteracting Green 
function ${\cal G}^0({\bf r},{\bf r'};\omega)$ both in Eq. (\ref{eqb5}) and Eq. (\ref{eqb7})
($G^0W^0$ approximation) one obtains accurate results for the description of one-electron
properties such as the excitation energy and the quasiparticle lifetime.
However, self-consistency modifies the one-electron excitation
spectrum (excitation energies and lifetimes),\cite{holm,schone} as well as the calculated 
screening properties. Self-consistent calculations have been performed only very recently for
the homogeneous electron gas\cite{holm}, simple semiconductors, and metals.\cite{schone}

Discrepancies between $G^0W^0$ and self-consistent GW calculations seems to be originated
in the fact that the so-called vertex corrections,\cite{Hedin,GW} which go beyond the GW
approximation, need to be included as well. Inclusion of these corrections might cancel
out the effect of self-consistency, thereby full self-consistent self-energy calculations
yielding results that would be close to $G^0W^0$ results. The main outcome of
self-consistent GW calculations for the electron gas is that the total energy\cite{Migdal}
turns out to be strikingly close to the total energy calculated 
with use of quantum Monte Carlo techniques.\cite{CA} 
This result may be related to the fact that
the self-consistent GW scheme conserves electron-number,
energy, and total momentum, that is, fulfills the microscopic conservation laws.\cite{Baym}

The simplest improvement to the GW approximation is achieved by introducing a vertex
correction that is consistent with an LDA calculation 
of the one-electron orbitals,\cite{dft} the XC potential
being regarded as a self-energy  correction to the Hartree approximation.
Based on this idea, the vertex  $\Gamma$\cite{Sole} can be easily expressed in
terms of the static local field correction
of Eq. (\ref{eq31}). This
is the so-called ${\rm GW\Gamma}$ approximation.\cite{Rice,Sernelius2,Mahan1,Mahan2} In this
approximation, the polarization propagator is formally equivalent to the density-response 
function obtained within linear response theory in the framework of time-dependent 
density-functional theory.\cite{Runge,Gross2,Gross3,tddft} In the case of a homogeneous
electron gas, this approximation yields the test$\_$charge-electron dielctric function of
Eq. (\ref{eqa17}).
   
\section{Lifetimes of hot electrons near the Fermi level: Approximations}

The damping rate of hot electrons near the Fermi level ($E<<E-E_F$) is
obtained, within RPA, from Eq. (\ref{eq46}) with the inverse dielectric
function of Eq. (\ref{eqomega}), i.e.:
\begin{equation}\label{eq104p}
\tau^{-1}={2\over\pi}\,\int_0^{2q_F}{dq\over
q^4}\,\left|\epsilon_{q,0}\right|^{-2}\,{(E-E_F)^2\over k_i}.
\end{equation}
For small values of $q$ ($q<<2q_F$),
\begin{equation}\label{eq105p}
\epsilon_{q,0}\approx(1-\beta\alpha^2)+{\alpha^2\over z^2},
\end{equation}
with
\begin{equation}
\alpha={1\over\sqrt{\pi q_F}},
\end{equation}
$z=q/2 q_F$, and $\beta=1/3$, and one obtains
\begin{equation}\label{eq106}
\tau^{-1}={\sqrt\pi\,q_F^{-3/2}\over
8\sqrt{1-\beta\alpha^2}}\left[\arctan{1\over\alpha_0}+{\alpha_0\over
1+\alpha_0^2}\right]{(E-E_F)^2\over k_i},
\end{equation}
where
\begin{equation}\label{eq107}
\alpha_0={\alpha\over\sqrt{1-\beta\alpha^2}}.
\end{equation}
This is the expression first obtained by Ritchie\cite{Ritchie59} and by
Ritchie and
Ashley.\cite{Ashley}

In the high-density limit ($q_F\to\infty$), the static dielectric
function of Eq.(\ref{eq105p}) is
\begin{equation}\label{eq108}
\epsilon_{q,0}^{TF}\approx 1+{q_{TF}^2\over q^2},
\end{equation}
which can also be derived within the Thomas-Fermi statistical model,
$q_{TF}=\sqrt{4q_F/\pi}$ being the Thomas-Fermi momentum. By introducing 
Eq. (\ref{eq108}) into Eq.
(\ref{eq104p}) one obtains the expression derived by
Quinn,\cite{Quinn62} which can
also be obtained by just taking $\beta=0$ in Eq. (\ref{eq106}):
\begin{equation}\label{Quinn}
\tau^{-1}={\sqrt\pi\,q_F^{-3/2}\over
8}\left[\arctan{1\over\alpha}+{\alpha\over
1+\alpha^2}\right]{(E-E_F)^2\over k_i}.
\end{equation}

If one replaces the static dielectric function of Eq. (\ref{eq105p}) by
the
high-density limit ($q_F\to\infty$), as given by Eq. (\ref{eq108}), and
extends, at the
same time, the integration of Eq. (\ref{eq104p}) to infinity, one finds Eq.
(\ref{qf1}), which can also be obtained by just keeping the first-order term in
the expansion of Eq. (\ref{eq106}) in $q_F^{-1}$. If we further replace 
$k_i\to q_F$ in Eq. (\ref{qf1}), then the formula of Quinn and
Ferrell is obtained:\cite{QF}
\begin{equation}\label{eq112}
\tau_{QF}^{-1}=C(r_s)\,(E-E_F)^2,
\end{equation}
with
\begin{equation}\label{eq113}
C(r_s)={\pi^2\sqrt{3}\over 128}{\omega_p\over E_F^2},
\end{equation}
or, equivalently,
\begin{equation}\label{eq113p}
C(r_s)={(3\,\pi^2/2)^{1/3}\over 36}r_s^{5/2}.
\end{equation}

In Fig. 18, the ratio $\tau/\tau_{QF}$ is plotted against $r_s$ for hot
electrons in the immediate vicinity of the Fermi surface ($k_i \sim q_F$), 
as computed from either Eq. (\ref{eq46}) or Eq. (\ref{eq104p}) (solid line), and 
also from Eqs. (\ref{eq106}) (dotted line) and (\ref{Quinn}) (dashed-dotted line).
In the range
$0<r_s<6$, Eq. (\ref{eq106}) reproduces the full RPA calculation within
a $2\%$,
while either Eq. (\ref{qf1}) or Eq. (\ref{eq112}) reproduce the full
calculation within a $7\%$. Differences between the approximation
of Quinn\cite{Quinn62} (Eq. (\ref{Quinn})) and
the full RPA
calculation go up to $14\%$ at $r_s=6$.

Departures of the predictions of Eq. (\ref{eq106}) (dotted line) from
the full RPA
calculation (solid line) arise from small differences between the
dielectric function
of Eq. (\ref{eq105p}) and the actual static RPA dielectric function.
Further replacing
the dielectric function of Eq. (\ref{eq105p}) by the high-density limit
($q_F\to\infty$), as given by Eq. (\ref{eq108}), leads to a too-strong
Thomas-Fermi-like screening and results, therefore, in a large
overestimation of the
lifetime (dashed-dotted line). However, this overestimation is largely
compensated if one also takes $q_F\to\infty$ in the integration of Eq.
(\ref{eq104p}), thereby the lifetime of Quinn and Ferrell $\tau_{QF}$
(dashed line) being closer to the full RPA calculation than the lifetime derived from Eq.
(\ref{Quinn}) (dashed-dotted line).

\clearpage

\begin{table}[h]
\caption{\label{tabexp} Available experimental data for hot-electron 
lifetimes in metals, as obtained by time-resolved two-photon 
photoemission and ballistic electron emission microscopy 
(BEEM).}
\begin{tabular}{|c|c|}
{\bf Metal} & {\bf Reference (technique)} \\
\tableline
\tableline
Cu & [\onlinecite{Aes96}],[\onlinecite{Petek0}],
[\onlinecite{exp4}],[\onlinecite{exp6}] (TR-TPPE)\\
\tableline
Ag & [\onlinecite{Aes96}],[\onlinecite{Knoesel}] (TR-TPPE)\\
\tableline
Au & [\onlinecite{Aes96}],[\onlinecite{Cao}] (TR-TPPE); [\onlinecite{Reuter98}]
(BEEM)\\
\tableline
\tableline
Ta & [\onlinecite{Aes96}],[\onlinecite{Knoesel}] (TR-TPPE)\\
\tableline
Pd &  [\onlinecite{Ludeke93}] (BEEM)\\
\tableline
\tableline
Al &  [\onlinecite{Aes98}] (TR-TPPE)\\
\tableline
\tableline
Co &  [\onlinecite{exp5}], [\onlinecite{Aes98.2}] (TR-TPPE)\\
\tableline
Fe &  [\onlinecite{Aes98.2}] (TR-TPPE)\\
\end{tabular}
\label{tabla0}
\end{table}

\begin{table}[h]
\caption{Linewidth (inverse lifetime) of image states, in meV.}
\begin{tabular}{ |c|c|c|c|c| }
Surface & Image state & 2PPE & TR2PPE & Theory \\
     \hline
Cu(100) & n=1 & $28\pm6^a$ & $16.5\pm3/2^{b,c}$        & 
$18^a;26^d;22^e$ \\
        & n=2 &            &  $5.5\pm0.8/0.6^{b,c}$    & 
$5^{e}$        \\
        & n=3 &            &  $2.20\pm0.16/0.14^{b,c}$ & 
$1.8^e$          \\
     \hline
Cu(111) & n=1 & $16\pm4^f;85\pm10^{a,g}$ & $38\pm14/9^h;30^i$ &
$20^j;421^a;118^d;38^e$ \\
     \hline
Ag(100) & n=1 & $21\pm4^a$    & $26\pm18/7^k;12\pm1^c$         &
$22^a;25^d$ \\
        & n=2 & $3.7\pm0.4^a$ & $3.7\pm0.4^k;4.1\pm0.3/0.2^c;$ &
$5^d$       \\
        & n=3 &               & $1.83\pm0.08^c$                &  \\
     \hline
Ag(111) & n=1 & $45\pm10^a;55^l$ & $22\pm10/6^m$ & $58^i;123^a;110^d$ \\
     \hline
Au(111) & n=1 & $160\pm40^a$ &  & $617^a$ \\
     \hline
Pd(111) & n=1 & $70\pm20^a$  &  & $40^a;35^n$ \\
     \hline
Ni(100) & n=1 & $70\pm8^a$ &  & $24^a$ \\
     \hline
Ni(111) & n=1 & $84\pm10^a$ &  & $40^a$ \\
     \hline
Co(0001) & n=1 & $95\pm10^a$ &  & $40^a$ \\
     \hline
Fe(110) & n=1 & $130\pm30^a$ &  & $95^a$ \\
\end{tabular}

\begin{tabbing}
$^{a}$Ref.\onlinecite{Fauster}, Th. Fauster and W. Steinmann.     \\
$^{b}$Ref.\onlinecite{Hofer}, U. H\"ofer {\em et al.}      \\
$^{c}$Ref.\onlinecite{Shumay}, I. L. Shumay {\em et al.}   \\
$^{d}$Ref.\onlinecite{Chulkov2}, E. V. Chulkov {\em et al.}   \\
$^{e}$Ref.\onlinecite{Chulkov1}, E. V. Chulkov {\em et al.}  \\
$^f$Ref.\onlinecite{scfiprb92}, S. Schuppler {\em et al.}    \\
$^g$Ref.\onlinecite{wafass97}, W. Wallauer and Th. Fauster.    \\
$^h$Ref.\onlinecite{Wolf}, M. Wolf {\em et al.}         \\
$^i$Ref.\onlinecite{Knoesel0}, E. Knoesel {\em et al},
at low temperature, $T=25\,{\rm K}.$         \\
$^j$Ref.\onlinecite{Echenique3}, P. L. de Andres {\em et al.} \\
$^k$Ref.\onlinecite{Fujimoto}, R. W. Schoenlein {\em et al.} \\
$^l$Ref.\onlinecite{mejoss93}, W. Merry {\em et al.}         \\
$^m$Ref.\onlinecite{Harris}, J. D. McNeil {\em et al.}     \\
$^n$Present work.                                      \\
\end{tabbing}
\label{tabla1}
\end{table}

\clearpage

\begin{table}[h]
\caption{Calculated contributions to the linewidth, in meV, of the $n=1$
image state on Cu surfaces.}
\begin{tabular}{ |c|c|c|c|c| }
Surface & $\Gamma_{bulk}$ & $\Gamma_{vac}$ & $\Gamma_{inter}$ & $\Gamma$
\\
     \hline
Cu(100) & $24$ & $14$ & $-16$ & $22$ \\
Cu(111) & $44$ & $47$ & $-54$ & $37$ \\
\end{tabular}
\label{tabla2}
\end{table}

\begin{table}[h]
\caption[]{Linewidths $\Gamma_1$ and $\Gamma_2$, in meV, of the $n=1$
image state on Cu(111), as calculated for non-zero momenta parallel to the
surface and with use of two models for the $n=1$ image-state wave function
(see text and Ref.{\onlinecite{Osma}}).}
\begin{tabular}{ |c|c|c| }
$k$ (a.u.) & $\Gamma_1$ & $\Gamma_2$ \\
     \hline
$0.0570$ & $38.1$ & $38.9$ \\
$0.0912$ & $38.5$ & $43.6$ \\
$0.1026$ & $38.7$ & $47.0$ \\
\end{tabular}
\label{tabla3}
\end{table}

\begin{table}[h]
\caption{Lifetimes of image states on Cu surfaces, in fs.}
\begin{tabular}{ |c|c|c|c| }
Surface & Image state & TR2PPE & Model potential \\
        &             &        & calculation     \\
     \hline
Cu(100) & n=1 & $40\pm6^{a,b}$          & $30^c;38^d$   \\
        & n=2 & $120\pm15^{a,b}$        & $132^c;168^d$       \\
        & n=3 & $300\pm20^{a,b}$        & $367^c;480^d$       \\
     \hline
Cu(111) & n=1 & $18\pm5^e;22\pm5^f$ & $17.5^c;22.5^{d,g}$ \\
\end{tabular}
\begin{tabbing}
$^a$Ref.\onlinecite{Hofer}, U. H\"ofer {\em et al.}    \\
$^b$Ref.\onlinecite{Shumay}, I. L. Shumay {\em et al.}   \\
$^c$Ref.\onlinecite{Chulkov1}, E. V. Chulkov {\em et al.}  \\
$^d$Ref.\onlinecite{Sarria}, I. Sarria {\em et al.} \\
$^e$Ref.\onlinecite{Wolf}, M. Wolf {\em et al.} \\
$^f$Ref.\onlinecite{Knoesel0}, E. Knoesel {\em et al}, at low temperature, $T=25\,{\rm K}$.
\\
$^g$Ref.\onlinecite{Osma}, J. Osma {\em et al.} \\
\end{tabbing}
\label{tabla4}
\end{table}

\clearpage
\newpage

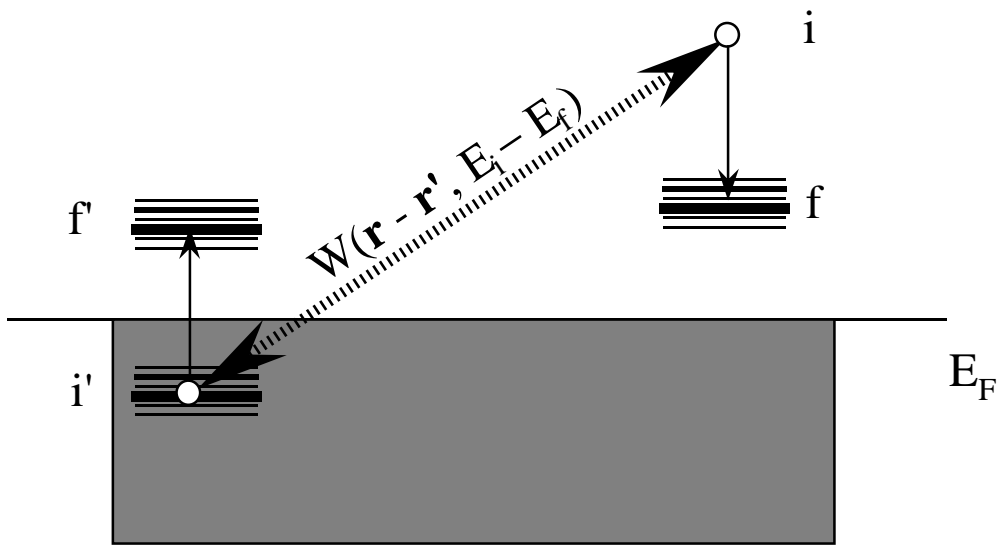
\begin{figure}
\caption[]{
\label{figsch}Scattering of an excited electron with the Fermi sea. The probe
electron is scattered from a state $\phi_i({\bf r})$ of energy $E_i$ to
some other state $\phi_f(\bf r)$ of energy $E_f$, by carrying one
electron of the Fermi sea from an initial state $\phi_{i'}(\bf r)$
of energy $E_{i'}$ to a final state $\phi_{f'}(\bf r)$ of energy
$E_{f'}$, according to a dynamic screened interaction
$W({\bf r}-{\bf r'},E_i-E_f)$. $E_F$ represents the Fermi level.}
\end{figure}

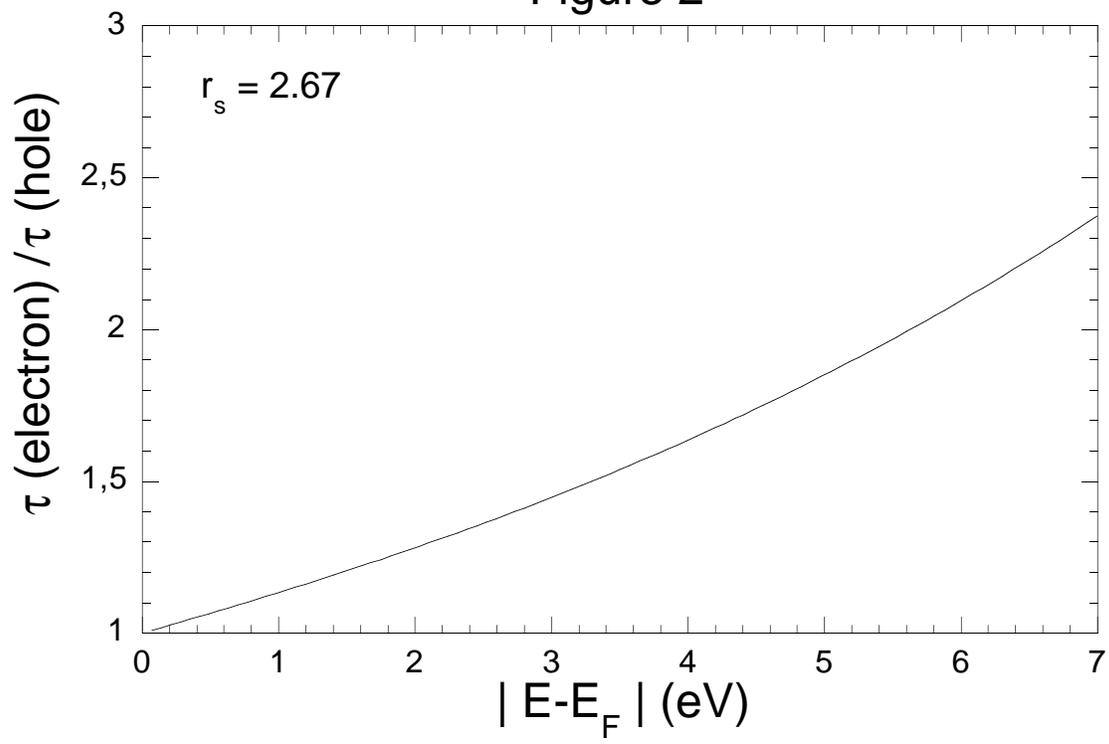
\begin{figure}
\caption[]
{\label{fighol}Ratio of the lifetime of electrons above the Fermi level
($E>E_F$) to the lifetime of holes below the Fermi level
($E<E_F$), as a function of $|E-E_F|$, calculated within RPA for an electron
density equal to that of valence ($4s^1$) electrons in copper, i.e., $r_s=2.67$.}
\end{figure}

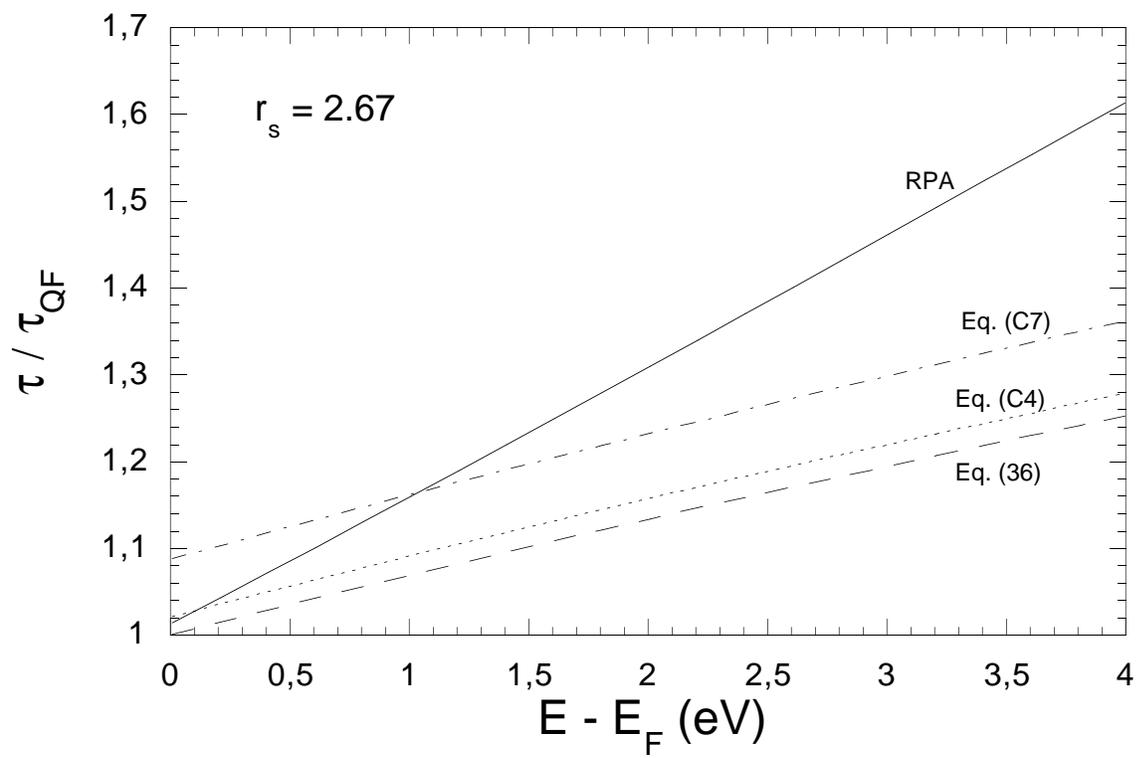
\begin{figure}
\caption[]
{Ratio $\tau/\tau_{QF}$ between the lifetime $\tau$ evaluated in
various approximations and the lifetime $\tau_{QF}$ of Eq. (\ref{eq114}), versus
$E-E_F$, as obtained for hot electrons in a homogeneous electron gas
with $r_s=2.67$. The solid line represents the result
obtained from Eq. (\ref{eq46}), within RPA. Results
obtained from Eqs. (\ref{eq106}) and (\ref{Quinn}) are represented
by dotted and dashed-dotted lines, respectively. The dashed line 
represents the result obtained from Eq. (\ref{qf1}).}
\end{figure}

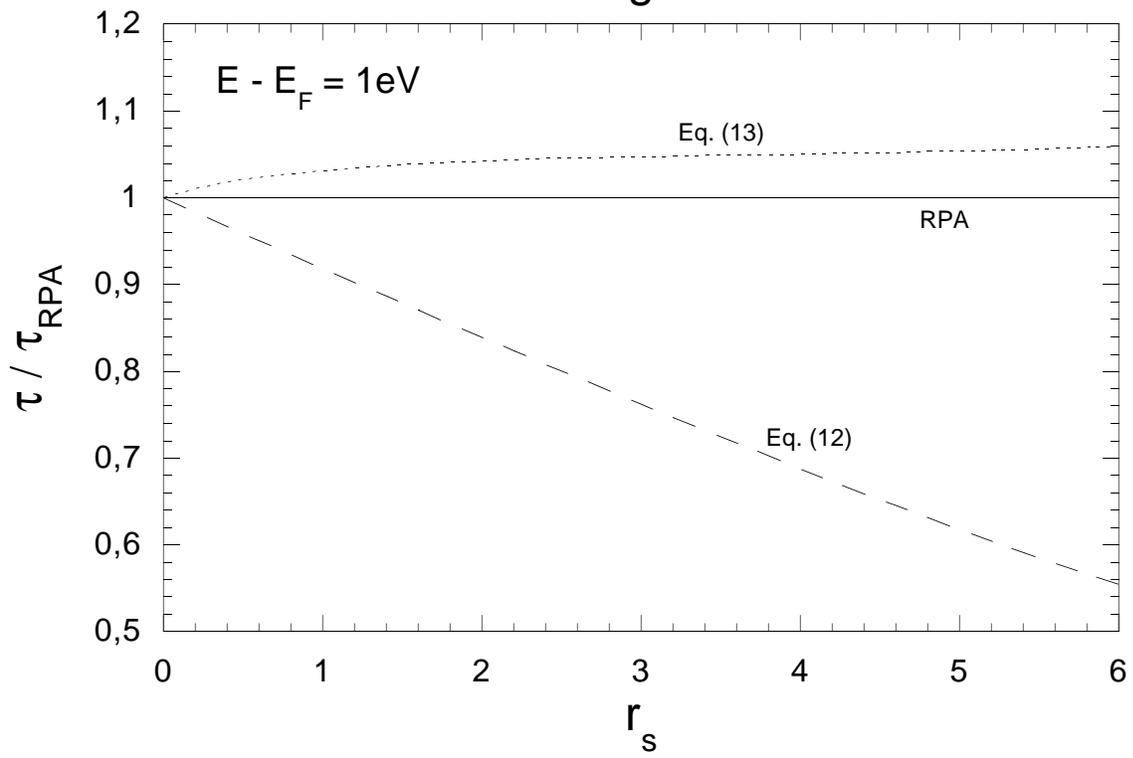
\begin{figure}
\caption[]
{\label{figexc_r}Exchange and correlation effects on the lifetime of hot
electrons with $E-E_F=1\,{\rm eV}$. The dashed line represents, as a function of
$r_s$, the ratio between lifetimes derived from Eq. (\ref{eq46}) with use of the dielectric
function of Eq. (\ref{di1}) with ($G_{\qb,\omega}\neq 0$) and without
($G_{\qb,\omega}=0$) local-field corrections. The dotted line represents, as a function
of $r_s$, the ratio between lifetimes derived from Eq. (\ref{eq46}) with use of the
dielectric function of Eq. (\ref{di2}) with ($G_{\qb,\omega}\neq 0$) and without
($G_{\qb,\omega}=0$) local-field corrections. If the local-field factor
$G_{\qb,\omega}$ is taken to be zero, both Eqs. (\ref{di1}) and (\ref{di2})
give the same result (solid line).}
\end{figure}

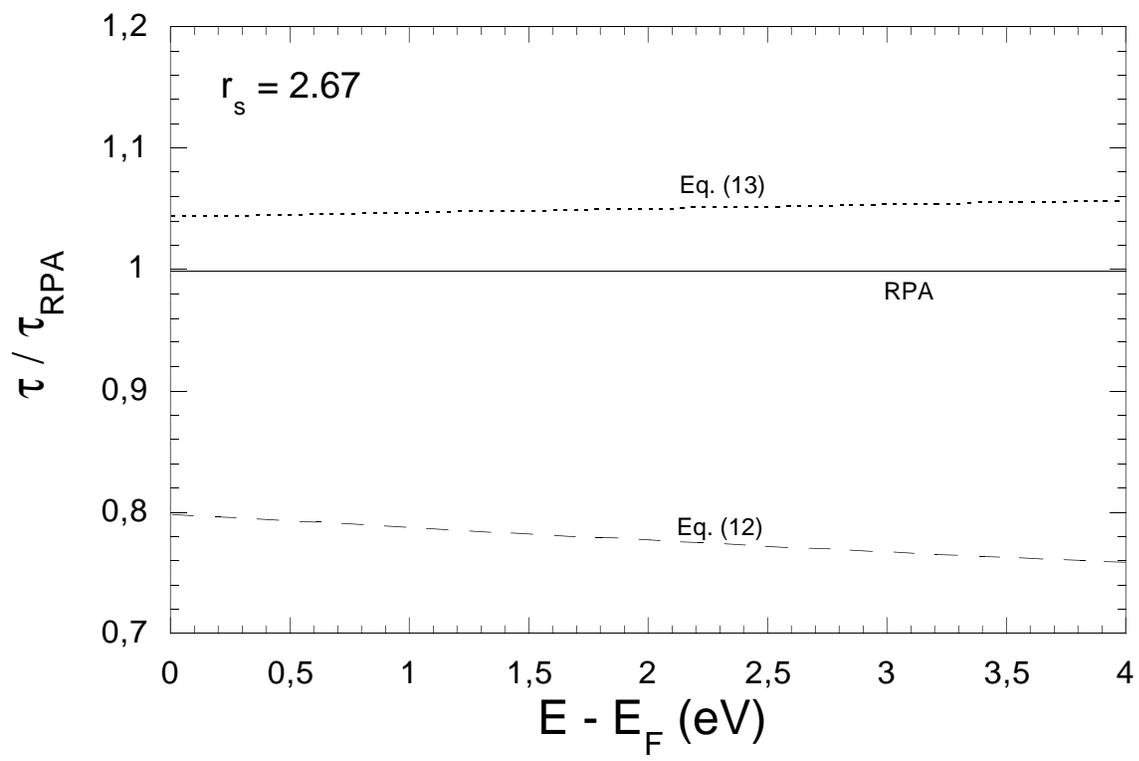
\begin{figure}
\caption[]
{\label{figexc_e}As in Fig. 4, but for hot electrons in a homogeneous electron
gas with $r_s=2.67$, and as a function of the electron energy $E-E_F$ with
respect to the Fermi level.}
\end{figure}

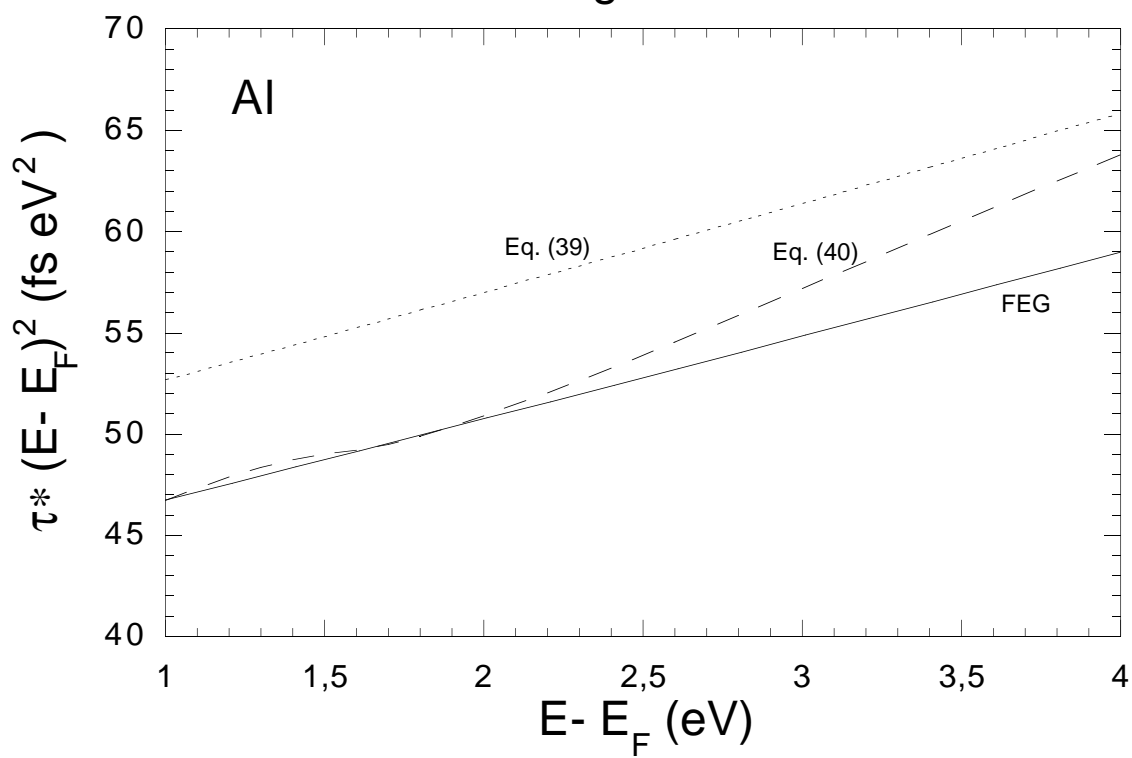
\begin{figure}
\caption[]
{\label{figcustat}Averaged lifetimes of hot electrons in Al, 
versus $E-E_F$, as obtained
from Eq. (\ref{eq117}) with the local electron density of
Ref.{\onlinecite{Tung3}} (dotted line), and from Eq. (\ref{eq46}) 
with the model dielectric function of Eq. (\ref{s6}) and the recipe described by
Salvat {\it et al\,}{\cite{Salvat}} to obtain the optical energy-loss
function (dashed line). The solid line represents the result obtained from Eq.
(\ref{eq46}) with use of the free-electron gas RPA energy-loss function and $r_s=2.07$.}
\end{figure}

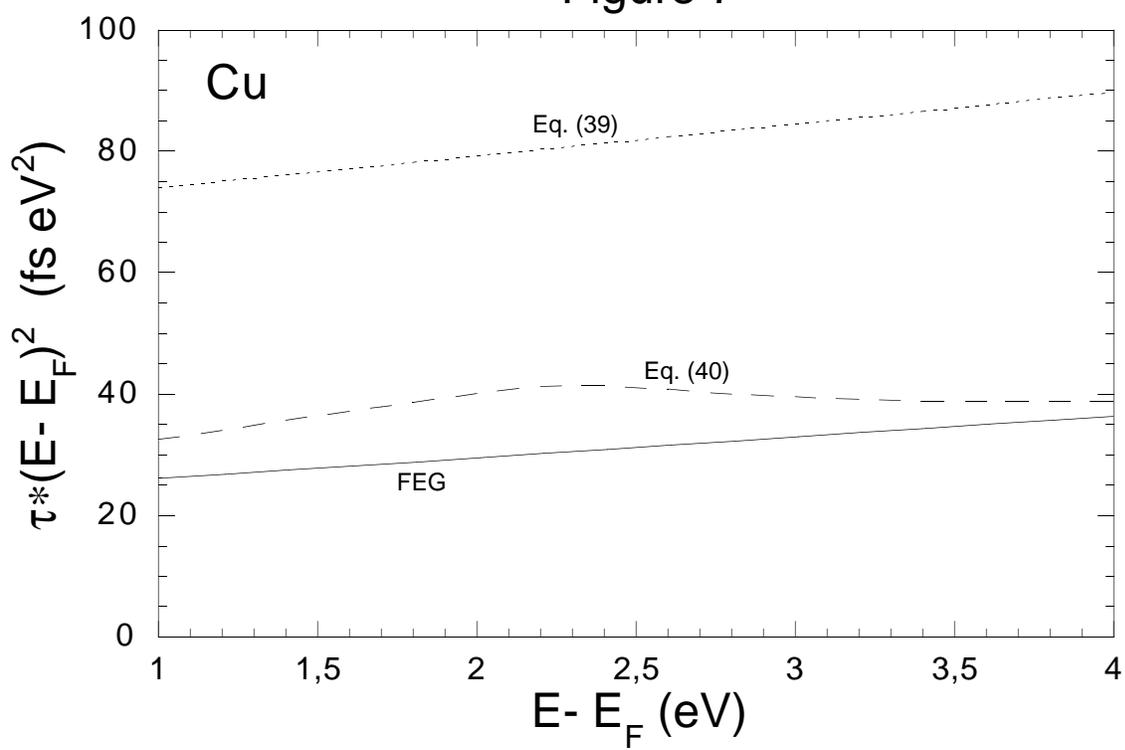
\begin{figure}
\caption[]
{\label{figalstat}Averaged lifetimes of hot electrons in Cu, 
versus $E-E_F$, as obtained
from Eq. (\ref{eq117}) with the local electron density of
Ref.{\onlinecite{Tung3}} (dotted line), and from Eq. (\ref{eq46})
with the model dielectric function of Eq. (\ref{s6}) and the recipe described by
Salvat {\it et al\,}{\cite{Salvat}} to obtain the optical energy-loss
function (dashed line). The solid line represents the result obtained from Eq.
(\ref{eq46}) with use of the free-electron gas RPA energy-loss function and
$r_s=2.67$.}
\end{figure}

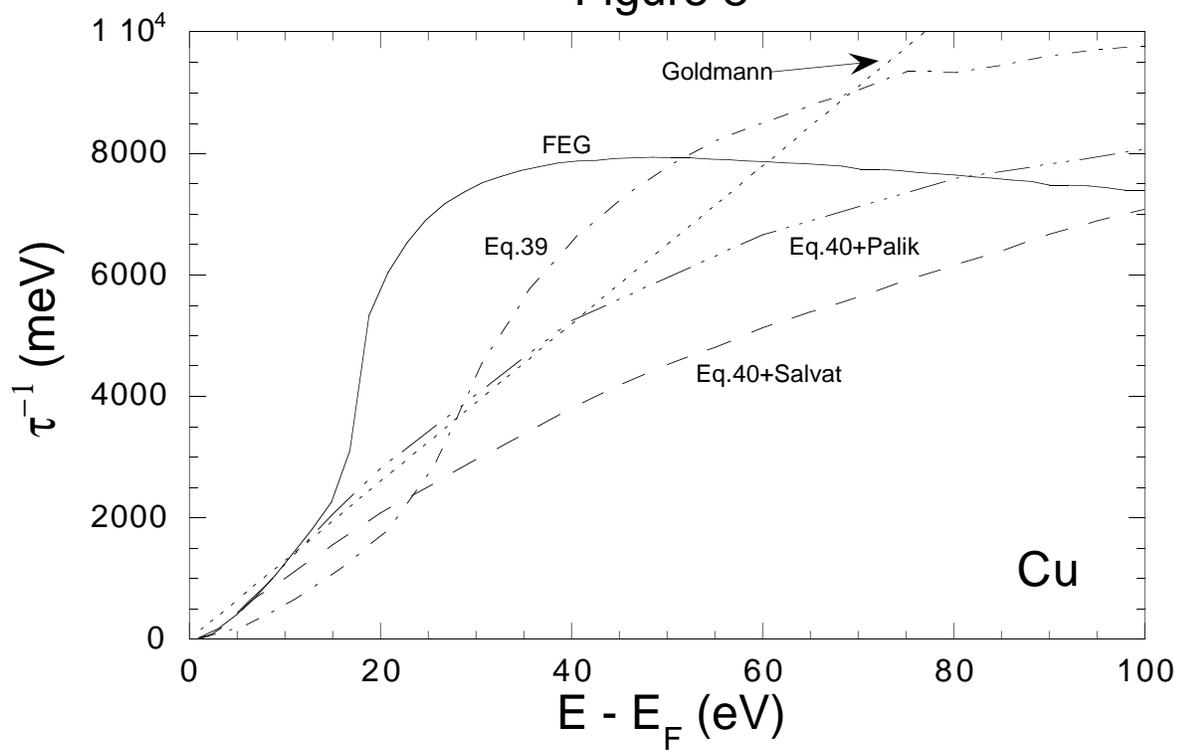
\begin{figure}
\caption[]
{\label{figcuhigh}Averaged lifetime-widths $\tau^{-1}$ of excited electrons in Cu,
versus $E-E_F$, as obtained from Eq. (\ref{eq117}) with the local electron density
of Ref.{\onlinecite{Tung3}} (dotted line), and from Eq. (\ref{eq46})
with the model dielectric function of Eq. (\ref{s6}) and the recipe described by
Salvat {\it et al\,}{\cite{Salvat}} to obtain the optical energy-loss
function (dashed line). The dashed-dotted-dotted-dotted line represents the result
obtained from Eq. (\ref{eq46}) with use of the model dielectric function of Eq. (\ref{s6})
and the experimental optical energy-loss function of
Ref.{\onlinecite{Palik}}. The dotted
line represents the result of using the 'universal' relationship
$\tau^{-1}=0.13\,(E-E_F)$ proposed by Goldmann {\it et al\,}.
{\cite{Goldmann}}}
\end{figure}

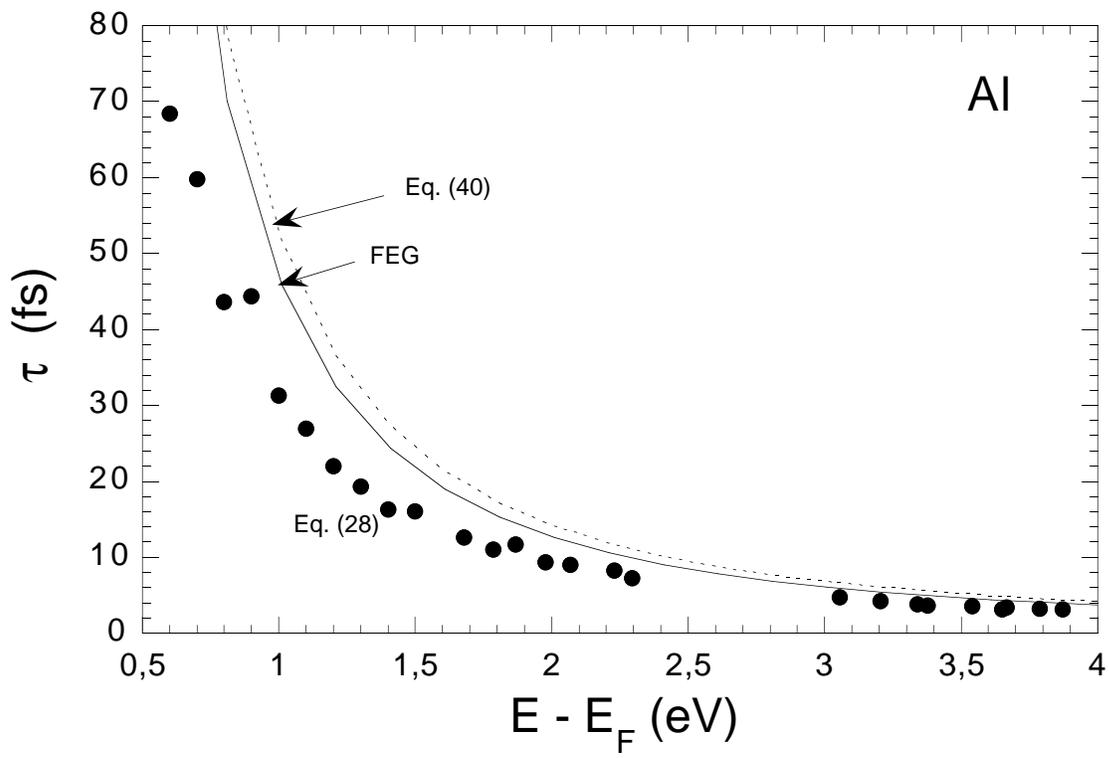
\begin{figure}
\caption[]
{\label{figetal}Hot-electron lifetimes in Al. Solid circles represent the full
{\it ab initio} calculation of $\tau(E)$, as obtained after averaging
$\tau$ of Eq. (\ref{eq105}) over wave vectors and over the band 
structure for each wave vector. The solid line represents the lifetime 
of hot electrons in a FEG with $r_s=2.07$,
as obtained from Eq. (\ref{eq46}). The dotted line represents the statistically
averaged lifetime, as obtained from Eq. (\ref{eq117}) by following the 
procedure of Tung {\it et al\,}.{\cite{Tung3}}}
\end{figure}

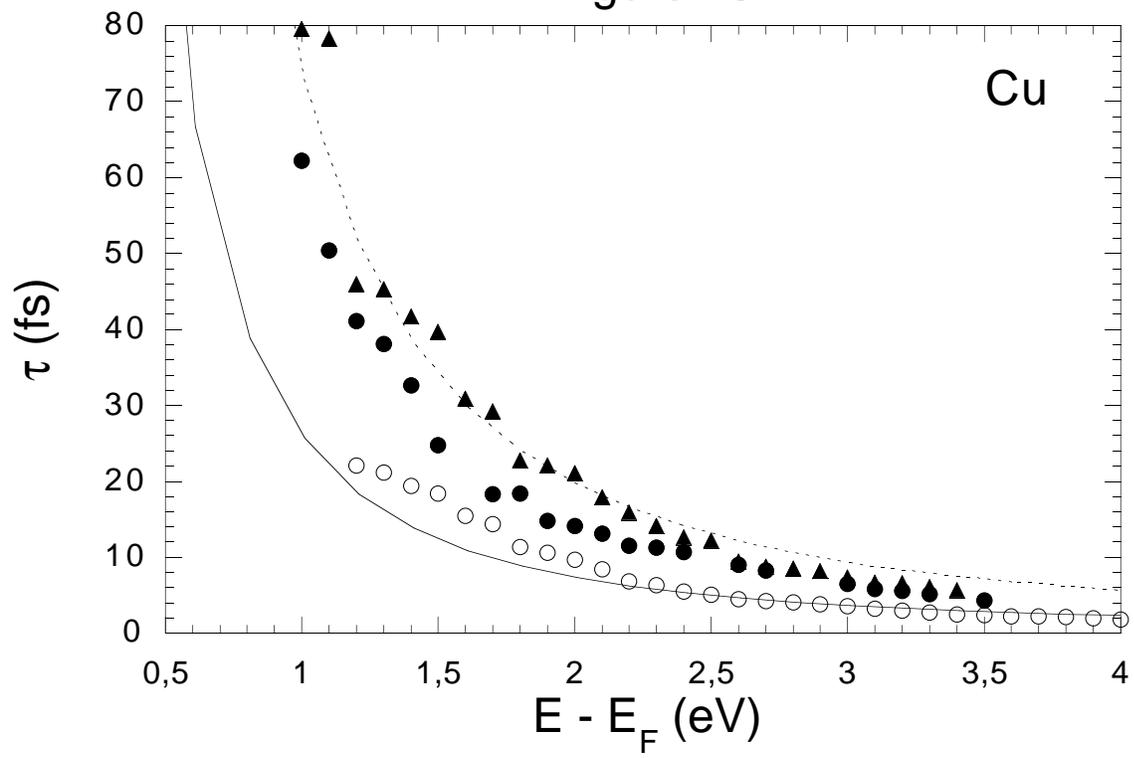
\begin{figure}
\caption{\label{figetcu} Hot-electron lifetimes in Cu. Solid circles represent
the full {\it ab initio} calculation of $\tau(E)$, as obtained after averaging
$\tau$ of Eq. (\ref{eq105}) over wave vectors and over the band 
structure for each wave vector. The solid line represents the lifetime 
of hot electrons in a FEG with $r_s=2.67$, as
obtained from Eq. (\ref{eq46}). The dotted line represents the statistically
averaged lifetime, as obtained from Eq. (\ref{eq117}) by following the 
procedure of Tung {\it et al\,}.\protect\cite{Tung3}\protect 
Open circles represent the result obtained from
Eq. (\ref{eq7p}) by replacing hot-electron initial and final states in
$\left|B_{if}({\bf q}+{\bf G})\right|^2$ by plane waves and the 
dielectric function in 
$\left|\epsilon_{{\bf G},{\bf G}}({\bf q},\omega)\right|^{-2}$ by that
of a FEG with $r_s=2.67$, but with full inclusion of the band 
structure in the calculation of
${\rm Im}\left[\epsilon_{{\bf G},{\bf G}}({\bf q},\omega)\right]$. 
Full triangles represent the result obtained from Eq. (\ref{eq7p}) by 
replacing hot-electron initial and final states in 
$\left|B_{if}({\bf q}+{\bf G})\right|^2$ by plane waves,
but with full inclusion of the band structure in the evaluation of both
${\rm Im}\left[\epsilon_{{\bf G},{\bf G}}({\bf q},\omega)\right]$ and
$\left|\epsilon_{{\bf G},{\bf G}}({\bf q},\omega)\right|^{-2}$.}
\end{figure}

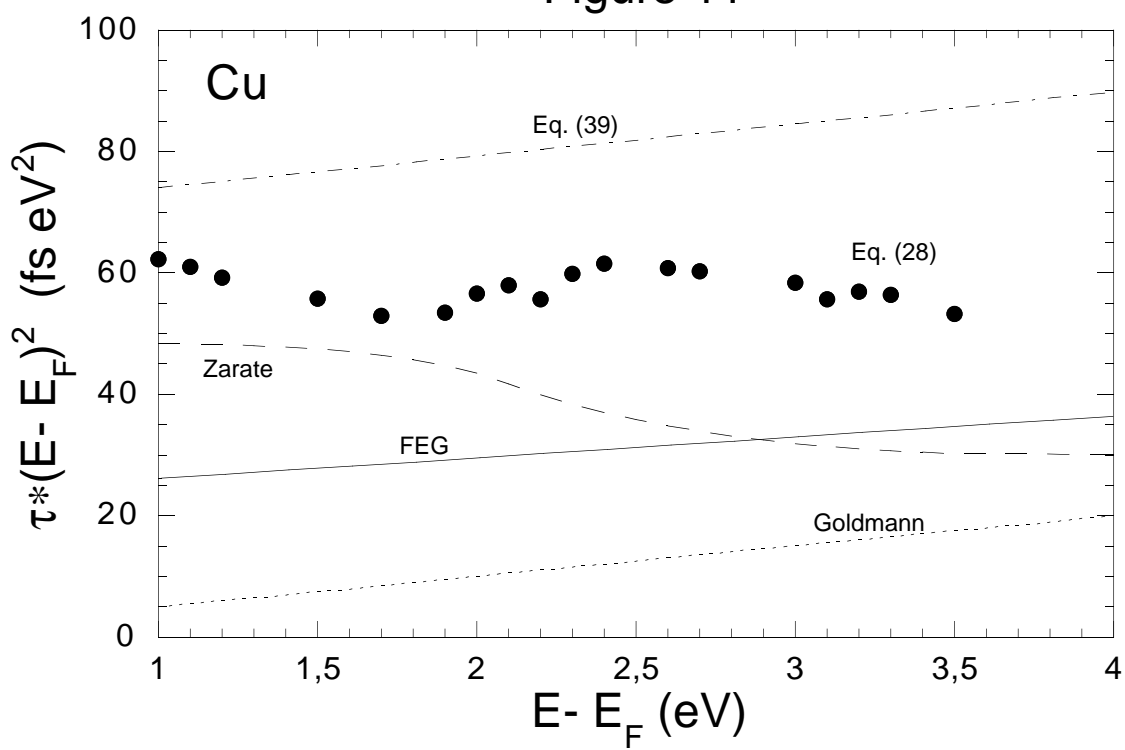
\begin{figure}
\caption[]
{\label{fig_igor_nor} Scaled hot-electron lifetimes in Cu. Solid circles
represent the full {\it ab initio} calculation of $\tau(E)$, as obtained after
averaging $\tau$ of Eq. (\ref{eq105}) over wave vectors and over the band 
structure for each wave vector.  The solid line represents the lifetime 
of hot electrons in a FEG with $r_s=2.67$, as
obtained from Eq. (\ref{eq46}). The dashed-dotted line represents the statistically
averaged lifetime, as obtained from Eq. (\ref{eq117}) by following the 
procedure of Tung {\it et al\,}.{\cite{Tung3}} The dashed line
represents the result of following the procedure described in
Ref.{\onlinecite{Zarate}}, and the dotted
line is the result of using the 'universal' relationship
$\tau^{-1}=0.13\,(E-E_F)$ proposed by Goldmann {\it et al\,}.
{\cite{Goldmann}}}
\end{figure}

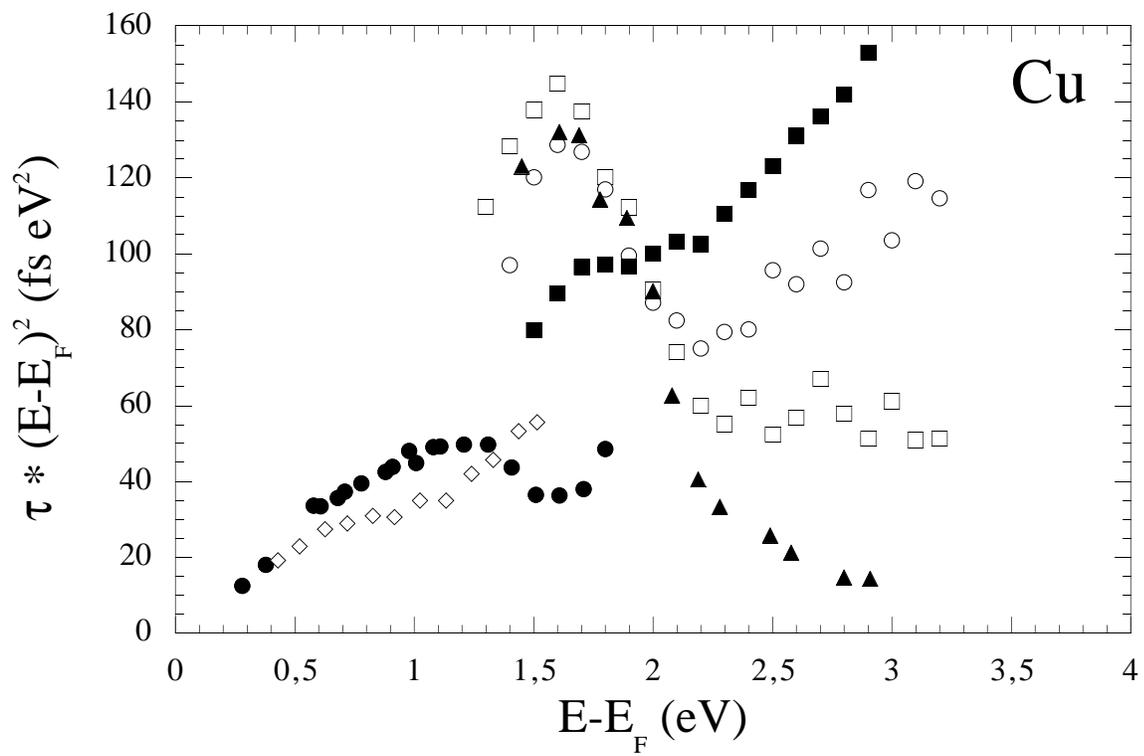
\begin{figure}
\caption[]{
\label{figcuexp}Experimental lifetimes of low-energy electrons in Cu, as taken
from Knoesel {\it et al\,}{\cite{exp6}} (solid circles), from Ogawa {\it et
al\,}{\cite{Petek0}} (Cu[100]: open circles, Cu[110]: open squares, Cu[111]:
solid squares), from Aeschlimann {\it et al\,}{\cite{Aes96}} (solid triangles), and
from Cao {\it et al\,}{\cite{exp4}} with
$\omega_{fot.}={\rm 1.63 eV}$ (open diamonds)}.
\end{figure}

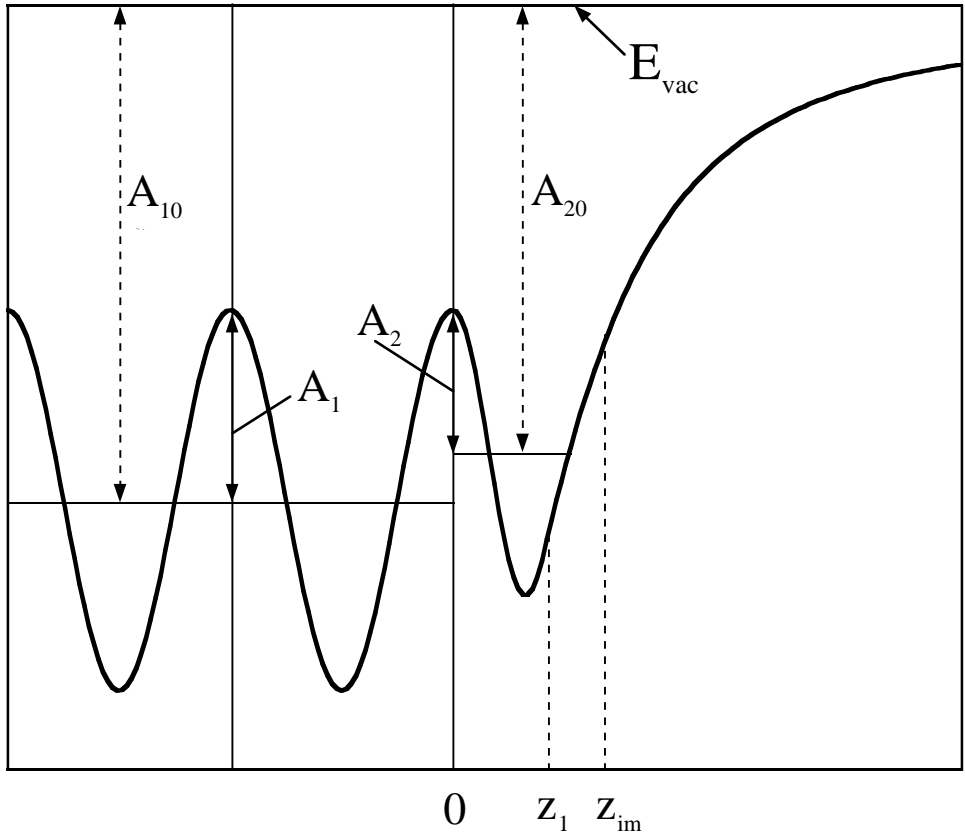
\begin{figure}
\caption[]{
\label{figpot} Schematic plot of the model potential of Eq. (\ref{Vz}).
Vertical solid lines represent the position of atomic layers.}
\end{figure}

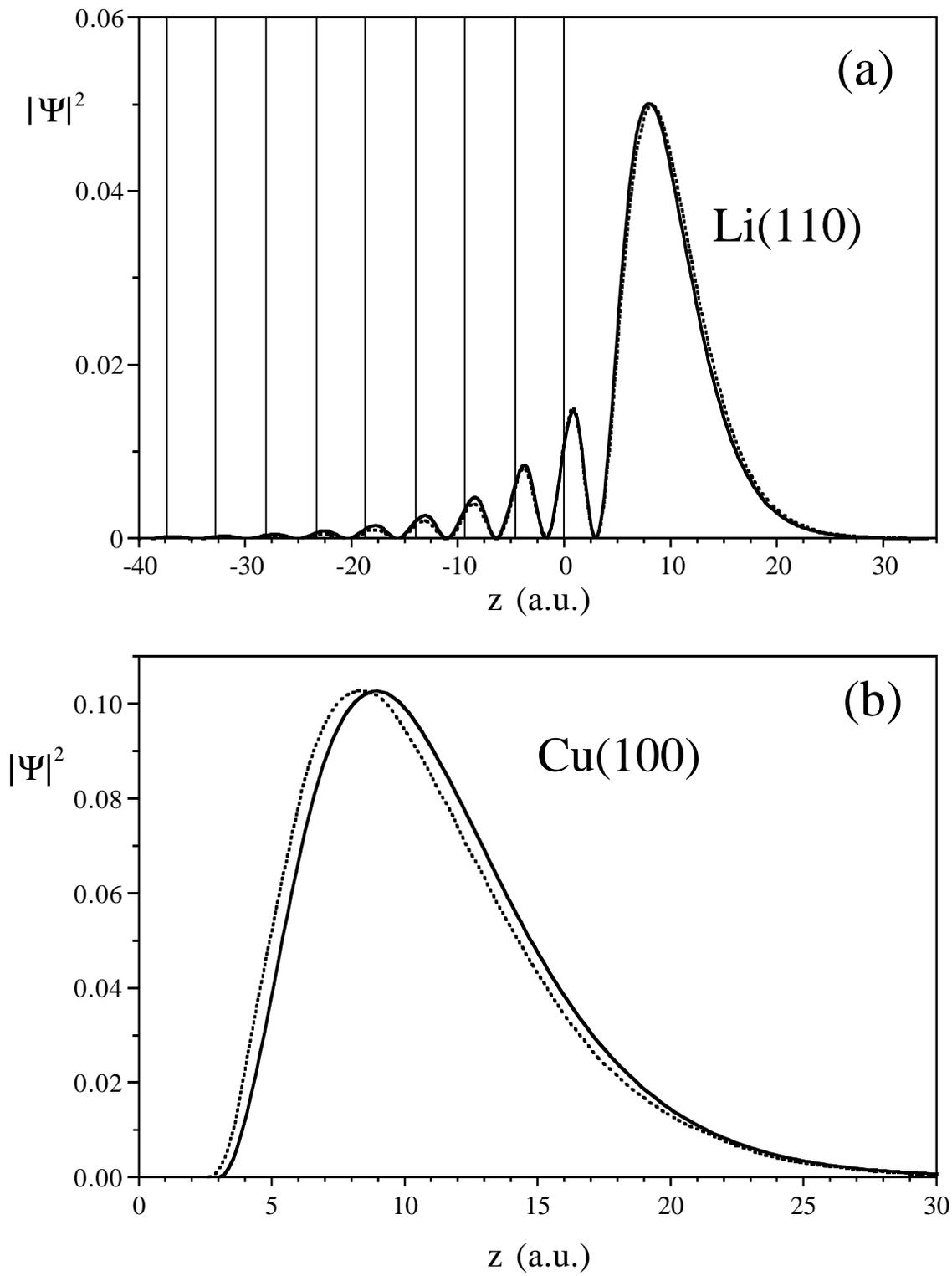
\begin{figure}
\caption[]{\label{figpsi} The probability amplitude of the $n=1$ image
state on (a) the (110) surface of Li, as obtained from the
model potential of Eq. (\ref{Vz}) (solid line) and from pseudopotential
calculations (dotted line), and (b) the (100) surface of Cu, as obtained from
the model potential of Eq. (\ref{Vz}) (solid line) and from linear augmented
plane-wave calculations (dotted line). Vertical solid lines represent the position
of atomic layers.}
\end{figure}

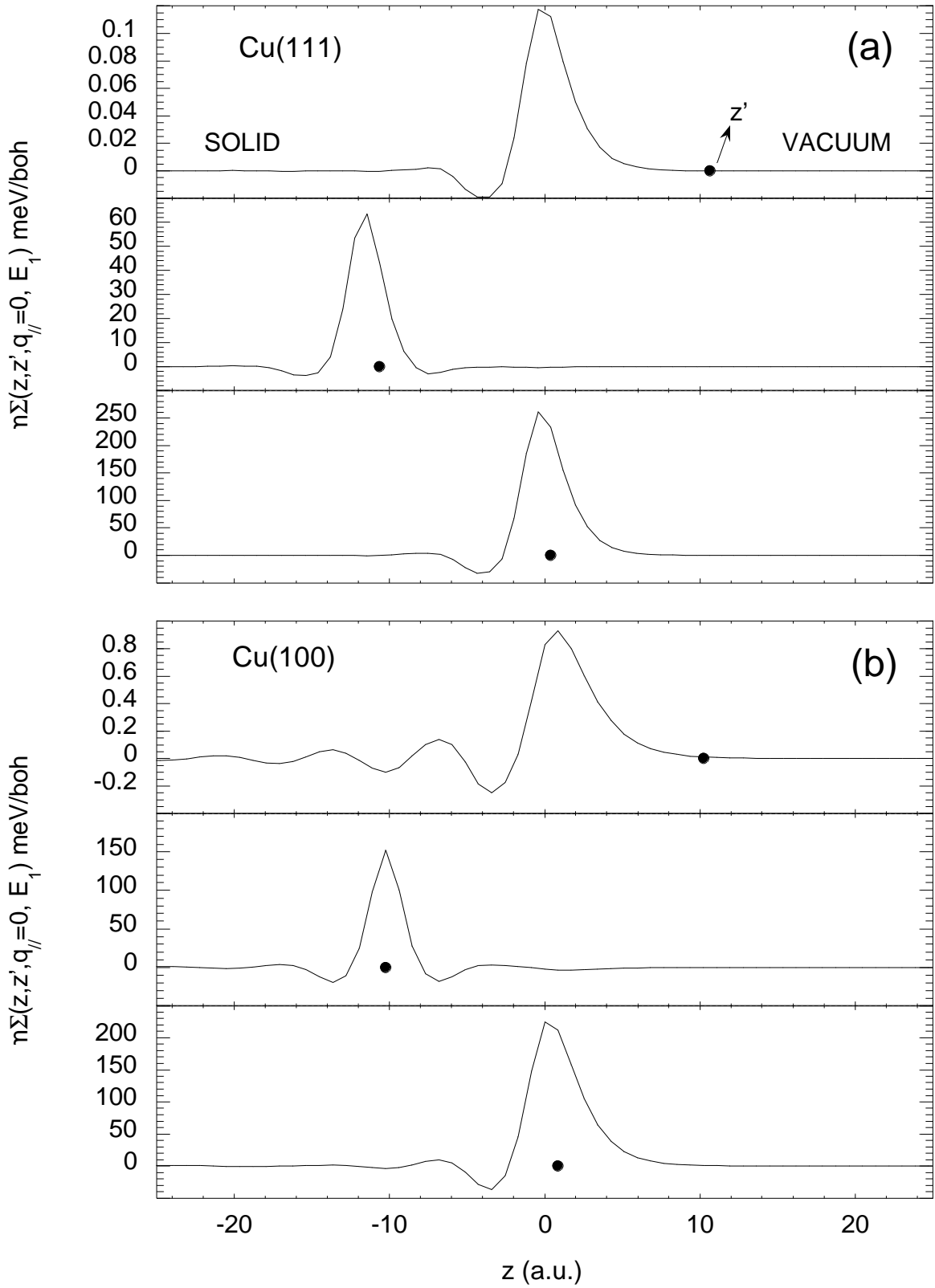
\begin{figure}
\caption[]{\label{figself}
Imaginary part of the electron self-energy, versus $z$,
for three fixed values of $z'$ (solid circles), as calculated for the $n=1$
image state on (a) Cu(111) and (b) Cu(100).}
\end{figure}

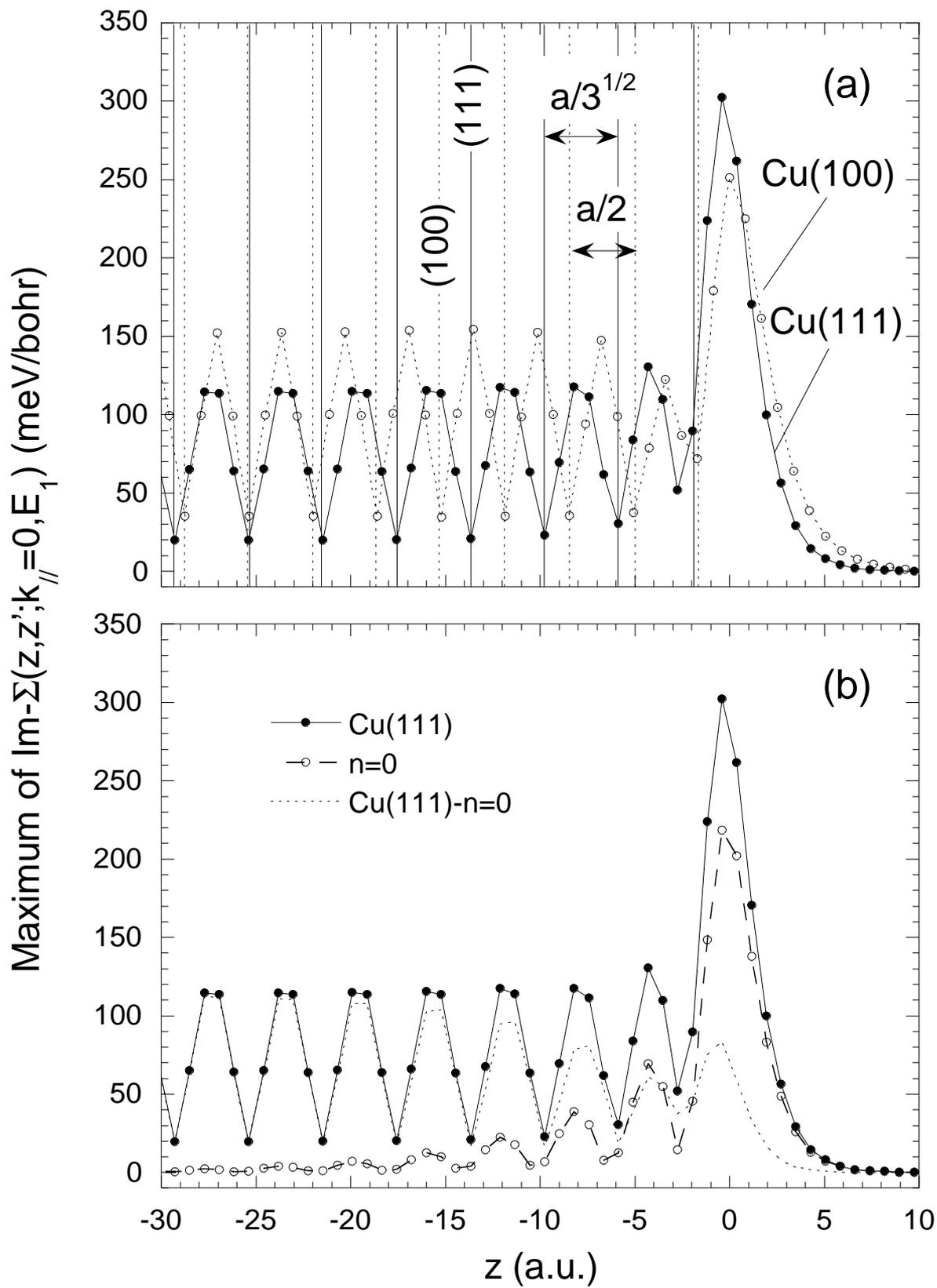
\begin{figure}
\caption[]{\label{figmax}
(a) Maximum value of ${\rm Im}\left[-\Sigma(z, z';{\bf k}_\parallel=0;E_1)\right]$
(see the text) for the $n=1$ image state on Cu(111) (solid line) and Cu(100)
(dotted line). Vertical lines represent the position of atomic layers in
Cu(111) (solid lines) and Cu(100) (dotted lines). (b) As in (a), for the
separate contributions to the  $n=1$ image state on Cu(111). The solid line with circles
describes the total maximum value of ${\rm Im}\left[-\Sigma(z, z';{\bf
k}_\parallel=0;E_1)\right]$, the dashed line represents
the contribution coming from the decay into the intrinsic surface state, and
the dotted line, the contribution from the decay into bulk states.
The solid line represents the result of replacing the realistic model-potential
$\phi_f(z)$ final wave functions entering Eq. (\ref{b1}) by the self-consistent {\it
jellium} LDA eigenfunctions of the one-electron Kohn-Sham hamiltonian but with the
restriction that only final states with energy $\varepsilon_f$ lying below the projected
band gap are allowed.}
\end{figure}

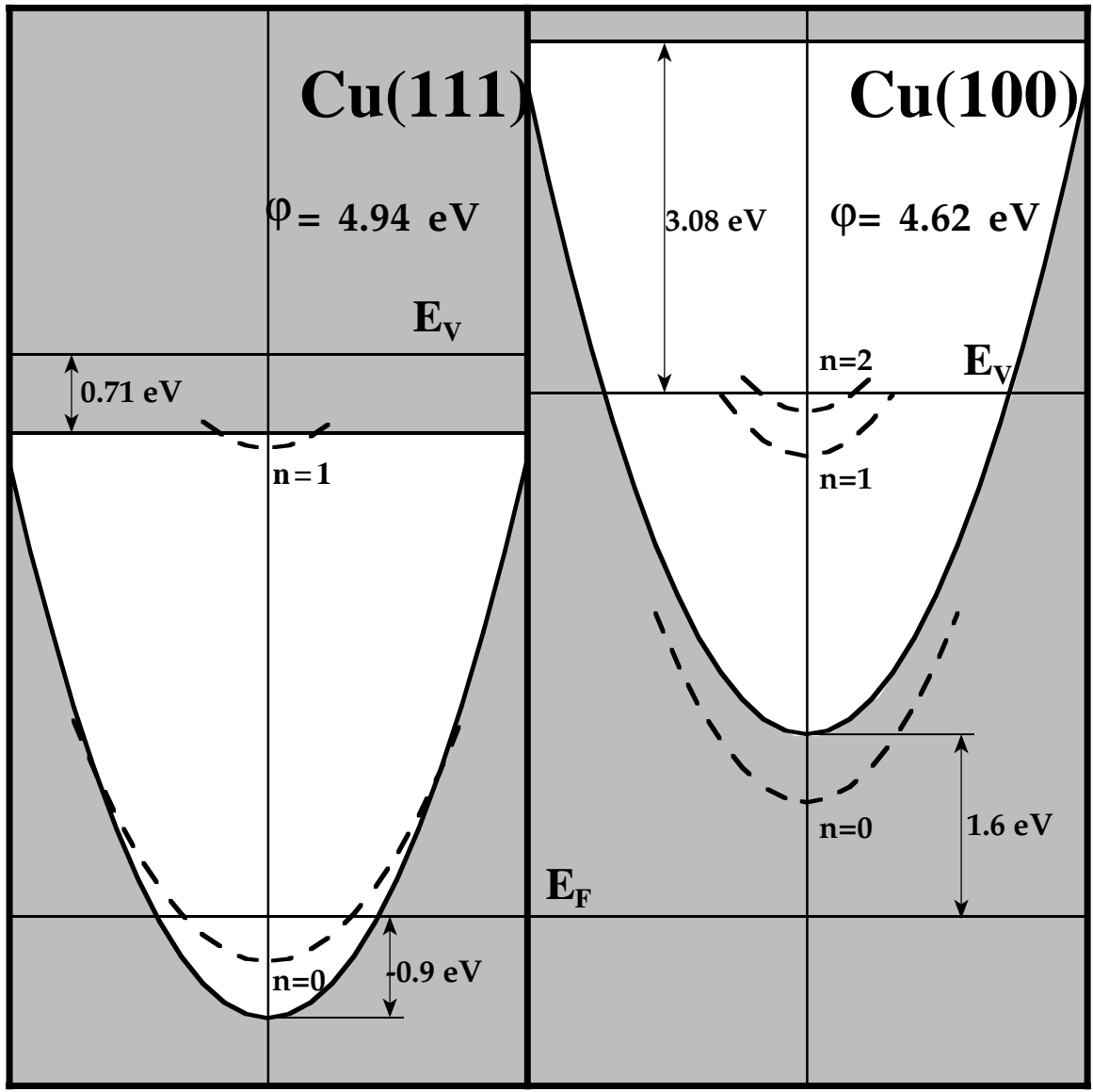
\begin{figure}
\caption[]{\label{figstr}
Schematical representation of the electronic structure of Cu(111)
and Cu(100).}
\end{figure}

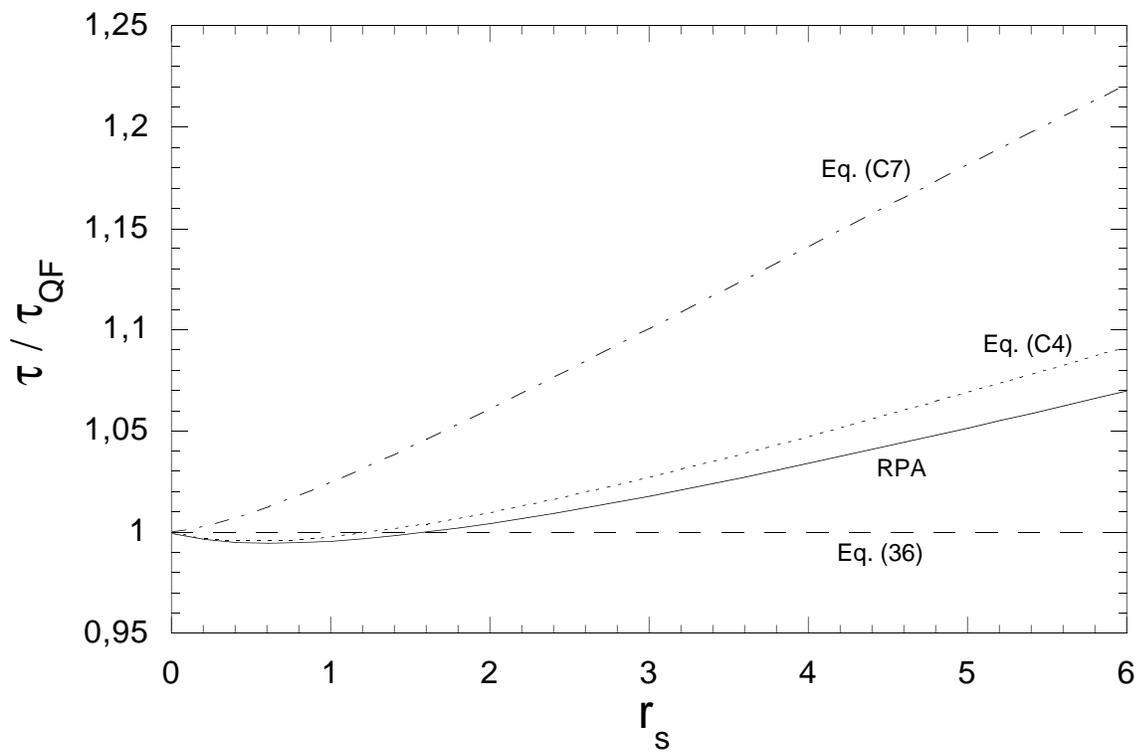
\begin{figure}
\caption[]
{\label{figt_r}Ratio $\tau/\tau_{QF}$ between the lifetime $\tau$ evaluated in
various approximations and the lifetime $\tau_{QF}$ of Eq. (\ref{eq114}), versus
the electron-density parameter $r_s$, as obtained for hot electrons in the
vicinity of the Fermi surface ($E\sim E_F$). The solid line represents the result
obtained, within RPA, from either Eq. (\ref{eq46}) or Eq. (\ref{eq104p}). Results
obtained from Eqs. (\ref{eq106}) and (\ref{Quinn}) are represented
by dotted and dashed-dotted lines, respectively. The dashed line 
represents the result obtained from Eq. (\ref{qf1}).}
\end{figure}


\begin{references}

\bibitem{Ritchie1} R. H. Ritchie, F. W. Garber, M. Y. Nakai, and R. D.
Birkhoff,
in {\it Advances in Radiation Biology}, Vol. 3, p. 1, edited by L. G. Augenstein, R.
Mason, and M.
Zelle (Academic Press, New York, 1969). The factor
$(1-\gamma^2/3)$ in the  numerator of $l_{eo}$
of Eq. (2) of this reference must be replaced by a factor
$\sqrt{1-\gamma^2/3}$.
\bibitem{Ritchie2} R. H. Ritchie, C. J. Tung, V. E. Anderson, and J. C.
Ashley, Radiat. Res. {\bf 64}, 181 (1975).
\bibitem{Petek1} H. Petek and S. Ogawa, Prog. Surf. Sci. {\bf 56}, 239
(1998).
\bibitem{Kanter} H. Kanter, Phys. Rev. B {\bf 1}, 522 (1970).
\bibitem{Powell1} C. J. Powell, Surf. Sci. {\bf 44}, 29 (1974);
Scanning Electron. Microsc. {\bf 184}, 1649 (1984). Surf. Interface Anal. {\bf 7}, 263
(1985). 
\bibitem{Sernelius1} L. I. Johansson and B. E. Sernelius, Phys. Rev. B
{\bf 50},
16817 (1994).
\bibitem{Pendry} J. B. Pendry, in {\it Photoemission and the Electronic
Properties of
Surfaces}, edited by B. Feuerbacher, B. Fitton, and R. F. Willis (Wiley, 1978).
\bibitem{Plummer0} W. Eberhardt and E. W. Plummer, Phys. Rev. B {\bf
21}, 3245
(1980).
\bibitem{Levinson} H. J. Levinson, F. Greuter, and E. W. Plummer,
Phys. Rev. B {\bf 27}, 727 (1983).
\bibitem{Goldmann} A. Goldmann, W. Altmann, and V. Dose, Solid State
Commun. {\bf 79}, 511 (1991).
\bibitem{Himpsel0} A. Santoni and F. J. Himpsel, Phys. Rev. B {\bf 43},
1305 (1991).
\bibitem{Ortega} D. Li, P. A. Dowben, J. E. Ortega, and F. J. Himpsel,
Phys. Rev. B {\bf 47}, 12895 (1993).
\bibitem{Bokor} J. Bokor, Science {\bf 246}, 1130 (1989).
\bibitem{Haight} R. Haight, Surf. Sci. Rep. {\bf 21}, 275 (1995).
\bibitem{exp1} C. A. Schmutenmaer, M. Aeschlimann, H. E. Elsayed-Ali, R.
J. D. Miller, D. A. Mantell, J. Cao, and Y. Gao, Phys. Rev. B {\bf 50},
8957 (1994).
\bibitem{Knoesel} E. Knoesel, A. Hotzel, T. Hertel, M. Wolf, and G.
Ertl, Surf. Sci. {\bf 368}, 76 (1996).
\bibitem{Aes96} M. Aeschlimann, M. Bauer, S. Pawlik, Chem. Phys. $\bf205$, 127
(1996).
\bibitem{Petek0} S. Ogawa, H. Nagano and H. Petek, Phys. Rev. B {\bf 55},
1 (1997).
\bibitem{exp4} J. Cao, Y. Gao, R. J. D. Miller, H. E. Elsayed-Ali, D. A.
Mantell, Phys. Rev. B {\bf 56}, 1099 (1997).
\bibitem{exp6} E. Knoesel, A. Hotzel and M. Wolf, Phys. Rev. B {\bf 57},
12812 (1998).
\bibitem{Goldm} A. Goldmann, R. Matzdorf, and F. Theilmann, Surf. Sci. {\bf 414}, L932
(1998).
\bibitem{Cao} J. Cao, Y. Gao, H. E. Elsayed-Ali, R. J. D. Miller, and D.
A. Mantell, Phys. Rev. B {\bf 58}, 10948 (1998).
\bibitem{Reuter98} K. Reuter, P. L. de Andr\' es, F. J. Garc\'\i a-Vidal, F. Flores, U. 
Hohenester, and P. Kocevar, Europhys. Lett. {\bf 45}, 181 (1999).
\bibitem{Ludeke93}
R. Ludeke and A. Bauer, Phys. Rev. Lett. $\bf71$, 1760 (1993).
\bibitem{Aes98}
M. Bauer, S. Pawlik, M. Aeschlimann, Proc. of the SPIE 3272, 201 (1998).
\bibitem{exp5} M. Aeschlimann, M. Bauer, S. Pawlik, W. Weber, R.
Burgermeister, D. Oberli, and H. C. Siegmann, Phys. Rev. Lett. {\bf 79}, 5158
(1997).
\bibitem{Aes98.2}
M. Aeschlimann,  R. Burgermeister, S. Pawlik, M. Bauer, D. Oberli, W. Weber,
J. Elec. Spec. Rel. Phen. {\bf 88}-{\bf 91}, 179 (1998).
\bibitem{Petek2} W. Nessler, S. Ogawa, H. Nagano, H. Petek, J.
Shimoyama, Y. Nakayama, and K. Kishio, Phys. Rev. Lett. {\bf 81}, 4480 (1998).
\bibitem{Hertel} T. Hertel, E. Knoesel, M. Wolf, and G. Ertl, Phys. Rev.
Lett. {\bf 76}, 535 (1996).
\bibitem{Wolf} M. Wolf, E. Knoesel, and T. Hertel, Phys. Rev. B {\bf 54} 5295 (1996); M.
Wolf, Surf. Sci. {\bf 377-379}, 343 (1997).
\bibitem{Knoesel0} E. Knoesel, A. Hotzel, and M. Wolf, J. Electron.
Spectrosc. Relat. Phenom. {\bf 88-91}, 577 (1998).
\bibitem{Lingle} R. L. Lingle Jr., N. H. Ge, R. E. Jordan, J. D.
McNeil, and C. B. Harris, Chem. Phys. {\bf 205}, 191 (1996).
\bibitem{Hofer} U. H\"ofer, I. L. Shumay, Ch. Reuss, U. Thomann, W.
Wallauer, and Th. Fauster, Science {\bf 277}, 1480 (1997).
\bibitem{Harris} J. D. McNeil, R. L. Lingle Jr., N. H. Ge,
C. M. Wong, R. E. Jordan, and C. B. Harris, Phys. Rev. Lett.
{\bf 79}, 4645 (1997); C. B. Harris, N.-H. Ge, R. L. Lingle Jr., J. D.
McNeill, and C.  M. Wong, Annu. Rev. Phys. Chem. {\bf 48}, 711 (1997).
\bibitem{Shumay} I. L. Shumay, U. H\"ofer, Ch. Reuss, U. Thomann, W.
Wallauer, and Th. Fauster,  Phys. Rev. B {\bf 58}, 13974 (1998).
\bibitem{Plummer} W. Plummer, Science {\bf 277}, 1447 (1997).
\bibitem{note1} Electron relaxation times due to coupling with the
lattice are found
to be on a picosecond scale (see, e.g., W. S. Fann, R. Storz, and H. W.
K. Tom,
Phys. Rev. B {\bf 46}, 13592 (1992)).
\bibitem{QF} J. J. Quinn and R. A. Ferrell, Phys. Rev. {\bf 112}, 812
(1958).
\bibitem{Ritchie59} R. H. Ritchie, Phys. Rev. {\bf 114}, 644 (1959). The
$1/2$
factor in
front of $z^2$ in the expansion of $f_1$ just before Eq. (6.15) of this
reference must be
replaced by $1/3$, as done in a subsequent paper [R. H. Ritchie and J.
C.
Ashley,
J. Phys. Chem. Solids {\bf 26}, 1689 (1963)].
\bibitem{Quinn62} J. J. Quinn, Phys. Rev. {\bf 126}, 1453 (1962).
\bibitem{Adler} S. L. Adler, Phys. Rev. {\bf 130}, 1654 (1963).
\bibitem{Quinn63} J. J. Quinn, Appl. Phys. Lett. {\bf 2}, 167 (1963).
\bibitem{Ashley} R. H. Ritchie and J. C. Ashley, J. Phys. Chem. Solids
{\bf
26}, 1689 (1963).
\bibitem{Kleinman} L. Kleinman, Phys. Rev. B {\bf 3}, 2982 (1971).
\bibitem{Penn0} D. R. Penn, Phys. Rev. B {\bf 13}, 5248 (1976).
\bibitem{kk} C. A. Kukkonen and A. W. Overhauser, Phys. Rev. B {\bf 20},
550
(1979).
\bibitem{Penn1} D. R. Penn, Phys. Rev. B {\bf 22}, 2677 (1980).
\bibitem{Lundqvist} B. I. Lundqvist, Phys. Stat. Sol. {\bf 32}, 273
(1969).
\bibitem{Shelton} J. C. Shelton, Surf. Sci. {\bf 44}, 305 (1974).
\bibitem{Ashley0} J. C. Ashley and R. H. Ritchie, Phys. Status Solidi B
{\bf
62}, 253 (1974);
{\bf 83}, K159 (1977).
\bibitem{Tung2} J. C. Ashley, C. J. Tung, and R. H. Ritchie, Surf. Sci.
{\bf
81}, 409 (1979).
\bibitem{Tung1} C. J. Tung and R. H. Ritchie, Phys. Rev. B {\bf 16},
4302
(1977).
\bibitem{Spicer} W. F. Krolikowski and W. E. Spicer, Phys. Rev. {\bf
185}, 882
(1969).
\bibitem{Tung3} C. J. Tung, J. C. Ashley, and R. H. Ritchie, Surf. Sci.
{\bf
81}, 427 (1979).
\bibitem{Ashley1} J. C. Ashley, J. Electron Spectrosc. {\bf 28}, 177
(1982); {\bf 46}, 199 (1988); {\bf 50}, 323 (1990);
J. Phys-Condens. Mat. {\bf 3}, 2741 (1991).
\bibitem{Leckey} J. Szajman and R. C. G. Leckey {\bf 23}, 83 (1981).
\bibitem{Liljequist} D. Liljequist, J. Phys. D: Appl. Phys. {\bf 16},
1567 (1983).
\bibitem{Salvat} F. Salvat, J. D. Mart\'\i nez, R. Mayol, and J.
Parellada, J. Phys. D:
Appl. Phys. {\bf 18}, 299 (1985); J. M. Fern\'andez-Varea, R. Mayol, F.
Salvat, and
D.  Liljequist, J. Phys.: Condens. Matter {\bf 4}, 2879 (1992).
\bibitem{Penn2} D. R. Penn, Phys. Rev. B {\bf 35}, 482 (1987).
\bibitem{Tanuma} S. Tanuma, C. J. Powell, and D. R. Penn, Surf.
Interface Anal. {\bf
11}, 577 (1988); {\bf 17}, 911 (1991); {\bf 17}, 927 (1991); {\bf 20},
77 (1993);
{\bf 21}, 165 (1994).
\bibitem{Ding} C. M. Kwei, Y. F. Chen, C. J. Tung, and J. P. Wang, Surf.
Sci. {\bf
293}, 202 (1993); Z.-J. Ding and R. Shimizu, Scanning {\bf 18}, 92
(1996); T.
Boutboul, A. Akkerman, A. Breskin, and R. Chechik, J. Appl. Phys. {\bf
79}, 6714
(1996); A. Akkerman, T. Boutboul, A. Breskin, R. Chechik, A.
Gibrekhterman, and Y.
Lifshitz, Phys. Stat. Sol. (b) {\bf 198}, 769 (1996).
\bibitem{Powell2} C. J. Powell and A. Jablonski, to appear in J. Phys.
Chem.
Ref. Data.
\bibitem{Igor0} I. Campillo, PhD. Thesis, University of the Basque
Country, 1999 (unpublished); I. Campillo, J. M. Pitarke, A. Rubio, and P. M. Echenique, to be
published.
\bibitem{Igor} I. Campillo, J. M. Pitarke, A. Rubio, E. Zarate, and P.
M. Echenique, submitted to Phys. Rev. Lett..
\bibitem{Echenique2} P. M. Echenique, F. Flores, and F. Sols, Phys. Rev.
Lett.
{\bf 55}, 2348 (1985).
\bibitem{Echenique3} P. L. de Andr\'es, P. M. Echenique, and F. Flores,
Phys.
Rev. B {\bf 35}, 4529 (1987); Phys. Rev. B {\bf 39}, 10356 (1989).
\bibitem{Uranga} M. E. Uranga, A. Rivacoba, and P. M. Echenique, Progr.
Surf. Sci.
{\bf 42}, 67 (1993).
\bibitem{Rundgren} J. Rundgren and G. Malmstrom, J. Phys. C {\bf 10},
4671 (1977).
\bibitem{Echenique0} P. M. Echenique, PhD. Thesis, University of Cambridge, 1976
(unpublished); P. M. Echenique and J. B. Pendry, J. Phys. C {\bf 11},
2065 (1978).
\bibitem{Smith} N. V. Smith, Phys. Rev. {\bf 32}, 3549 (1985); N. V.
Smith, Rep. Progr. Phys. {\bf 51}, 1227 (1988).
\bibitem{Himpsel} F. J. Himpsel, Comments Cond. Matter Phys. {\bf 12},
199 (1986).
\bibitem{Borstel} G. Borstel and G. Th\"{o}rner, Surf. Sci. Rep.
{\bf 8}, 1 (1988).
\bibitem{Echenique90} P. M. Echenique and J. B. Pendry, Prog. Surf. Sci.
{\bf 32}, 111
(1990).
\bibitem{Fauster} Th. Fauster and W. Steinmann, in {\it Electromagnetic
Waves: Recent Development in Research}, Vol. 2, p. 350, edited by P. Halevi
(Elsevier, Amsterdam, 1995).
\bibitem{Osgood} R. M. Osgood Jr. and X. Wang, Solid State Phys. {\bf
51}, 1
(1997).
\bibitem{Gao} S. Gao and B. I. Lundqvist, Prog. Theor. Phys. Suppl. {\bf
106},
405 (1991);
Solid State Commun. {\bf 84}, 147 (1992).
\bibitem{Chulkov1} E. V. Chulkov, I. Sarria, V. M. Silkin, J. M.
Pitarke, and P. M. Echenique, Phys. Rev. Lett. {\bf 80}, 4947 (1998).
\bibitem{Osma} J. Osma, I. Sarria, E. V. Chulkov, J. M. Pitarke, and
P. M. Echenique, Phys. Rev. B {\bf 59}, 10591 (1999).
\bibitem{Silkin} E. V. Chulkov, J. Osma, I. Sarria, V. M. Silkin, and J.
M. Pitarke, to appear in Surf. Sci..
\bibitem{Sarria} I. Sarria, J. Osma, E. V. Chulkov, J. M. Pitarke,
and P. M.
Echenique, submitted to Phys. Rev. B.
\bibitem{Chulkov2} E. V. Chulkov, V. M. Silkin, and P. M. Echenique,
Surf. Sci.
{\bf 391}, L1217 (1997).
\bibitem{Sung} C. C. Sung and R. H. Ritchie, Phys. Rev. A {\bf 28}, 674
(1983).
\bibitem{Zaremba} C. D. Hu and E. Zaremba, Phys. Rev. B {\bf 37}, 9268
(1988).
\bibitem{Esbensen} H. Esbensen and P. Sigmund, Ann. Phys. (N. Y.) {\bf
201}, 152 (1990).
\bibitem{Pitarke1} J. M. Pitarke, R. H. Ritchie, P. M. Echenique, and E.
Zaremba, Europhys. Lett. {\bf 24}, 613 (1993); J. M. Pitarke, R. H. Ritchie, and
P. M. Echenique, Nucl. Instrum. Methods B {\bf 79}, 209 (1993).
\bibitem{Pitarke2} J. M. Pitarke, R. H. Ritchie, and P. M. Echenique,
Phys. Rev. B {\bf 52}, 13883 (1995).
\bibitem{Bergara} J. M. Pitarke, A. Bergara, and R. H. Ritchie, Nucl. Instrum. Methods B {\bf
99}, 87 (1995); A. Bergara, I. Campillo, J. M. Pitarke, and P. M. Echenique, Phys. Rev. {\bf
56}, 15654 (1997).
\bibitem{Wang} N.-P. Wang and J. M. Pitarke, Phys. Rev. A {\bf 56}, 2913
(1997); N.-P. Wang and J. M. Pitarke, Phys. Rev. A {\bf 57}, 4053 (1998); N.-P. Wang and J.
M. Pitarke, Nucl. Instrum. Methods B {\bf 135}, 92 (1998).
\bibitem{Barkas} W. H. Barkas, W. Birnbaum, and F. M. Smith, Phys. Rev.
{\bf 101}, 778 (1956); W. H. Barkas, N. J. Dyer, and H. H. Heckman, Phys. Rev.
Lett. {\bf 11}, 26 (1963); {\bf 11}, 138(E) (1963).
\bibitem{Andersen} L. H. Andersen, P. Hvelplund, H. Knudsen, S. P.
Moller, J. O. P. Pedersen, E. Uggerhoj, K. Elsener, and E. Morenzoni, 
Phys. Rev. Lett. {\bf 62}, 1731 (1989).
\bibitem{Shiff} L. I. Shiff, {\it Quantum Mechanics} (McGraw-Hill,
London, 1985); A. Galindo and P. Pascual, {\it Quantum Mechanics} 
(Springer-Verlag, Berlin, 1990).
\bibitem{Echenique1} P. M. Echenique, F. Flores, and R. H. Ritchie,
Solid State Phys. {\bf 43}, 229 (1990).
\bibitem{Pines} D. Pines and P. Nozieres, {\it The Theory of Quantum
Liquids, Volume I: Normal Fermi Liquids} (Addison-Wesley, New York, 1989).
\bibitem{Lindhard} J. Lindhard, K. Dan. Vidensk. Selsk. Mat.- Fys.
Medd. {\bf  28}, No. 8 (1954).
\bibitem{Pines2} D. Pines, {\it Elementary excitations in solids}
(Addison-Wesley, New York, 1963).
\bibitem{Fetter} A. L. Fetter and J. D. Walecka, {\it Quantum Theory of
Many-Particle Systems} (McGraw-Hill, New York, 1971); A. A. Abrikosov, L. P. Gorkov, and I.
E. Dzyaloshinski, {\it Methods of Quantum Field Theory in Statistical Physics} (Dover
Publications, New York, 1975).
\bibitem{Hubbard} J. Hubbard, Proc. R. Soc. London, Ser. A {\bf 240},
539 (1957); {\bf 243}, 336 (1957).
\bibitem{Kleinman68} L. Kleinman, Phys. Rev. {\bf 172}, 383 (1968).
\bibitem{HL} L. Hedin and B. I. Lundqvist, J. Phys. C {\bf 4}, 2064
(1971).
\bibitem{Hedin} L. Hedin, Phys. Rev. {\bf 139}, A796 (1965); L. Hedin
and S. Lundqvist, Solid State Phys. {\bf 23}, 1 (1969).
\bibitem{GW} F. Aryasetiawan and O. Gunnarsson, Rep. Prog. Phys. {\bf
61}, 237 (1998).
\bibitem{Rice} T. M. Rice, Ann. Phys. (N.Y.) {\bf 31}, 100 (1965).
\bibitem{Sernelius2} G. D. Mahan and B. E. Sernelius, Phys. Rev. Lett.
{\bf 62}, 2718
(1989).
\bibitem{Mahan1} G. D. Mahan, {\it Many-Particle Physics}, 2nd ed.
(Plenum, New
York, 1990).
\bibitem{Mahan2} G. D. Mahan, Comments Cond. Mat. Phys. {\bf 16}, 333
(1994).
\bibitem{note2} If Eq. (\ref{eq94}) for the screened interaction is taken in combination
with the RPA density-response function of Eq. (\ref{eq98}), then one obtains Eq.
(\ref{eq46}) with the RPA dielectric function. 
\bibitem{note3} $1\,{\rm a. u.}=658\,{\rm meV}\,{\rm fs}$.
\bibitem{Lindhard2} J. Lindhard and M. Scharff, K. Dan. Vidensk. Selsk.
Mat.-Fys. Medd {\bf 27}, No. 15 (1953); J. Lindhard, M. Scharff, and H. E. Schiott, {\it
ibid.} {\bf 33}, No. 14 (1963).
\bibitem{Ashcroft0} N. W. Ashcroft and N. D. Mermin, {\it Solid State Physics}
(Saunders, Philadelphia, 1976).
\bibitem{Palik} E. D. Palik, {\it Handbook of Optical Constants of
Solids} (Academic Press, New York, 1985); E. D. Palik, {\it Handbook 
of Optical Constants of Solids II} (Academic Press, New York, 1991).
\bibitem{Zarate} E. Zarate, PhD. Thesis, University of the Basque 
Country, 1999 (unpublished); E. Zarate, P. Apell, and P. M. Echenique, submitted to Phys.
Rev. B.
\bibitem{Berglund64} C. N. Berglund and W. E. Spicer, Phys. Rev. {\bf 136}, 1030
(1964).
\bibitem{Kane67} E. O. Kane, Phys. Rev. {\bf159}, 624 (1967).
\bibitem{Penn85} D. R. Penn, S. P. Apell, and S. M. Girvin, Phys. Rev. Lett.
{\bf 55}, 518 (1985), {\it ibid.}  Phys. Rev. B {\bf32}, 7753 (1985).
\bibitem{Drouhin97} H.-J. Drouhin, Phys. Rev. B, {\bf56}, 14886 (1997).
\bibitem{Kohn1} P. Hohenberg and W. Kohn, Phys. Rev. {\bf 136}, B864
(1964).
\bibitem{Kohn2} W. Kohn and L. Sham, Phys. Rev. {\bf 140}, A1133 (1965).
\bibitem{Troullier} N. Troullier and J. L. Martins, Phys. Rev. B {\bf
43}, 1993 (1991).  
\bibitem{note4} These authors\cite{Petek0} approximated the FEG dielectric function in
$|\epsilon_{\Gb,\Gb}(\qb,\omega)|^{-2}$ within the static Thomas-Fermi model, as in
Eq. (\ref{eq108}), and with the screening length $q_{TF}^{-1}=0.47\AA$ taken from the
actual DOS at the Fermi level. Though the actual DOS at the Fermi level being larger
than the corresponding DOS from the FEG model makes the screening stronger
($q_{TF}>q_{TF}^{FEG}$, $q_{TF}^{FEG}$ being the screening length from the FEG
model), the increase in the actual number of states available for real transitions
yields lifetimes that are below the FEG calculation.    
\bibitem{tazp32} I. E. Tamm, Z. Phys. {\bf 76}, 849 (1932).
\bibitem{shpr39} W. Shockley, Phys. Rev. {\bf 56}, 317 (1939).
\bibitem{josmprb83} P. D. Johnson and N. V. Smith, Phys. Rev. B {\bf 27},
2527 (1983).
\bibitem{doalprl84} V. Dose, W. Altmann, A. Goldmann, U. Kolac,
and J. Rogozik, Phys. Rev. Lett. {\bf 52}, 1919 (1984).
\bibitem{sthiprl84} D. Straub and F. J. Himpsel, Phys. Rev.
Lett. {\bf 52}, 1922 (1984).
\bibitem{hiorprb92} F. J. Himpsel and J. E. Ortega, Phys. Rev. B
{\bf 46}, 9719 (1992).
\bibitem{gihaprl85} K. Giesen, F. Hage, F. J. Himpsel, H. J. Riess,
and W. Steinmann, Phys. Rev. Lett. {\bf 55}, 300 (1985).
\bibitem{Fujimoto} R. W. Schoenlein, J. G. Fujimoto, G. L. Eesley, and
T. W. Capehart, Phys. Rev. Lett. {\bf 61}, 2596 (1988); R. W. Schoenlein, J.
G. Fujimoto, G. L. Eesley, and T. W. Capehart, Phys. Rev. B {\bf 43}, 4688 (1991).
\bibitem{scfiprb92} S. Schuppler, N. Fischer, Th. Fauster, and
W. Steinmann, Phys. Rev. B {\bf 46}, 13539 (1992).
\bibitem{wafass97} W. Wallauer and Th. Fauster, Surf. Sci. {\bf 374}, 44
(1997).
\bibitem{wehuprl85} M. Weinert, S. L. Hulbert, and P. D. Johnson,
Phys. Rev. Lett. {\bf 55}, 2055 (1985).
\bibitem{orecprb86} M. Ortu\~{n}o and P. M. Echenique, Phys. Rev. B
{\bf 34}, 5199 (1986).
\bibitem{pelass86} J. B. Pendry, C. G. Larsson, and P. M. Echenique,
Surf. Sci. {\bf 166}, 57 (1986).
\bibitem{lesuprb87} Z. Lenac, M. \v{S}unji\'{c}, H. Conrad, and
M. E. Kordesch, Phys. Rev. B {\bf 36}, 9500 (1987).
\bibitem{liwaprb89} S. \AA. Lindgren and L. Walld\'{e}n, Phys. Rev. B
{\bf 40}, 11546 (1989).
\bibitem{smchprb89} N. V. Smith, C. T. Chen, and M. Weinert,
Phys. Rev. B {\bf 40}, 7565 (1989).
\bibitem{faap94} Th. Fauster, Appl. Phys. A {\bf 59}, 639 (1994).
\bibitem{justss97} L. Jurczyszyn and M. St\c{e}\'{s}licka, Surf. Sci.
{\bf 376}, L424 (1997).
\bibitem{chsitbp1} E. V. Chulkov, V. M. Silkin, and P. M. Echenique, to
be published.
\bibitem{hujoprb86} S. L. Hulbert, P. D. Johnson, M. Weinert, and
R. F. Garrett, Phys. Rev. B {\bf 33}, 760  (1986).
\bibitem{neinel92} M. Nekovee and J. E. Inglesfield, Europhys.
Lett. {\bf 19}, 535 (1992).
\bibitem{egheprl92} A. G. Eguiluz, M. Heinrichsmeier, A. Fleszar,
and W. Hanke, Phys. Rev. Lett. {\bf 68}, 1359 (1992).
\bibitem{necrprl93} M. Nekovee, S. Crampin, and J. E. Inglesfield,
Phys. Rev. Lett. {\bf 70}, 3099 (1993).
\bibitem{sichpss94} V. M. Silkin and E. V. Chulkov, Phys. Solid
State {\bf 36}, 404 (1994) [Rus. Fiz. Tverd. Tela {\bf 36}, 736
(1994)].
\bibitem{ligaprb94} Z. Li and S. Gao, Phys. Rev. B {\bf 50}, 15349
(1994).
\bibitem{heflprb98} M. Heinrichsmeier, A. Fleszar, W. Hanke, and
A. Eguiluz, Phys. Rev. B {\bf 57}, 14974 (1998).
\bibitem{Kevan} S. D. Kevan (Ed.), {\it Angle-Resolved Photoemission,
Theory and Current  Applications} (Elsevier, Amsterdam, 1992).
\bibitem{wereap} M. Weinelt, Ch. Reuss, M. Kutschera, U. Thomann, I. L.
Shumay, Th. Fauster,
U. H\"ofer, F. Theilmann, and A. Goldmann, to appear in Appl. Phys. B.
\bibitem{mcbaprb95} B. A. McDougall, T. Balasubramanian, and E. Jensen,
Phys. Rev. B {\bf 51}, 13891
(1995).
\bibitem{thmaprb97} F. Theilmann, R. Matzdorf, G. Meister, and A.
Goldmann, Phys. Rev.
B {\bf 56}, 3632 (1997).
\bibitem{reshprl99} Ch. Reuss, I. L. Shumay, V. Thomann, M. Kutschera,
M. Weinelt, Th. Fauster, and U. H\"{o}fer, Phys. Rev. Lett. {\bf 82},
153 (1999).
\bibitem{chsitbp2} E. V. Chulkov, V. M. Silkin, and P. M. Echenique, to
be published.
\bibitem{degprb97} J. J. Deisz and A. G. Eguiluz, Phys. Rev. B {\bf 55}, 9195 
(1997).
\bibitem{note5} The oscillatory behaviour within the bulk is dictated by the
periodicity of the amplitude of final-state wave functions $\phi_f(z)$ in periodic
crystals.
\bibitem{note6} The intraband contribution to the linewidth, coming from transitions
between the states $\phi_{\kb_\parallel,n=1}$ and $\phi_{\kb_\parallel',n=1}$ with
$\kb_\parallel\neq\kb_\parallel'$ has not been included in this calculation.
\bibitem{Padowitz} D. F. Padowitz, W. R. Merry, R. E. Jordan, and C. B. Harris,
Phys. Rev. Lett. {\bf 69}, 3583 (1992).
\bibitem{Fischer94}
N. Fischer, S. Schuppler, Th. Fauster, and W. Steinmann, Surf. Sci. {\bf 314}, 89
(1994).
\bibitem{Fischer93}
N. Fischer, S. Schuppler, R. Fischer, Th. Fauster, and W. Steinmann, Phys. Rev.
B {\bf 47}, 4705 (1993).
\bibitem{Fischer96}
R. Fischer and Th. Fauster, Surf. Rev. Lett. {\bf 3}, 1783 (1996).
\bibitem{mejoss93} W. Merry, R. E. Jordan, D. E. Padowitz, and
C. B. Harris, Surf. Sci. {\bf 295}, 393 (1993).
\bibitem{Neill96}
J. D. McNeill, R. L. Lingle Jr., R. E. Jordan, D. F. Padowitz, and 
C. B. Harris, J. Chem. Phys. {\bf 105}, 3883 (1996).
\bibitem{Wolf97}
M. Wolf, Surf. Sci. {\bf 377}-{\bf 379}, 343 (1997).
\bibitem{Harris97}
C. B. Harris, N. H. Ge, R. L. Lingle Jr., J. D. McNeill, and C. M. Wong, 
Annu. Rev. Phys. Chem. {\bf 48}, 711 (1997).
\bibitem{Berthold}
W. Berthold, I. L. Shumay, P. Feulner, and U. Hofer, to be published.
\bibitem{Donath94}
M. Donath, Surf. Sci. Rep. {\bf 20}, 251 (1994).
\bibitem{Passek95}
F. Passek, M. Donath, K. Ertl, and V. Dose, Phys. Rev. Lett. {\bf 75}, 2746
(1995).
\bibitem{Nekovee93}
M. Nekovee, S. Crampin, and J. E. Inglesfield, Phys. Rev. Lett. {\bf 70}, 3099
(1993).
\bibitem{Rossi96}
S. De Rossi, F. Cicacci, and S. Crampin, Phys. Rev. Lett. {\bf 77}, 908 (1996).
\bibitem{Jacinto} J. Osma (unpublished).
\bibitem{Diau} E. W. G. Diau, J. L. Herek, Z. H. Kim, and  A. H. Zewail,
Science {\bf 279}, 847 (1998).
\bibitem{Petaccia} L. Petaccia, L. Grill, M. Zangrando, and S. Modesti, Phys. Rev.
Lett. {\bf 82}, 386 (1999).
\bibitem{Reiners} T. Reiners and H. Haberland, Phys. Rev. Lett. {\bf
77}, 2440 (1996).
\bibitem{Domps} A. Domps, P. G. Reinhard, and E. Suraud, Phys. Rev. Lett. {\bf 81},
5524 (1998).
\bibitem{BLA} B. L. Altshuler, Y. Gefen, A. Kamenev, and L. S. Levitov, Phys. Rev.
Lett. {\bf 78}, 2803 (1997).
\bibitem{excitons} M. Rohfling and S. G. Louie, Phys. Rev. Lett. {\bf
81}, 2312 (1998); L. X. Benedict, E. L. Shirley, and R. B. Bohn,
Phys. Rev. Lett. {\bf 80}, 4514 (1998); S. Albretcht, L. Reining, R. Del
Sole, and G. Onida, Phys. Rev. Lett. {\bf 80}, 4510 (1998).
\bibitem{shirley} E. Shirley, Phys. Rev. B {\bf 54}, 7758 (1996); {\bf 54}, 8411 (1996).
\bibitem{holm} B. Holm and U. Von Barth, Phys. Rev. B {\bf 57}, 2108
(1998).
\bibitem{schone} W.-D. Sch\"one and A. G. Eguiluz, Phys. Rev. Lett. {\bf
81}, 1662 (1998); A. G. Eguiluz and W.-D. Sch\"one, Mol. Phys. {\bf 94}, 87 (1998).
\bibitem{Aryasetiawan} F. Aryasetiawan, Phys. Rev. B {\bf 46}, 13051 (1992).
\bibitem{Gerhardt} W. Gerhardt, S. Marquardt, N. Schroeder, and S. Weis, Phys. Rev. B
{\bf 58}, 6877 (1998).
\bibitem{Louie} M. S. Hybertsen and S. G. Louie, Phys. Rev. Lett.
{\bf 55}, 1418 (1985); M. S. Hybertsen and S. G. Louie, Phys. Rev. B {\bf 34}, 5390 (1986);
R. W. Godby, M. Schl\"uter, and L. J. Sham,
Phys. Rev. Lett. {\bf 56}, 2415 (1986); Phys. Rev. B {\bf 37}, 10159
(1988).
\bibitem{Jorge} J. S. Dolado, M. A. Cazalilla, A. Rubio, and
P. M. Echenique (unplublished).
\bibitem{Rojas} H. N. Rojas, R. W. Godby and R. J. Needs, Phys. Rev. Lett.
{\bf 74}, 1827 (1995).
\bibitem{Blase} X. Blase, A. Rubio, M. L. Cohen, and S. G. Louie,
Phys. Rev. B {\bf 52}, 2225 (1995).
\bibitem{imtime} X. Blase, A. Rubio, M. L. Cohen, and S. G. Louie (unpublished).
\bibitem{Lutz} L. Steinbeck, A. Rubio, I. D. White, and R.W. Godby, to be published.
\bibitem{Xu} S. Xu, J. Cao, CC. Miller, D. A. Mantell, R. J. D. Miller, 
and Y. Gao, Phys. Rev. Lett. {\bf 76}, 483 (1996).
\bibitem{Zheng} L. Zheng and S. Das Sarma, Phys. Rev. Lett. {\bf 77},
1410 (1996); S. Xu, J. Cao, CC. Miller, D. A. Mantell, R. J. D. Miller, and Y.
Gao, Phys. Rev. Lett. {\bf  77}, 1411 (1996).
\bibitem{Guinea} J. Gonzalez, F. Guinea, and M. A. H. Vozmediano,
Phys. Rev. Lett. {\bf 77}, 3589 (1996).
\bibitem{Xu1} S. Xu, CC. Miller, S. J. Diol, Y. Gao, D. A. Mantell,
M. G. Mason, A. A. Muenter, L. I.  Sharp, B. A. Parkinson, and R. J. D. Miller,
Chem. Phys. Lett. {\bf 272}, 209 (1997).
\bibitem{MiguelAngel} M. A. Cazalilla, A. Rubio, and P.M. Echenique
(unpublished).
\bibitem{Palmer} P. Laitenberger and R. E. Palmer, Phys. Rev. Lett. {\bf 76},
1952 (1996).
\bibitem{Runge} E. Runge and E. K. U. Gross, Phys. Rev. Lett. {\bf 52},
997 (1984).
\bibitem{Gross2} E. K. U. Gross and W. Kohn, Phys. Rev. Lett. {\bf 55},
2850 (1985); {\bf 57}, 923 (E) (1986).
\bibitem{Gross3} M. Petersilka, U. J. Gossmann and E. K. U. Gross,
Phys. Rev. Lett. {\bf 76}, 1212 (1996).
\bibitem{tddft} E. K. U. Gross, F. J. Dobson, and M. Petersilka, {\it
Density Functional Theory} (Springer, New York, 1996).
\bibitem{Flocard} H. Flocard, S. E. Koonin, and M. S. Weiss, Phys. Rev. C
{\bf 17}, 1682 (1978).
\bibitem{review} A. Rubio, J. A. Alonso, X. Blase, and S. G. Louie,
Int. J. Mod. Phys. B {\bf 11}, 2727 (1997), and references therein.
\bibitem{Yabana} K. Yabana and G. F. Bertsch, Phys. Rev. B {\bf 54}, 4484
(1996).
\bibitem{rby} A. Rubio and G. F. Berstch (unpublished).
\bibitem{Wigner} E. P. Wigner, Phys. Rev. {\bf 46}, 1002 (1934); Trans.
Faraday Soc. {\bf 34}, 678 (1938).
\bibitem{Vosko} S. H. Vosko, L. Wilk, and M. Nusair, Can. J. Phys. {\bf
58}, 1200 (1980).
\bibitem{Perdew} J. Perdew and A. Zunger, Phys. Rev. B {\bf 23}, 5048
(1981).
\bibitem{CA} D. M. Ceperley and B. J. Alder, Phys. Rev. Lett. {\bf 45},
1196 (1980).
\bibitem{Alder} C. Bowen, G. Sugiyama, and B. J. Alder, Phys. Rev. {\bf 50}, 14838 (1994).
\bibitem{Ceperley} S. Moroni, D. M. Ceperley, and G. Senatore, Phys.
Rev. Lett. {\bf 75}, 689 (1995).
\bibitem{Singwi} K. S. Singwi, M. P. Tosi, R. H. Land, and A.
Sjolander,  Phys.
Rev. {\bf 176}, 589 (1968); K. S. Singwi, A. Sjolander, M. P. Tosi, and
R. H. Land,
{\it ibid.}, {\bf 1}, 1044 (1970); K. S. Singwi and M. P. Tosi, Solid
State Phys. {\bf
36}, 177  (1981).
\bibitem{Ichimaru} K. Utsumi and S. Ichimaru, Phys. Rev. B {\bf 22},
5203 (1980); S.
Ichimaru, Rev. Mod. Phys. {\bf 54}, 1017 (1982).
\bibitem{Gold} A. Gold and L. Camels, Phys. Rev. B {\bf 48}, 11622
(1993).
\bibitem{Brosens} F. Brosens, L. F. Lemmens, and J. T. Devreese, Phys.
Status Solid B {\bf 74}, 45 (1976); J. T. Devreese, F. Brosens, 
and L. F. Lemmens, Phys. Rev. B {\bf 21}, 1349 (1980).
\bibitem{Ashcroft} C. F. Richardson and N. W. Ashcroft, Phys. Rev. B {\bf 50},
8170 (1994).
\bibitem{Larson} B. C. Larson, J. Z. Tischler, E. D. Isaacs, P. Zschack,
A. Fleszar, and A. G. Eguiluz, Phys. Rev. Lett. {\bf 77}, 1346 (1996).
\bibitem{Nozieres} P. Nozieres, {\it Theory of Interacting Fermi Liquids}
(Benjamin, New York, 1962).
\bibitem{Migdal}
V. M. Galitski and A. B. Migdal, Sov. Phys. JEPT {\bf 7}, 96 (1958).
\bibitem{Baym} G. Baym and L. P. Kadanoff, Phys. Rev. {\bf 124}, 287
(1961); L. P. Kadanoff and G. Baym, {\it Quantum Statistical Mechanics}
(Benjamin, New York, 1962).
\bibitem{dft} R. M. Dreizler and E. K. U. Gross, {\it Density Functional Theory,
an Approach to the Quantum Many Body Problem} (Springer, Berlin, 1990).
\bibitem{Sole} R. Del Sole, L. Reining, and R. W. Godby, Phys. Rev. B {\bf 49}, 8024
(1994).
\end{references}
\end{document}